\documentclass[pre,showpacs,floatfix,twocolumn,superscriptaddress]{revtex4}
\usepackage{amsmath}
\usepackage{amssymb}
\usepackage{latexsym}
\usepackage{multirow}
\usepackage{color}
\usepackage[hidelinks]{hyperref}
\usepackage{mathtools}
\usepackage{comment}
\usepackage{graphicx}
\usepackage{float}
\usepackage[scr=boondox]{mathalfa}

\newcommand{\ess}{\mathscr{s}}

\begin{document}

\title{Jamming and flocking in the restricted active Potts model}
%Characterization and control of jamming transition for self-propelled particles with discrete symmetry: The restricted active Potts model}

\author{Mintu Karmakar}
\email{mcsmk2057@iacs.res.in}
\affiliation{School of Mathematical \& Computational Sciences, Indian Association for the Cultivation of Science, Kolkata -- 700032, India.}

\author{Swarnajit Chatterjee}
\email{swarnajit.chatterjee@uni-saarland.de}
\affiliation{Center for Biophysics \& Department for Theoretical Physics, Saarland University, D-66123 Saarbr{\"u}cken, Germany.}

\author{Matthieu Mangeat}
\email{mangeat@lusi.uni-sb.de}
\affiliation{Center for Biophysics \& Department for Theoretical Physics, Saarland University, D-66123 Saarbr{\"u}cken, Germany.}

\author{Heiko Rieger}
\email{heiko.rieger@uni-saarland.de}
\affiliation{Center for Biophysics \& Department for Theoretical Physics, Saarland University, D-66123 Saarbr{\"u}cken, Germany.}
\affiliation{INM – Leibniz Institute for New Materials, Campus D2 2, D-66123 Saarbrücken, Germany.}

\author{Raja Paul}
\email{raja.paul@iacs.res.in}
\affiliation{School of Mathematical \& Computational Sciences, Indian Association for the Cultivation of Science, Kolkata -- 700032, India.}

%%%%%%%%%%%%%%%%%%%%%%%%%%%%%%%%%
%%%%%%%%%%%% ABSTRACT %%%%%%%%%%%%%%%%
%%%%%%%%%%%%%%%%%%%%%%%%%%%%%%%%
\begin{abstract}
We study the active Potts model with either site occupancy restriction or on-site repulsion to explore jamming and kinetic arrest in a flocking model. The incorporation of such volume exclusion features leads to a surprisingly rich variety of self-organized spatial patterns. While bands and lanes of moving particles commonly occur without or under weak volume exclusion, strong volume exclusion along with low temperature, high activity, and large particle density facilitates jams due to motility-induced phase separation. Through several phase diagrams, we identify the phase boundaries separating the jammed and free-flowing phases and study the transition between these phases which provide us with both qualitative and quantitative predictions of how jamming might be delayed or dissolved. We further formulate and analyze a hydrodynamic theory for the restricted APM which predicts various features of the microscopic model.
\end{abstract}

\maketitle 
%%%%%%%%%%%%%%%%%%%%%%%%%%%%%%%%%
%%%%%%%%%%%% INTRODUCTION %%%%%%%%%%%%%%%%
%%%%%%%%%%%%%%%%%%%%%%%%%%%%%%%%

\section{Introduction}
Flocking is a collective phenomenon of active matter \cite{Marchetti2013,Shaebani2020}, occurs in ensembles of self-propelled particles and denotes the emergence of ordered motion of large clusters, called flocks. It plays a significant role in a wide range of systems across disciplines including physics, biology, ecology, social sciences, and neurosciences and is abundantly observed in nature: from human crowds \cite{Bottinelli2016,Helbing1995}, mammalian herds \cite{Garcimartin2015}, bird flocks \cite{Ballerini2008}, and fish schools \cite{Beccoa2006,Calovi2014} to unicellular organisms such as amoebae, bacteria \cite{Steager2008,Peruani2012}, collective cell migration in dense tissues \cite{Giavazzi2018}, and subcellular structures including cytoskeletal filaments and molecular motors \cite{Schaller2010,Sumino2012,Sanchez2012}.

The paradigmatic model to describe the flocking transition is the Vicsek
model (VM) \cite{viscek,toner,toner-2,toner-3,solon-vicsek} in which  individual particles tend to align with the average direction of the motion of their  neighbors. At low noise and high density, the particles cluster and move collectively in a common direction, which is the landmark of flocking. The transition from the gas phase at high noise and low density to the polar ordered Toner-Tu phase at low noise and high density, displaying  long-range order by coherent motion of all particles, is first order~\cite{Chate2004}. But, in contrast to conventional first-order phase transition scenarios, the coexistence phase of the VM shows either multiple bands of collectively moving particles denoted as microphase separation \cite{solon-vicsek,solon-vm2} or a polar-ordered cross sea phase \cite{ihle}. A large number of variations of the original VM has been studied (see Ref.~\cite{Shaebani2020} for an overview), and recently discretized versions of it have been analyzed: the active Ising model (AIM)~\cite{AIM,AIM-2,ishibashi2022}, active Potts model (APM)~\cite{APM,APM-2,ishibashi2022}, and active clock model (ACM) \cite{ACM,ACM-2}, in which particles move on a lattice in two (AIM) or more (APM, ACM) 
discrete directions. Common to those is that the microphase separation 
in the coexistence region is replaced by macrophase separation - a single collectively 
moving band \cite{AIM,AIM-2,APM,APM-2,ACM}.

Self-propelled particles with repulsive interactions, such as active Brownian particles (ABPs)~\cite{Romanczuk2012}, show at high density and high P\'eclet number another cluster state, different from flocking via alignment, denoted as motility-induced phase separation (MIPS) \cite{mips}.
Consequently, the interplay between alignment and repulsive 
interactions could lead to complex phase diagrams, as was
demonstrated for the VM with repulsive particle interactions~\cite{chate2003} or ABPs with alignment interactions~\cite{ignacio,sese2018}.

In this paper we address the question of what happens to the phase  diagram of the APM when on-site interactions between the particles are present, either in the form of hard or soft core repulsion, or in the form of a maximal occupancy larger than one of each site, the restricted APM (rAPM). In analogy with the VM with particle-particle repulsion one expects the phase diagram for the rAPM to be enriched by at least one MIPS state, but also other kinetically arrested states can occur as active lattice gas models with repulsive interactions \cite{peruani2011}. These are sometimes denoted as ``jammed states'', not to be confused with jamming occurring in active glasses~\cite{Berthier2019}.

The paper is organized as follows. In Sec.~\ref{s2}, we define the restricted active Potts model and provide details of the simulation protocols. Section.~\ref{s3} presents our results for three different versions of the on-site
repulsion: (1) the `maximum particle per site' (MPS) is restricted to one akin to the active lattice gas (ALG)~\cite{ALG}, (2) hard-core restriction or ${\rm MPS} > 1$, and (3) soft-core repulsion. 
In Sec.~\ref{secHydro}, we present the hydrodynamic description of our model. Finally, we conclude this paper with a summary and discussion of the results in Sec.~\ref{s4}.

%%%%%%%%%%%%%%%%%%%%%%%%%%%%%%%%%
%%%%%%%%%%%% MODEL & SIMULATION DETAILS %%%%%%%%%%%%%%%%
%%%%%%%%%%%%%%%%%%%%%%%%%%%%%%%%
\section{Modeling and simulation details}
\label{s2}
\begin{figure}[t]
\centering
\includegraphics[width=\columnwidth]{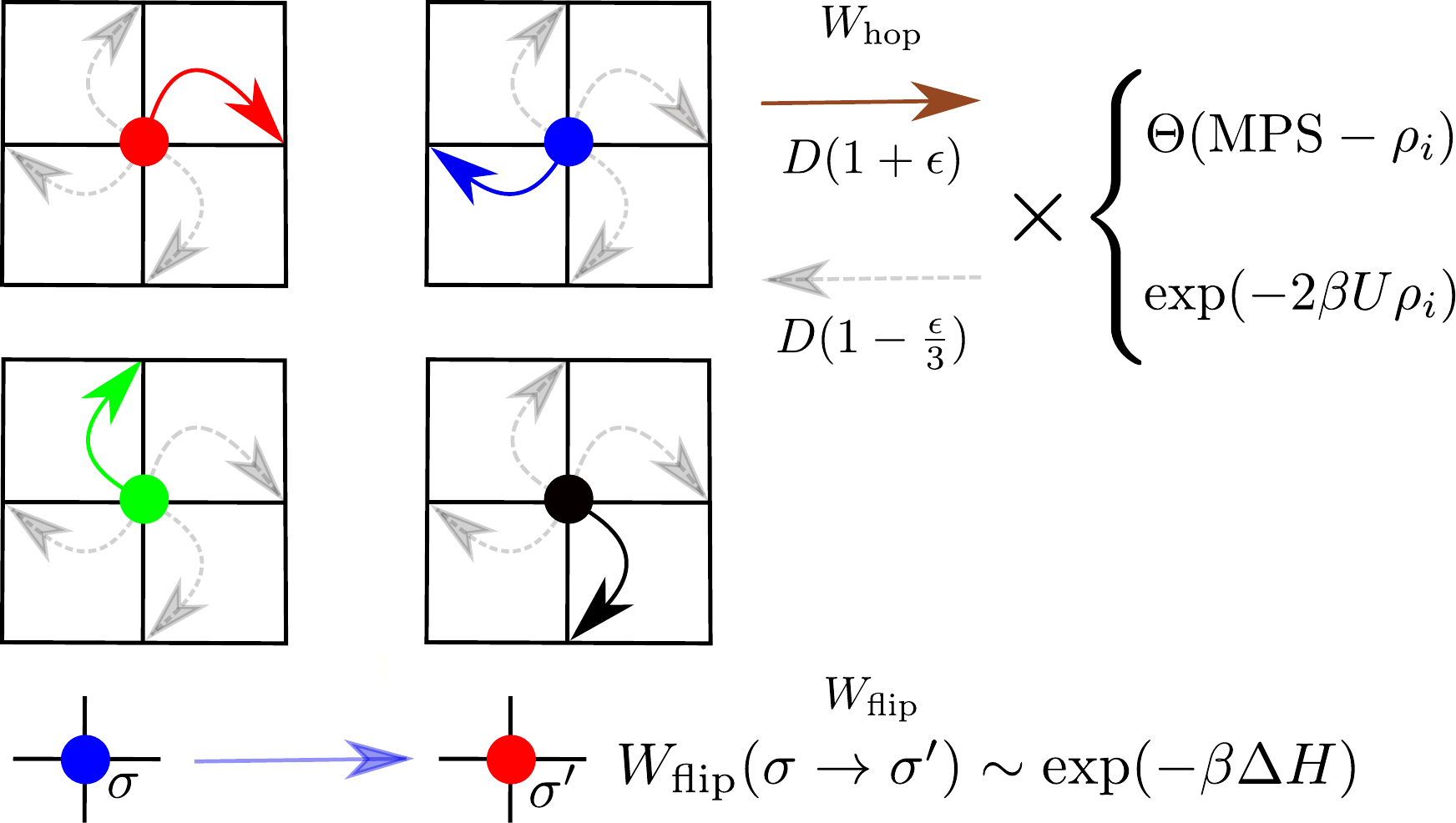}
\caption{(color online) (above) Sketch of the $q=4$ rAPM showing biased hopping (solid bent arrow) to a neighboring lattice site with rate $D(1+\epsilon)$ and unbiased hopping (faint bent arrow) along the remaining directions with rate $D(1-\epsilon/3)$. Hard-core restriction is represented by a Heaviside step function $\Theta({\rm MPS}-\rho_i)$, and soft-core repulsion is represented by the probability $\exp(-2\beta U \rho_i)$. Red, green, blue, and black circles represent particles of state $\sigma = \{1, 2, 3, 4\}$, respectively. (below) Illustration of on-site flipping of a particle from $\sigma=3$ to $\sigma=1$ with flipping rate $W_{\rm flip}$.}
\label{fig1}
\end{figure}

We consider an ensemble of $N$ particles defined on a two-dimensional square lattice of size $L^2$ with periodic boundary conditions applied on both sides, where $L$ is the linear lattice dimension. The average particle density $\rho_0$ is then defined as $\rho_0=N/L^2$. The model is built upon the APM~\cite{APM,APM-2} in which the dynamics is governed either by the on-site flipping of the internal spin state or by nearest-neighbor hopping. We also now propose restrictions on particle hopping. We suggest three types of mutually exclusive restrictions: a particle is allowed to hop to its neighbor if (1) that neighbor is empty or (2) the population of the neighboring site is less than the maximum occupation per site (hard-core restriction), or (3) the hopping is allowed with a probability (soft-core repulsion). A schematic diagram of this arrangement is shown in Fig.~\ref{fig1}. 

The spin state of the $k$-th particle on lattice site $i$ is denoted by $\sigma^k_i$, with an integer value in $[1,q]$, while the number of particles in state $\sigma$ on-site $i$ is $n^\sigma_i$. The local density on-site $i$ is then defined by $\rho_i=\sum_{\sigma=1}^q n^\sigma_i$, counting the total number of particles on the site. The Hamiltonian of a $q$-state APM is defined as $H_{\rm APM}=\sum_i H_i$ decomposed as the sum of local Hamiltonian $H_i$ \cite{APM,APM-2}:
\begin{equation}
\label{Hapm}
H_i=-\frac{J}{2\rho_i}\sum_{k=1}^{\rho_i}\sum_{l\ne k}(q\delta_{\sigma_i^k,\sigma_i^l}-1),
\end{equation}
where $J=1$ is the coupling between the neighboring sites. We take $q=4$. A particle at site $i$ with state $\sigma$ either flips to another state $\sigma^\prime$ or hops to any of the neighboring sites (as permitted by the different restriction protocols). The local magnetization corresponding to state $\sigma$ at site $i$ is defined as $m_i^{\sigma}$:
\begin{equation}
\label{Hapm_mag}
m_i^{\sigma}=\sum_{j=1}^{\rho_i}\frac{q\delta_{\sigma,\sigma_i^j}-1}{q-1} \, .
\end{equation}

\subsection{On-site alignment or flipping dynamics}
A particle at site $i$ with state $\sigma$ can flip and align with another state $\sigma^\prime$, and therefore, flipping is a purely on-site phenomenon. From Eq.~\eqref{Hapm}, one can calculate the local energy difference before and after the flipping. From Refs.~\cite{APM,APM-2}, the expression of the energy difference reads
\begin{equation}
\label{delH}
\Delta H_i=H_{i}^{\rm new}-H_i^{\rm old}=\frac{qJ}{\rho_i}(n_i^{\sigma}-n_i^{\sigma^\prime}-1) \, .
\end{equation}
The flipping is then accepted with the rate
\begin{align}
\label{flipeq}
W_{\rm flip}(\sigma\to\sigma^\prime)&=\gamma \exp(-\beta \Delta H_i) \nonumber \\
&=\gamma \exp\left[-\frac{q\beta J}{\rho_i}(n_i^\sigma-n_i^{\sigma^\prime}-1)\right] ,
\end{align}
where $\beta=T^{-1}$ is the inverse temperature. It should be noted that for MPS = 1, only one particle is allowed per site and consequently, on-site alignment interaction is absent for this limit of the model. From Eq.~\eqref{flipeq}, $n_i^{\sigma}=1$ and $n_i^{\sigma^\prime}=0$ leads to $\Delta H_i=0$, and hence we have $W_{\rm flip}(\sigma\to\sigma^\prime)= \gamma$ as the flipping rate of particles for MPS = 1. For hard-core restriction and soft-core repulsion, we take $\gamma=1$. 

\subsection{Biased diffusion or hopping dynamics}
The biased diffusion mechanism is similar to the process described in Refs. \cite{APM,APM-2}. A particle with state $\sigma$ hops to a direction $p$ with rate
\begin{equation}
\label{whop}
W_{\rm hop}(\sigma,p)=D\left(1+\epsilon\frac{q\delta_{\sigma,p}-1}{q-1}\right) \, ,
\end{equation}
%where $\epsilon$ is the self-propulsion parameter. $\epsilon=0$ signifies pure diffusion and $\epsilon=q-1$ signifies complete self-propulsion. 
where $\epsilon$ ($0 \leqslant \epsilon \leqslant q-1$) is the self-propulsion parameter. At $\epsilon=q-1$, particles move purely ballistically, resulting in complete self-propulsion, while $\epsilon=0$ corresponds to the absence of self-propulsion. However, with $\epsilon = 0$, particles are not passive and can still diffuse on the lattice [see Eq.~(\ref{whop})] but without any bias. This differs from the VM where the zero-velocity limit corresponds to immobile particles with a dynamics reminiscent of the XY model. Let us also mention that for $\epsilon>0$, the system is out-of-equilibrium but the model is not at equilibrium even when $\epsilon=0$. Following Ref. \cite{AIM-2}, it can be shown using Kolmogorov’s criterion~\cite{kolmogorov} that the system does not satisfy detailed balance with respect to any distribution, i.e., the product of the forward transition rates differs from that of the reverse order. The system would be effectively an equilibrium system only when the particles would cease to move (being jammed), for instance, when the density is high and hopping restrictions are strong.

Under the purely repulsive hard-core exclusion, biased diffusion is subjected to the maximum number of particles allowed per site set by the parameter MPS, and Eq.~\eqref{whop} gets modified in the following way:
\begin{equation}
\label{HC_eq}
W^{\rm HC}_{\rm hop}(\sigma,p)=W_{\rm hop}(\sigma,p) \Theta({\rm MPS}-\rho_i) \, ,
\end{equation}
where $\Theta({\rm MPS}-\rho_i)$ is a Heaviside step function and is defined as:
\[
\Theta({\rm MPS}-\rho_i) =
\begin{cases}
1, & \text{for} \quad \rho_i < \text{MPS} \\
0, & \text{otherwise} \, .
\end{cases}
\]
$\rho_i$ is the particle number at a neighboring site $i$ to which a hopping is attempted. MPS = 1 is a special case under the hard-core exclusion category where a move to the neighbor is possible only if that site is empty. Therefore, unlike hard-core and soft-core repulsions, on-site interactions between the particles are absent for MPS = 1. This can also be thought of as an asymmetric simple exclusion process (ASEP) in which particles perform biased random walks under the hard-core repulsion that two particles cannot occupy the same site at a given time (MPS $>1$ is the non-lattice-gas variety of the ASEP). Therefore, MPS = 1 constructs the simple-exclusion ``active lattice gas''~\cite{ALG} version of the APM.

A soft-core repulsion would allow a particle hop to a neighboring site $i$ from a randomly chosen site depending on the change in the local field. The local field is defined by: 
\begin{equation}
\label{BH}
V(\rho_i)=U\rho_i(\rho_i -1) \, ,
\end{equation}
where $U$ is an interaction coefficient that can be attractive ($U<0$) or repulsive ($U>0$). After hopping, the local field with $\rho_i+1$ particles at site $i$ becomes
$V(\rho_i+1)=U\rho_i(\rho_i +1)$.
Particle hopping to site $i$ is then accepted with probability:
\begin{align}
P &= \min[1,\exp(-\beta \Delta V)] \\ \nonumber 
&=\min[1,\exp(-2\beta U \rho_i)] \, ,
\end{align}
where $\Delta V=V(\rho_i+1)-V(\rho_i)=2U\rho_i$. Then, for soft-core repulsion, the modified form of Eq.~\eqref{whop} can be written as:
\begin{equation}
\label{SC_eq}
W^{\rm SC}_{\rm hop}(\sigma,p)= W_{\rm hop}(\sigma,p) \exp(-2\beta U \rho_i) \, .
\end{equation}
$U$ symbolizes the restriction strength that regulates particle accumulation on a lattice site. Note that, $U \leqslant 0$ denotes $\exp(-\beta\Delta V) \geqslant 1$ which physically signifies an attractive field where particles can freely crowd into a site similar to the APM \cite{APM}. In this paper, we consider only repulsive interactions, $U>0$, acting in the limit $U\to\infty$ 
like volume exclusion. 

\subsection{Simulation Details}
\label{simul}
Simulation evolves in the unit of Monte Carlo steps (MCSs) $\Delta t$ resulting form a microscopic time $\Delta t/N$, $N$ being the total number of particles. During $\Delta t/N,$ a randomly chosen particle either updates its spin state with probability $p_{\rm flip}=W_{\rm flip}\Delta t$ or hops to one of the neighboring sites with probability $p_{\rm hop}=W_{\rm hop}\Delta t$. For $q$-state APM, an expression for $\Delta t$ can be obtained by minimizing the probability of nothing happens $p_{\rm wait}=1-(p_{\rm hop}+p_{\rm flip})$ \cite{APM}, 
\begin{equation}\label{delt}
\Delta t=[qD+\exp(q\beta J)]^{-1} \, .
\end{equation}
This hybrid Monte Carlo dynamics was used previously in the simulations of the AIM~\cite{AIM,AIM-2}, APM~\cite{APM,APM-2}, and ACM~\cite{ACM}. Instead of computing $\Delta t$ from the minimum of $p_{\rm wait}$ for systems that have small transition probabilities and therefore large $p_{\rm wait}$ (one has to generate random numbers until the chosen transition is accepted), one can also apply a Gillespie-like algorithm where one computes the time at which the next event will take place in the system.

\section{Numerical Results}
\label{s3}
In this section, we present the numerical simulation results of the $q = 4$ rAPM with MPS = 1, hard-core restriction (${\rm MPS}>1$), and soft-core repulsion. The models are simulated on a square lattice of linear size $L = 100$ with periodic boundary conditions, where individual particle states $\sigma = \{1, 2, 3, 4\}$ correspond to the movement directions right, up, left, and down, respectively. Simulations are performed for various control parameters: $D = 1$ is kept constant throughout the simulations, $\beta=1/T$ regulates the noise in the system, $\rho_0=N/L^2$ defines the average particle density, and self-propulsion parameter $\epsilon$ dictates the effective velocity of the particles. Starting from a homogeneous initial condition, the Monte Carlo algorithm (Sec.~\ref{simul}) evolves the system under various control parameters until the stationary distribution is reached. Following this, measurements are carried out and thermally averaged data are recorded.
%%%%%%%%%%%%%%%%%%%%%%%%%%%%%%%%%
%%%%%%%%%%%% MPS 1 %%%%%%%%%%%%%%%%
%%%%%%%%%%%%%%%%%%%%%%%%%%%%%%%%

\subsection{MPS = 1 (ALG version of the rAPM)}
\label{ALG}
In this segment, we present the results for $q=4$ state APM with MPS = 1. Following Ref.~\cite{ALG} and Eq.~\eqref{flipeq}, $\gamma$ represents the flipping parameter with flipping probability $\gamma\Delta t$. We then define the P\'eclet number Pe as
\begin{equation} \label{peclet}
{\rm Pe} = \frac{v}{\sqrt{D\gamma}} \, ,
\end{equation} 
%{\color{blue}(The definition of ${\rm Pe}$ is independent of lattice size. When we describe data, however, P\'eclet is multiplied by lattice size.)}
where $v=4 D \epsilon/3$ is the self-propulsion velocity in the hydrodynamic limit of the four-state APM \cite{APM}, and we get ${\rm Pe}=(4\epsilon/3)\sqrt{D/\gamma}$. As Pe is proportional to $\epsilon$, for small Pe, diffusion dominates, and the effect of self-propulsion becomes negligible. Conversely, the effect of activity gets more and more pronounced as Pe increases.

\begin{figure}[t]
\centering
\includegraphics[width=\columnwidth]{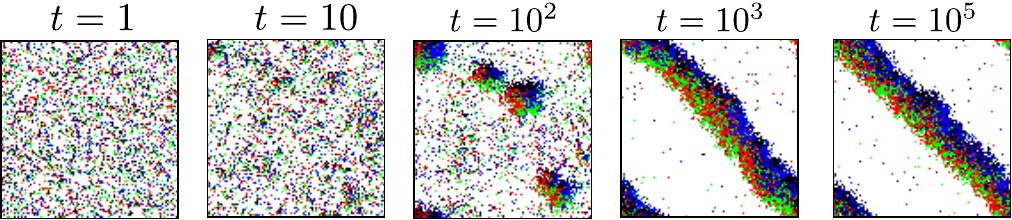}
\caption{(color online) Time evolution snapshots of the rAPM with MPS = 1 displaying MIPS. The right boundary of the diagonal high-density band is due to particles of state $\sigma=3$ (blue) and $\sigma=4$ (black) and the left boundary is due to particles of state $\sigma=1$ (red) and $\sigma=2$ (green). Parameters: Pe = 50 and $\rho_0=0.3$.}
\label{fig2}
\end{figure}

In Fig.~\ref{fig2}, snapshots demonstrate MIPS via the time evolution of the
rAPM starting from a random initial configuration. Initially, the self-propelled particles (SPPs) nucleate stable clusters (where domains of all the four states can be visible) and coarsen and coalesce at later times ($t=10^5$) to phase separate into a diagonal solid phase that stabilizes in a steady state and a gas phase, a consequence of MIPS. A careful examination of the diagonal domain ($t=10^5$) reveals that the right (upper) and left (lower) domain boundaries are formed by multiple opposite spin states ($e.g.$ for a diagonal band spanning from the bottom-right corner to the top-left corner, the right domain boundary is always formed by particles with $\sigma=3$ and 4, and the left domain boundary is formed by particles of $\sigma=1$ and 2). A two-state variant of this model having $\sigma=1$ and 3 would result in a vertically jammed band~\cite{ALG} and a combination of $\sigma=2$ and 4 would result in a horizontally jammed band. Therefore, the high-density diagonal band arises when the steady state culminates into orthogonally directed clusters intercepted by oppositely directed clusters. See the supplemental movie~\cite{SM}, which demonstrates the formation of a diagonal band (parameters: $\rho_0=0.45$, $\epsilon=2.7$, Pe = 113).

\begin{figure}[!htbp]
\centering
\includegraphics[width=\columnwidth]{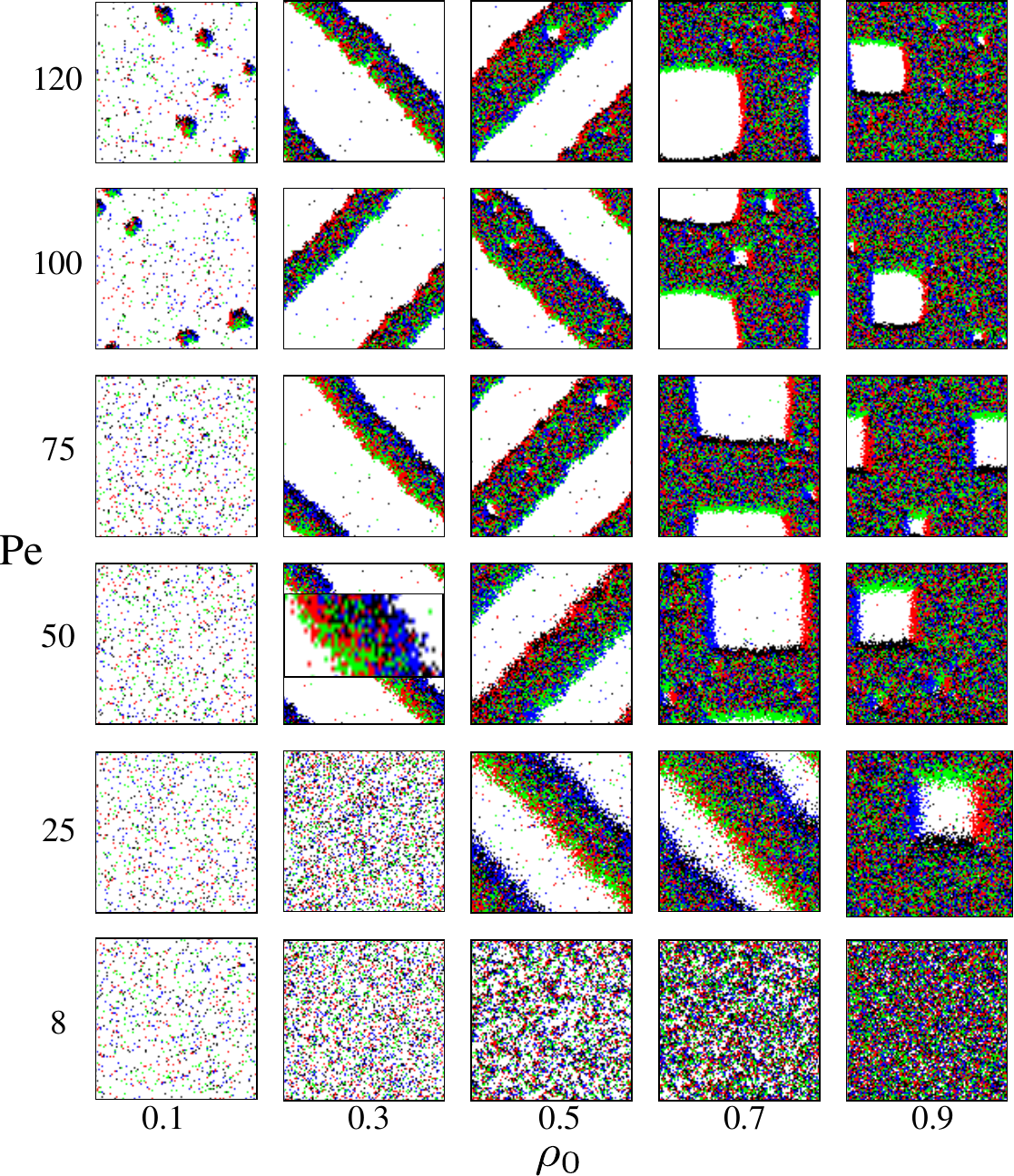}
\caption{(color online) ${\rm Pe}-\rho_0$ phase diagram (for $\gamma=10$ and varying $\epsilon$) for MPS=1 illustrated by snapshots at time $t=10^5$. As density increases, the system undergoes a transition from the disordered gas phase to a MIPS state~\cite{mips} for ${\rm Pe}>8$. At the intermediate densities, we observe solid diagonal bands where a small section of the band (Pe = 50, $\rho_0=0.3$) is magnified to provide a detailed structure of the domain.}
\label{fig3}
\end{figure}

The steady-state behavior of the rAPM with MPS = 1 is illustrated in Fig.~\ref{fig3} by representative late-stage snapshots as a function of propulsion strength Pe and average particle density $\rho_0$. As a function of density, the system undergoes a transition from a disordered gaseous phase to the MIPS state at both intermediate and large Pe. At intermediate densities, we observe diagonal high-density bands on a disordered background, which we discussed in detail in the context of Fig.~\ref{fig2}. As density is increased, the area of the high-density solid phase also increases by shrinking the area of the gaseous phase. The boundaries of such a square gaseous region are always formed by particles of oppositely moving states ($\sigma=1$: right vertical boundary, $\sigma=3$: left vertical boundary, $\sigma=2$: up horizontal boundary, and $\sigma=4$: down horizontal boundary) whose preferred hopping directions are unavailable due to the restriction. Nevertheless, density fluctuation happens from the domain boundaries to the square void space (therefore, the center of mass of the jammed clusters shifts position with time, though slowly). The internal structure of the high-density phases in both these MIPS states are similar, i.e., orientationally disordered (due to the lack of the alignment interactions), and can be described as amorphous solid. As mentioned before, the formation of these jammed amorphous solid phases is a consequence of MIPS. MIPS refers to the spontaneous phase separation of a system of SPPs with purely repulsive interactions (and without any attractive interaction) into coexisting dense and dilute phases. The physics of MIPS can be understood as slowing of SPPs due to enhanced crowding when the local density of SPPs increases in some part of the system due to fluctuation, as shown in Fig.~\ref{fig3}.

In Ref.~\cite{peruani2011}, a related model with alignment interactions 
extending to the nearest neighbors was studied and shown to exhibit jams, gliders, and bands, which we do not observe here since the MPS=1 case only has hard-core interactions. In addition, the hopping dynamics in our model depends on the self-propulsion parameter $\epsilon$ ($0 \leqslant \epsilon \leqslant q-1$) and for $\epsilon<q-1$, we always have a nonzero hopping rate along the nonpreferred directions, whereas in Ref.~\cite{peruani2011}, a particle can hop only in the preferred direction.

\begin{figure}[t]
\centering
\includegraphics[width=\columnwidth]{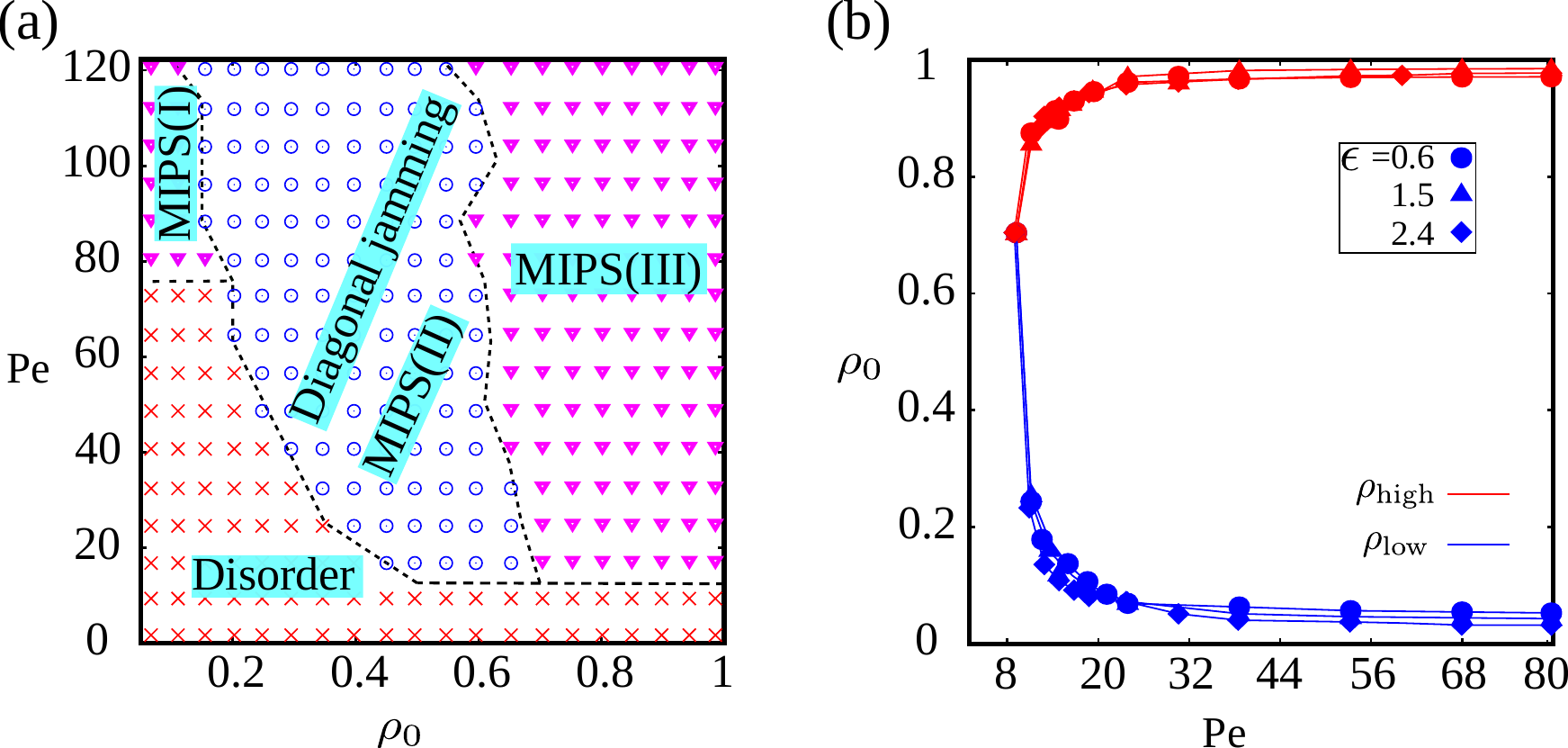}
\caption{(color online) (a) ${\rm Pe}-\rho_0$ phase diagram for MPS=1 for the snapshots in Fig.~\ref{fig3}. (b) MIPS phase diagram (for $\epsilon=0.6$, 1.5, and 2.4) showing that the binodals $\rho_{\rm low}$ and $\rho_{\rm high}$ are independent of $\epsilon$. The critical Pe is estimated as ${\rm Pe}_c \simeq 8$ above which the phase separation occurs.}
\label{fig4}
\end{figure}

Phase diagrams of the rAPM with MPS = 1 are shown in Fig.~\ref{fig4}. Figure~\ref{fig4}(a) shows the quantitative analog of the diagram shown in Fig.~\ref{fig3} where three different shades signify the three MIPS states having similar internal domain morphology but different shapes. We categorized these three MIPS states as MIPS (I) (small cluster state at large Pe and small $\rho_0$), MIPS (II) (diagonal high-density state at intermediate $\rho_0$), and MIPS (III) (high-density state with a square gaseous domain at large $\rho_0$). At large Pe, as the average density increases, the system transitions from MIPS (I) state to MIPS (III) state via the MIPS (II) state of jammed diagonal stripes. The system remains in the gaseous phase at small Pe because of low activity (diffusion dominates self-propulsion) and then transitions to the MIPS (III) state at large densities.

We further compute the binodals $\rho_{\rm low}$ and $\rho_{\rm high}$ for several $\epsilon$ and plot the resulting phase diagram in Fig.~\ref{fig4}(b). The binodals are the coexisting densities and physically signify the average densities of the gas and ordered phases at a given Pe and are estimated by calculating the average densities in different square boxes inside the high and low-density regions. From the diagram we observe that the binodals are independent of $\epsilon$ up to fluctuations and the critical Pe is estimated as ${\rm Pe}_c \simeq 8$ above which phase separation proceeds via spinodal decomposition. The shape of the phase diagram and the qualitative nature of the coexistence lines are similar to the diagram obtained for the active lattice gas~\cite{ALG} where the critical Pe was estimated as ${\rm Pe}_c \simeq 4$.

%%%%%%%%%%%%%%%%%%%%%%%%%%%%%%%%%
%%%%%% hard-core %%%%%%%%%%%
%%%%%%%%%%%%%%%%%%%%%%%%%%%%%%%
\subsection{Hard-core restriction (${\rm MPS} > 1$)}
\label{HC}

This section presents our findings of the rAPM with hard-core restriction where we restrict the number of allowed particles on a site (${\rm MPS} > 1$). A lower MPS signifies higher restriction on particle movement.

\begin{figure*}[t]
\centering
\includegraphics[width=\textwidth]{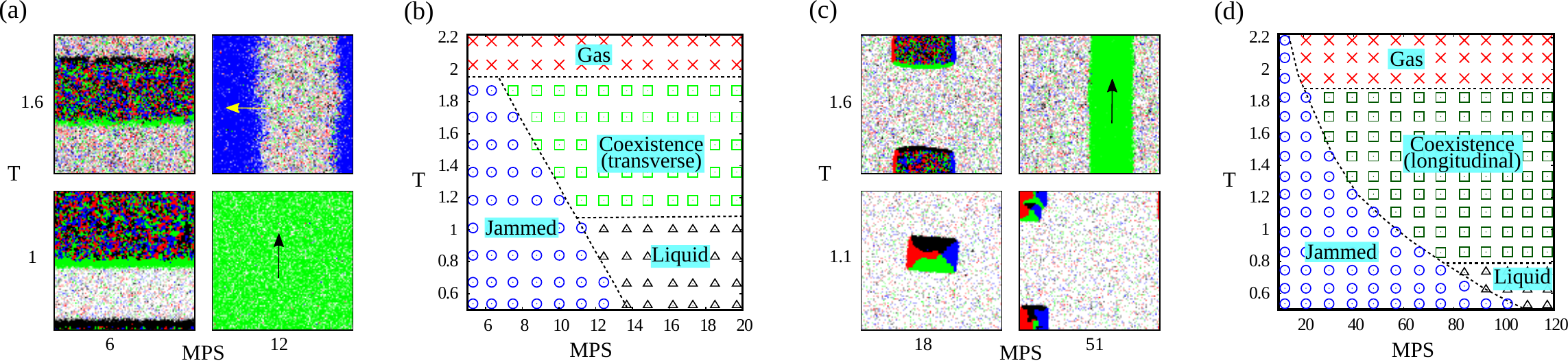}
\caption{(color online) (a, c) Steady-state snapshots of the four-state rAPM 
with MPS$>$1 in the $T-{\rm MPS}$ plane. (a) For $\epsilon=0.9$, an increase in ${\rm MPS}$ induces a transition from the jammed MIPS state to the liquid phase at lower $T$ and to the coexistence region with transversely propagating liquid band at a higher $T$. (c) Snapshots for $\epsilon=2.7$ where a jam-to-lane transition occurs at large ${\rm MPS}$ and higher $T$. Here the coexistence band is longitudinal. Legend: red ($\sigma=1$): right; green ($\sigma=2$): up; blue ($\sigma=3$): left; black ($\sigma=4$): down; white: empty. Arrows indicate the direction of motion. (b, d) $T-{\rm MPS}$ phase diagrams showing the four phases of rAPM where phase transition happens from the jammed state to the free-flowing state as either MPS or $T$ is increased. Parameter: $\rho_0=4$.}
\label{fig5}
\end{figure*}

Figure~\ref{fig5}(a) and Fig.~\ref{fig5}(c), respectively, show steady-state snapshots as a function of temperature and MPS for small ($\epsilon=0.9$) and large ($\epsilon=2.7$) propulsion velocities. In Fig.~\ref{fig5}(a), the system exhibits phase-separated orientationally disordered jammed domains with well-defined ordered domain boundaries for strong repulsion, whereas it shows features of unrestricted APM as relaxation on particle movement increases. By jamming we denote a transition from a free-flowing state to a high-density kinetically arrested solid configuration~\cite{nagel2010} due to MIPS. Although domain boundaries are ordered, the preferred directions of motion of the particles on the boundaries are inaccessible due to the hard cutoff and as a consequence, the jammed bands are almost immobile. With more freedom on movement, the system manifests a liquid phase at low temperature and a liquid-gas coexistence region (with transversely moving liquid band on gaseous background) at a relatively higher temperature typical of unrestricted flocking models~\cite{AIM,AIM-2,APM,ACM}. In this paper, by liquid we always mean a liquid phase with respect to orientational order. A further increase in temperature leads to a disordered gaseous phase (not shown in the snapshots). \\

Figure~\ref{fig5}(c) is analogous to Fig.~\ref{fig5}(a) but for higher particle velocity. Figure~\ref{fig5}(c) at low temperature demonstrates jammed clusters with four orientationally ordered subdomains (as $q=4$) in a gridlocked position (where the four quadrants orient against each other in such a way that the cluster is jammed), although the ordering does not extend over a long distance even at long times and the global polarization of such a cluster is zero (see the supplemental movie~\cite{SM}; parameters: $\rho_0=4$, $\epsilon=2.7$, ${\rm MPS} = 100$, $\beta=1.05$). Such a configuration is almost immobile although there are fluctuations in the boundaries. With their preferred hopping directions fully crowded, particles from the boundaries of such a cluster can hop to sites in the gaseous region but the probability of such a jump is much lower when the particle velocity is high [see Eq.~\eqref{whop}]. At high temperature, however, the morphology of the jammed cluster changes to a spatially and structurally disordered phase with well-defined domain boundaries which appears less congested compared to the low-temperature clusters. Both these less and strongly congested jammed clusters are examples of MIPS where the difference in the internal domain structure arises due to the effect of the temperature. MIPS with a phase-separated cluster of ordered domains having finite characteristic length scales  (similar to the snapshot corresponding to $T=1.1$ and ${\rm MPS}=18$) has also been observed for ABPs and active Janus colloids~\cite{odmips}. For ${\rm MPS}>1$ (and also for soft-core repulsion; see Sec.~\ref{SC}), we have local alignment interaction between particles, and although MIPS is traditionally defined as a phase separation of repulsively interacting particles without any alignment interaction, a phase separation where particle velocity reduces with increasing local density can also be attributed to MIPS~\cite{geyer2019}. It has also been shown for ABPs with alignment interaction that alignment promotes MIPS {\cite{sese2018,sese2021}}. 

In Fig.~\ref{fig5}(c), as restriction is relaxed, enhanced flipping at high temperature dissolves the congestion observed at low temperature, and the system makes a transition from the jammed phase to the coexistence phase exhibiting lane. During this process, the system also exhibits a band-to-lane reorientation transition as a function of the particle self-propulsion velocity, a novel feature of the flocking phenomenon in the unrestricted APM \cite{APM,APM-2}, where we observe a transverse band [blue band in Fig.~\ref{fig5}(a)] at small particle velocity whereas a longitudinally moving lane [green band in Fig.~\ref{fig5}(c)] at large velocity.\\

Figure~\ref{fig5}(b) and Fig.~\ref{fig5}(d) show the $T-{\rm MPS}$ phase diagrams for $\epsilon=0.9$ and $\epsilon=2.7$, respectively. In Fig.~\ref{fig5}(b), the combination of temperature and MPS determines four phases of the rAPM. At large MPS, which facilitates particle hopping to neighboring sites, the rAPM behaves like the unrestricted APM \cite{APM,APM-2}, and we observe the three phases of the unrestricted APM in the phase diagram: a gaseous phase at high temperatures, an ordered liquid phase at low temperatures, and a coexistence region at intermediate temperatures, where the motion of the ordered liquid band is transverse. At low MPS, the hard-core repulsion prevents collective motion, resulting in a jammed MIPS state for a large range of temperatures. As temperature is increased, the jammed region shrinks, and at high enough temperatures, the system always remains gaseous. These jammed clusters occur either in the coexistence region or in the liquid region and emerge due to MIPS.

Figure~\ref{fig5}(d) shows $T-{\rm MPS}$ phase diagram for large particle velocity. The coexistence region is now characterized by lanes \cite{APM}, and for small MPS, the jammed phase occurs even at very high temperature. As portrayed in both phase diagrams, fluctuations play a crucial role in the transition of the jammed phase to the coexistence or liquid phase by enhancing the probability of flipping. For $\epsilon=0.9$, the hopping rate to nonpreferred directions is substantial compared to $\epsilon=2.7$. Thus, moderate self-propulsion and thermal fluctuations help break the jammed configuration more efficiently. Another difference is the MPS range. Particles hop quickly at large $\epsilon$. At higher MPS values, tending toward unrestricted APM, the liquid band of the coexistence region becomes narrower and more populated with increasing $\epsilon$. The unrestricted APM $\epsilon-\rho_0$ phase diagram shows this \cite{APM}. As the liquid binodal value increases with $\epsilon$ (at a fixed $T$), so does the cutoff MPS.
%Note that we draw the smooth phase boundaries by taking $10-20$ random initial realizations of the system which minimizes the statistical fluctuation of the phase boundaries obtained for discrete parameter space.

In this paper our main goal is to investigate the effect of various repulsive interactions on particle hopping and on the consequent flocking dynamics. To do so, we will now compare the temperature-density and velocity-density phase diagrams of the unrestricted APM \cite{APM,APM-2} with the phase diagrams obtained with the current model.\\
\begin{figure*}[t]
\centering
\includegraphics[width=\textwidth]{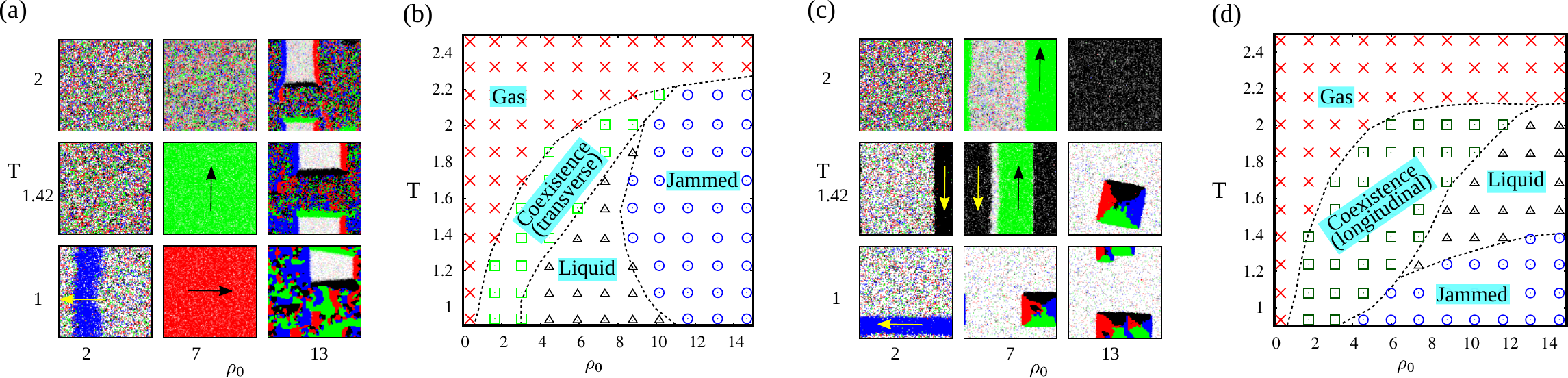}
\caption{(color online) (a, c) Steady-state snapshots for MPS$>$1 in the $T-\rho_0$ plane for (a) MPS $=15$ and $\epsilon=0.9$ and (c) MPS $=60$ and $\epsilon=2.4$. (b, d) $T-\rho_0$ phase diagrams of the rAPM with hard-core restriction for (b) $\epsilon=0.9$ and (d) $\epsilon=2.4$. In (b), the liquid phase appears for intermediate densities but makes a transition to the jammed MIPS state at higher densities whereas in (d), at large activity, jamming appears early and the liquid phase is observed only when the temperature is moderately high.}
\label{fig6} 
\end{figure*}

Fig.~\ref{fig6}(a) and Fig.~\ref{fig6}(c) show steady-state snapshots 
as a function of average particle density and temperature with hard-core repulsion for $\epsilon=0.9$ and $\epsilon=2.4$, respectively. In Fig.~\ref{fig6}(a), at smaller densities, as the temperature is increased, the system shows a transition from coexistence region and polar-ordered phase to the disordered gas phase. No jamming occurs beacuse the allowed MPS merely restricts particle hopping at these densities. At a larger density, however, jamming happens for all values of temperature because a larger density requires higher MPS to avoid jamming. The jamming observed for large densities shows a MIPS state of ordered domains at low temperature and a MIPS state of disordered internal domain with well-defined ordered domain boundaries at high temperatures. 

Figure~\ref{fig6}(c) shows the steady-state snapshots with faster moving particles compared to Fig.~\ref{fig6}(a) which was for slow particles. Notice the difference in the domain morphology of the MIPS state in Fig.~\ref{fig6}(a) and Fig.~\ref{fig6}(c) for low temperatures and high density. For $\epsilon=0.9$, hopping probability along the nonbiased directions is substantial compared to $\epsilon=2.4$ and therefore the high-density jammed area is larger and less congested for $\epsilon=0.9$. At large particle velocity ($\epsilon=2.4$), the transition to more crowded jamming is enhanced due to large MPS and high velocity where more particles can now gather at a site, and the activity along the nonpreferred directions decreases significantly. As a consequence, the jammed cluster of four orientationally ordered subdomains in a gridlocked position now occupy less area but the congestion is extremely strong.

The corresponding $T-\rho_0$ phase diagrams are shown in Fig.~\ref{fig6}(b) and Fig.~\ref{fig6}(d). Figure~\ref{fig6}(b) shows that at low temperature and activity, the system mimics the unrestricted APM behavior with transition from a gaseous to an ordered liquid phase via the liquid-gas coexistence region as density is increased. At higher densities, phase transition to the jammed phase happens from the coexistence region and the ordered liquid phase due to the hopping restriction through MPS.

With high particle velocity, however, low temperature facilitates jamming even at intermediate densities [Fig.~\ref{fig6}(d)]. Increasing the temperature helps the system to break the congestion, and free-flowing phases such as lane and ordered liquid phases emerge. A large $\epsilon$ allows particles to self-propel more along the preferred direction, which is unaltered at low temperatures, and particles of different states meeting at a point stay stuck for a long time, causing a jammed MIPS state at low temperature and high density. A temperature increase partially dissolves this situation by enhancing flipping as switching the state changes the preferred direction of propulsion. Both phase diagrams show a high-temperature gaseous phase unaffected by the control parameters beyond a critical temperature $T_c \sim 2.3$.
\begin{figure}
\centering
\includegraphics[width=\columnwidth]{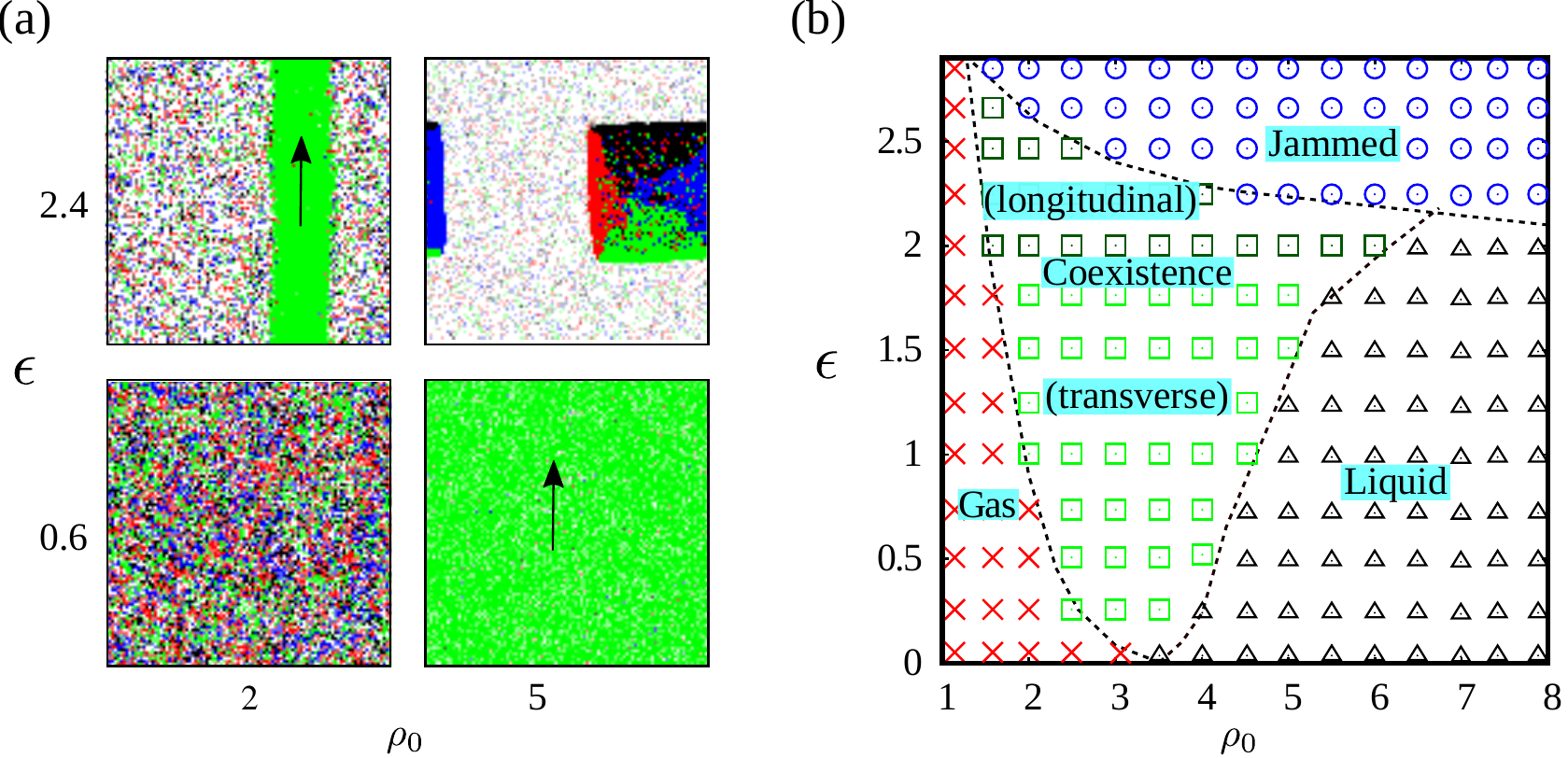}
\caption{(color online) (a) Steady-state snapshots and (b) $\epsilon-\rho_0$ phase diagram of the rAPM with MPS=20 for $T=1.42$ ($\beta=0.7$). (a) An increase in $\rho_0$ and $\epsilon$ drives the system towards a jammed state. (b) The phase diagram resembles the unrestricted APM~\cite{APM,APM-2} for $\epsilon<2$ but breaks down at larger $\rho_0$ and $\epsilon$. The reorientation transition (transverse to longitudinal band motion) happens at $\epsilon \sim 2.1$ and a liquid-gas phase transition takes place at $\rho_0 \sim 3.5$ for $\epsilon=0$ \cite{APM,APM-2}.}
\label{fig7} 
\end{figure}

Next, we will discuss the $\epsilon-\rho_0$ diagram of the rAPM by changing the strength of the self-propulsion $\epsilon$ while keeping the temperature fixed. The resultant steady-state snapshots and the corresponding phase diagram are shown in Fig.~\ref{fig7}(a) and Fig.~\ref{fig7}(b), respectively. The snapshots show that the system exhibits a jam of orientationally ordered subdomains due to MIPS at high density and motility. High particle motility promotes particle accumulation at a lattice site, and since these particles have a very small probability to hop toward the nonbiased directions, a higher MPS is required to avoid jamming. The snapshots also exhibit the three typical phases of the unrestricted APM at small density and velocity.
%The jammed cluster in Fig.~\ref{fig7}(a) for $\rho_0=5$ and $\epsilon=2.4$ is an example of the densest congestion. The appearance of such a cluster happens at low $T$ and large $\epsilon$ due to a lack of fluctuations and minimal hopping probabilities along non-preferred directions. A high particle density also enhances the possibility of such a jam. Particles belonging to each of the ordered domains in such a dense, locally ordered cluster are unlikely to flip and break the dense congestion when $T$ is low or $\rho_0$ is high.

The $\epsilon-\rho_0$ phase diagram in Fig.~\ref{fig7}(b) shows four regions similar to the $T-\rho_0$ phase diagrams shown in Fig.~\ref{fig6}(b) and Fig.~\ref{fig6}(d). Excluding the high velocity limit, the phase diagram resembles the unrestricted APM diagram~\cite{APM,APM-2} where the binodals $\rho_{\rm gas}$ and $\rho_{\rm liq}$ delimit the coexistence region from the gas and liquid phases. The reorientation transition, which is a novel feature of the APM and where the system exhibits a transverse band motion at small $\epsilon$ and a longitudinal lane motion at large $\epsilon$, is also observed. The conventional phase diagram, however, breaks down at large $\epsilon$ where a transition to the jammed MIPS state from the coexistence region and the liquid phase is realized. At $\epsilon=0$, similar to the unrestricted APM \cite{APM,APM-2}, the system exhibits a direct liquid-gas phase transition around $\rho_0 \sim 3.5$. 
\begin{figure}
\centering
\includegraphics[width=\columnwidth]{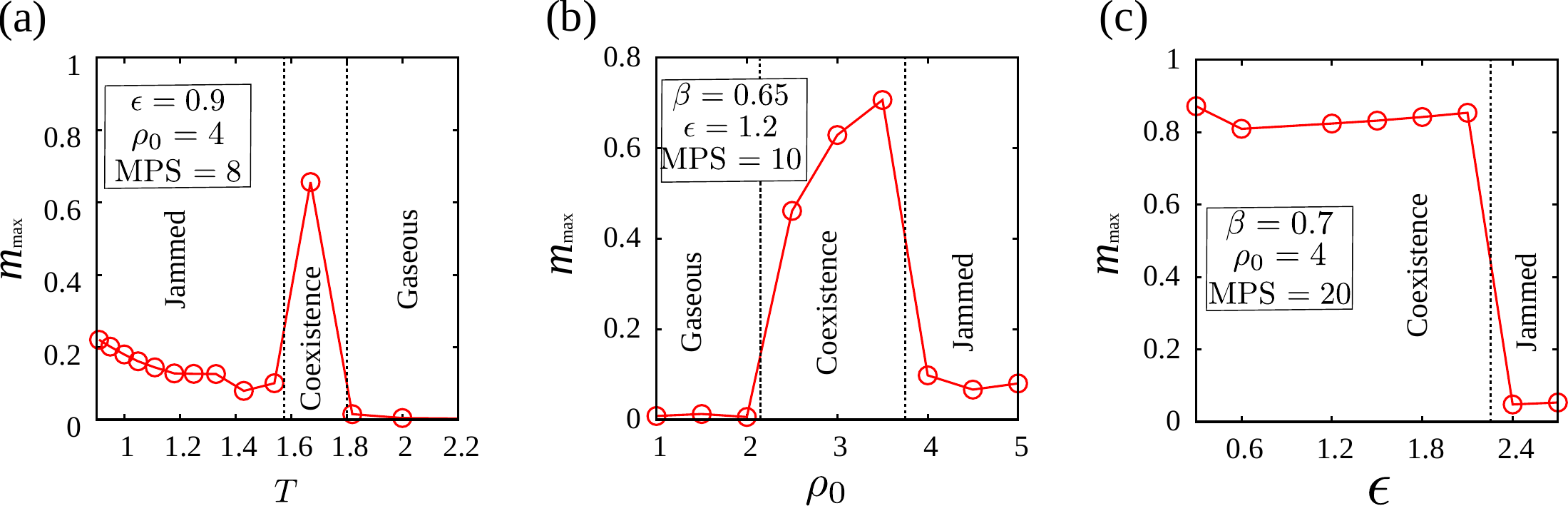}
\caption{(color online) Normalized maximal magnetization 
$m_{\rm max}$ for the rAPM with MPS$>$1 as a function of $T$, $\rho_0$, and $\epsilon$ in (a), (b), and (c), respectively. $m_{\rm max}$ of the coexistence region is larger than the jammed phase. For the disordered gas phase, $m_{\rm max}=0$ and $m_{\rm max} \simeq 1$ for the liquid phase.}
\label{fig8} 
\end{figure}
\begin{figure*}[t]
\centering
\includegraphics[width=\textwidth]{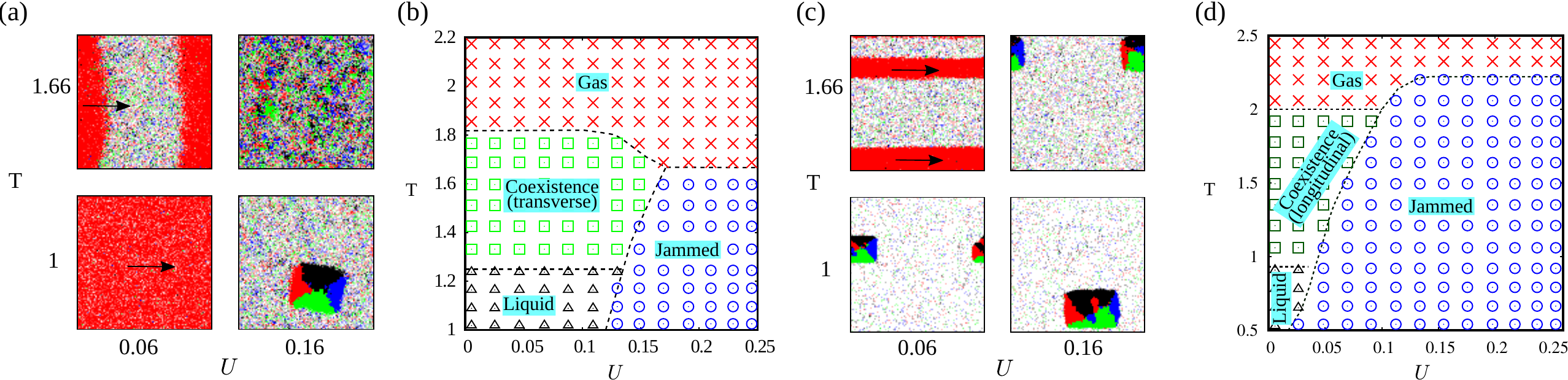}
\caption{(color online) (a, c) Steady-state snapshots for the rAPM with soft-core repulsion in the $T-U$ plane for (a) $\epsilon=0.9$ and (c) $\epsilon=2.7$ with fixed $\rho_0=4$. (b, d) $T-U$ phase diagrams for the parameters of (a) and (c), respectively. Transition to jamming happens at low $T$ and large $U$ with the fact that high propulsion facilitates jamming.}
\label{fig9} 
\end{figure*}

The systematic variation of magnetization is an indicator of symmetry breaking. Figure~\ref{fig8} shows the order parameter against different control parameters where we consider the maximal magnetization $m_{\rm max}$ among the four estimates [see Eq.(~\ref{Hapm_mag})]. It can be seen from the plots that magnetization changes abruptly across the phase boundaries denoted by dashed vertical lines. A fully ordered liquid state is characterized by $m_{\rm max}\simeq 1$, and $m_{\rm max}= 0$ signifies a disordered gaseous phase. Figure~\ref{fig8}(a) shows $m_{\rm max}$ versus temperature where the low-temperature region is characterized by a nonzero but small $m_{\rm max}$ signifying a MIPS state (locally ordered subdomains marginally raise the $m_{\rm max}$ to a nonzero value). As the temperature is increased, a phase transition from the MIPS state to the coexistence region happens with a sharp jump in the magnetization indicating a first-order phase transition. Further increase of the temperature shows another first-order phase transition from the coexistence to the disordered gas phase. $m_{\rm max}$ against density in Fig.~\ref{fig8}(b) also presents two first-order phase transitions as density is increased, gas to coexistence and coexistence to MIPS, respectively. The MIPS state occurs at a relatively high $\epsilon$ in Fig.~\ref{fig8}(c) because of the large MPS value. Figure~\ref{fig8}(c) also validates that the transition to the MIPS state is a first-order transition as demonstrated by the discontinuous jump of the order parameter at the transition point. 

As discussed so far, jamming in our model is a kinetically arrested state due to MIPS and MIPS signifies the coexistence of an active low density gas with a high density jammed cluster, which is reminiscent of equilibrium liquid-gas demixing and thus can be seen as a first-order transition. As shown in Fig.~\ref{fig8}, we also find the transition from the ordered or flocking phase to the jammed phase as a discontinuous first-order phase transition. At this point, the system undergoes a sudden jamming transition, leading to the formation of large-scale clusters that are kinetically trapped in a glassy state. It has already been shown in the context of active Brownian particles that MIPS can be described as an equilibrium-like phase transition (not just a dynamically trapped state), and MIPS verify the characteristic properties of first-order liquid–gas phase transitions ~\cite{levis2017}. That MIPS-like transitions between polar liquid and amorphous jammed solid is a first-order transition has also been shown in the context of motile colloids both experimentally and theoretically~\cite{geyer2019}. This observation is also consistent outside the field of active matter where experimental data of traffic flow on highways finds the phase transition from free flow to traffic jam as a first-order phase transition \cite{kerner1997}. 

One can also distinguish between a jammed state and a free-flowing state by 
the mean-square displacement (MSD) of the high-density clusters or by 
the number fluctuations~\cite {Marchetti2011}. 
Besides strongly suppressed number fluctuations a  jammed state can be characterize by the time dependence of the MSD: either oscillatory~\cite{Marchetti2011}, because jammed clusters oscillate around their mean position, or saturating at large times, because clusters or bands cease 
to move; see Appendix~\ref{appendix_msd}). For high-density MIPS clusters observed in Fig.~\ref{fig3}, we have noticed an oscillatory nature in the MSD (data not shown).

%%%%%%%%%%%%%%%%%%%%%%%%%%%%%
%%%%%%%%%% soft-core %%%%%%%%%%%
%%%%%%%%%%%%%%%%%%%%%%%%%%%%
\subsection{Soft-core restriction}
\label{SC}
In this section, we present the results of the rAPM with soft-core repulsion paramterized by the value of $U>0$ in Eq.~\eqref{SC_eq}. Figure~\ref{fig9}(a) and Fig.~\ref{fig9}(c) depict the late-stage coarsening of the four-state rAPM in the $T-U$ plane for (a) $\epsilon=0.9$ and (c) $\epsilon=2.7$. For small propulsion, the system exhibits jammed clusters 
composed of four ordered subdomains at low temperature and large restriction similar to the hard-core repulsion (see the supplemental movie~\cite{SM}; parameters: $\rho_0=7$, $\epsilon=0.9$, $U = 0.07$, $\beta=1$). An increase in the temperature (and low $U$) helps to dissolve the jam and we observe the three phases of the unrestricted APM \cite{APM,APM-2}. Figure~\ref{fig9}(c) is analogous to Fig.~\ref{fig9}(a) but for a larger velocity. Due to the high propulsion and therefore suppressed particle motion along the nonpreferred directions, the liquid phase (observed with $\epsilon=0.9$) becomes a jammed MIPS state. A rise in the thermal fluctuation helps to break the clog for small $U$ but due to high motility, the jam persists even when the temperature is high for strong repulsion. As explained before, this jammed phase is a kinetically arrested jammed phase due to MIPS where we observe reduction of the active particle current as
density becomes sufficiently high. The structural transformation of such a kinetically jammed phase with temperature is demonstrated in Appendix~\ref{appendix_jam_structure}.
%At large $U(=0.16$), we see a locally ordered, high-density cluster for $T=1$, but a higher temperature destroys this cluster and the system exhibits orientational disorder, while a group of particles self-segregates into disordered particle aggregates. As explained in the context of hard-core restriction, an increased particle flipping probability with $T$ helps the system dissipate a jammed state. The heterogeneous, arrested, dense cluster at $T=1$ is the typical morphology of an extreme jam where the shape and internal structure of the cluster depend on $q$. 

\begin{figure*}[t]
\centering
\includegraphics[width=\textwidth]{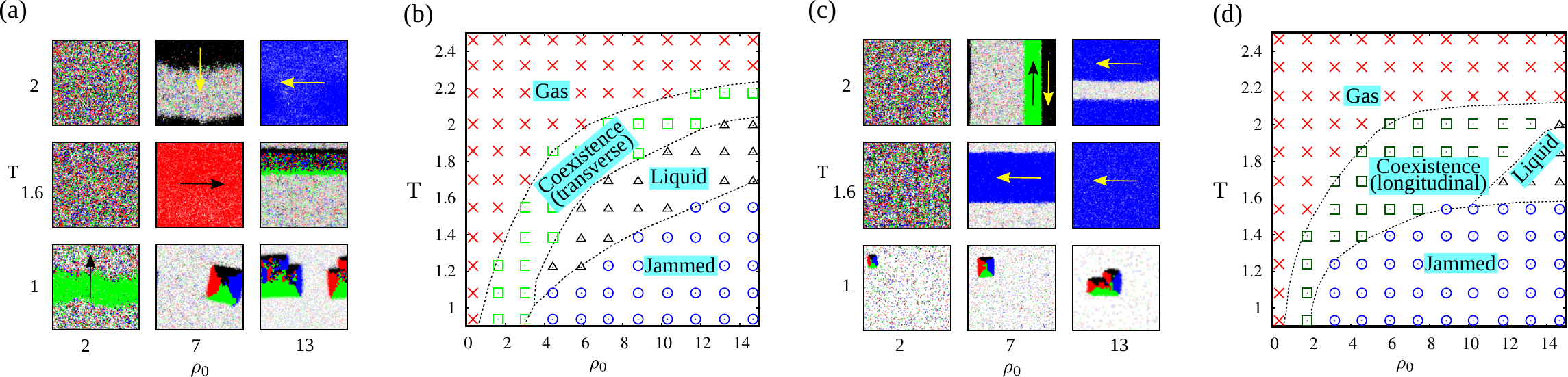}
\caption{(color online) (a, c) Snapshots of the rAPM with soft-core repulsion
illustrating the different self-organized patterns as a function of $T$ and $\rho_0$ for (a) small $\epsilon=0.9$ ($U=0.07$) and (c) large $\epsilon=2.7$ ($U=0.02$). The snapshots signify that low temperature, high density and large activity facilitates jamming. (b, d) $T-\rho_0$ phase diagrams of the rAPM with soft-core repulsion for parameters of (a) and (c), respectively. Apart from the usual gas and liquid binodals observed in the unrestricted APM, we observe a jammed phase adjoining both coexistence and liquid phases at large densities and low temperatures.}
\label{fig10}
\end{figure*}

We find, e.g., square or rectangular kinetically jammed clusters with our model for both hard-core restriction and soft-core repulsion [see Fig.~\ref{fig5}(c) (MPS = 18, $T=1.1$) and Fig.~\ref{fig9}(a) ($U=0.16$, $T=1$)]. The square structure is due to the four-state version of our model where particles can hop to only the four nearest neighbors of a square lattice, and it is through hopping that the neighbors are connected. Therefore, jammed clusters
have a square or rectangular geometry and contain four locally ordered subdomains for $q = 4$ and a hexagonal cluster with six locally ordered subdomains for $q = 6$ (simulated on a triangular lattice, data are not shown here). We have also investigated the AIM ($q = 2$) with soft-core repulsion and observe a vertically jammed MIPS state similar to active lattice gas \cite{ALG}. The effect of soft-core repulsion has also been investigated on off-lattice flocking models such as ACM and VM, which manifest MIPS (see Appendix~\ref{racm_vm}). 

%(a) $\epsilon=0.9$: As $U$ is increased, the system shows a transition from the liquid and coexistence region to the jammed phase. At very large $T$ system remains in the gaseous phase. (d) $\epsilon=2.7$: The system manifests a jammed phase at small $U$ due to large particle velocity. The jammed phase diminishes gradually with increasing temperature.}

The corresponding $T-U$ phase diagrams are shown in Fig.~\ref{fig9}(b) and Fig.~\ref{fig9}(d). Figure~\ref{fig9}(b) shows a liquid-to-gas transition via the coexistence region for small values of $U$ as temperature is raised. 
%There is a transition from the gaseous phase at low $T$ to an ordered liquid phase at high $T$ via a liquid-gas coexistence region at intermediate $T$; no jamming is observed. 
An increase of hopping restriction via $U$, however, changes this scenario, and we observe a transition to the jammed phase at low and intermediate temperatures. At very high temperatures, the system remains in the gaseous phase. In Fig.~\ref{fig9}(d), due to the high velocity of the particles, jamming dominates the phase diagram, and at low temperature, the system exhibits jamming even at small $U$.
%the system exhibits jamming at low temperatures even at small $U$ and the liquid phase is observed only for very small $U$. At high $T$, for small $U$, the coexistence region is observed due to the interplay between increased flipping at enhanced hopping of particles. At large $U$, the system always manifests a jammed phase, and at high $T$, a gaseous phase is observed analogous to Fig.~\ref{fig9}(b).

Now we will focus on the $T-\rho_0$ and $\epsilon-\rho_0$ phase diagrams of the rAPM with soft-core repulsion, and we will compare them with the similar diagrams obtained for hard-core rAPM and unrestricted APM.

The soft-core $T-\rho_0$ phase diagrams for low and high particle velocities along with the steady-state snapshots are shown in Fig.~\ref{fig10}. The snapshots in Fig.~\ref{fig10}(a) and Fig.~\ref{fig10}(c) tell a story of jamming transition that is consistent with the findings of hard-core rAPM where we observe jamming for low temperatures and high densities. Notice the difference in the internal structure of the jammed clusters in Fig.~\ref{fig10}(a) and Fig.~\ref{fig10}(c) where for small activity, we observe a MIPS state of orientationally disordered active particles having ordered domain boundaries whereas the MIPS state exhibits a gridlocking of ordered sub-domains at large activity. The system behaves as the unrestricted APM at high temperatures.
%(a) and Fig.~\ref{fig10}(c) show the representative snapshots of 4-state APM as a function of average particle density $\rho_0$ and temperature $T$ for (a) $U=0.07$ and $\epsilon=0.9$ and (c) $U=0.02$ and $\epsilon=2.7$. 
%Fig.~\ref{fig10}(a) indicates that the system behaves like the unrestricted APM~\cite{APM,APM-2} at high temperature ($T=2$), a gaseous phase at a density $\rho_0=2$, a liquid phase at $\rho_0=13$ via a coexistence region at intermediate $\rho_0=7$. Jammed clusters appear at a very low temperature, and their size increases with increasing density. At an intermediate temperature ($T=1.6$), a liquid phase forms at $\rho_0=7$ due to moderate flipping and freedom to hop in the non-preferred directions, and jams are observed only at very high densities. With the increase of particle velocity $\epsilon=2.7$, at low temperature ($T=1$), we find in Fig.~\ref{fig10}(c) dense locally ordered clusters of mixed states embedded in a low-density background. The clusters are jammed by domains of oppositely directed particles that face each other and cannot dissolve due to less flipping at low temperatures. An increase in $\rho_0$ widens the jammed domain. At high temperature, $T=2$ ($\beta=0.5$), the system is in the gaseous state at a small $\rho_0=2$. It transitions from the gaseous phase to the lane formation as density increases(unrestricted APM behavior). As the magnitude of $U$ is not large ($U=0.02$), jamming is not observed for high temperatures. 
\begin{figure}[t]
\centering
\includegraphics[width=\columnwidth]{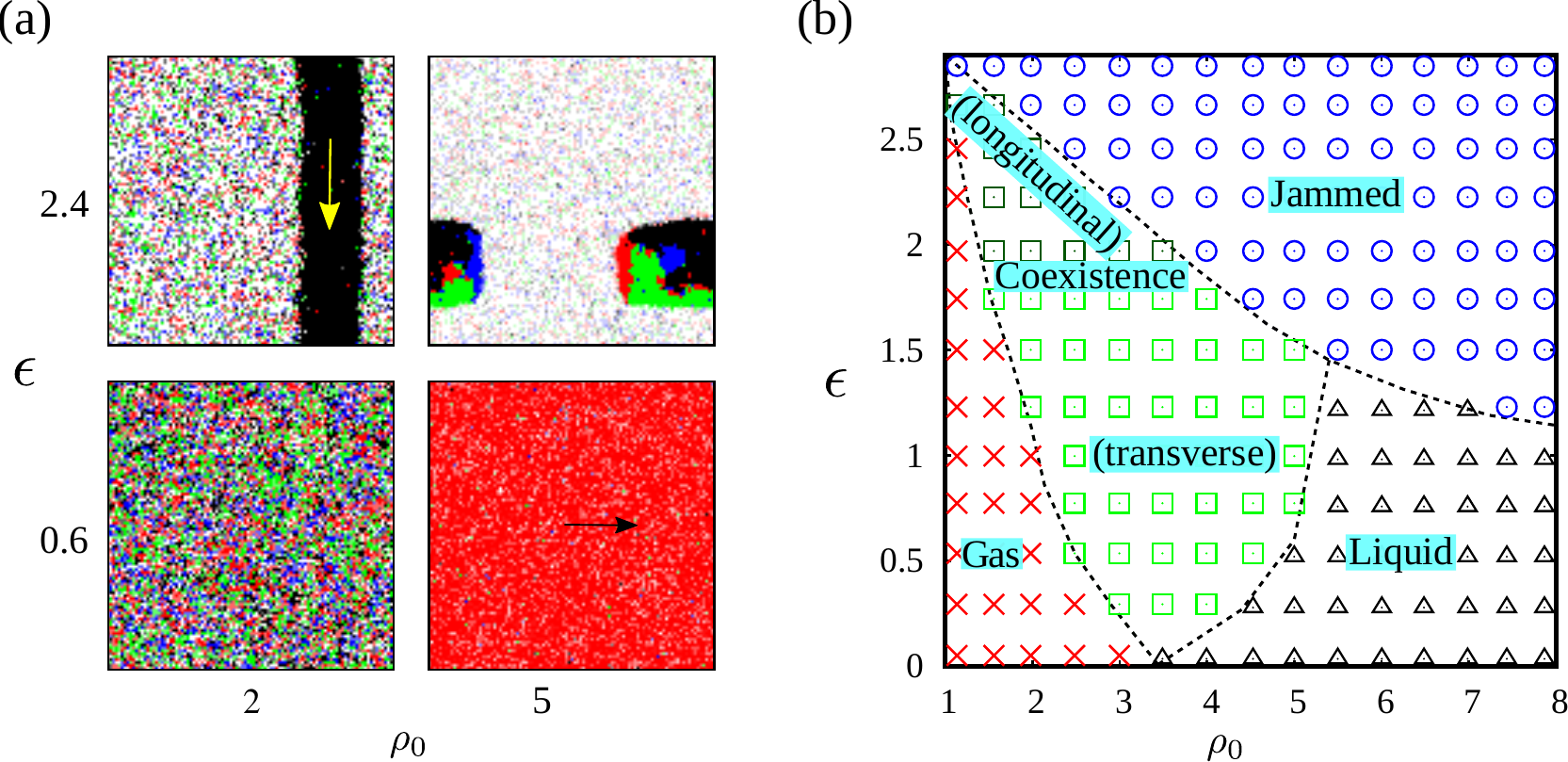}
\caption{(color online) (a) Steady-state snapshots of the rAPM with 
soft-core repulsion in the $\epsilon-\rho_0$ plane for $T=1.42$ ($\beta=0.7$) and $U=0.07$. (b) The soft-core $\epsilon-\rho_0$ phase diagram qualitatively resembles the hard-core phase diagram of Fig.~\ref{fig7}(b) showing that the jamming threshold decreases with density and activity.}
\label{fig11} 
\end{figure}
\begin{figure}[t]
\centering
\includegraphics[width=\columnwidth]{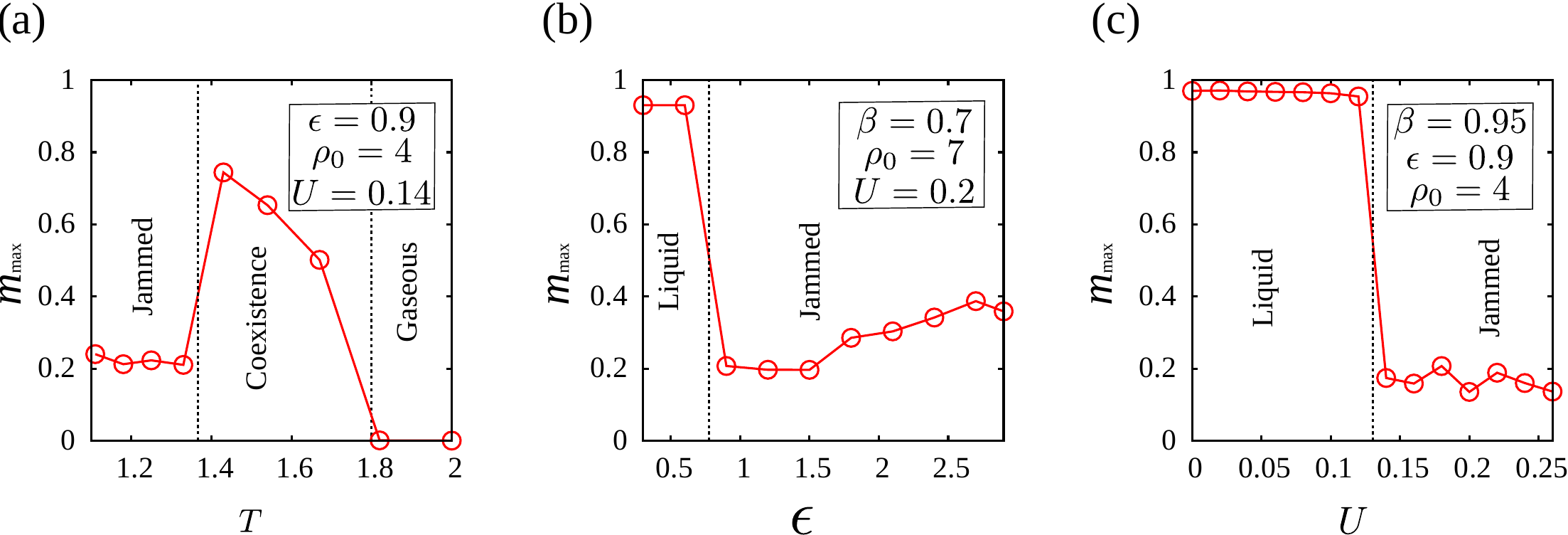}
\caption{(color online) (a--c) 
Normalized maximal magnetization $m_{\rm max}$ of the rAPM with soft-core repulsion as a function of $T$, $\epsilon$, and $U$, respectively. A small nonzero $m_{\rm max}$ is indicative of a jammed phase and a comparatively high $m_{\rm max}<1$ characterizes a coexistence region.}
\label{fig12} 
\end{figure}
These observations are presented in the form of phase diagrams in Fig.~\ref{fig10}(b) and Fig.~\ref{fig10}(d) where we notice that the system replicates unrestricted APM behavior \cite{APM,APM-2} for intermediate and large temperatures. Temperature reduction, on the other hand, leads to an increase in the jammed region with increasing densities. 
%Earlier studies have shown that reducing $T$ for a fixed $\rho_0$ leads to a liquid phase from a coexisting phase. As a result, at these high densities, the local density of clusters of individual states increases during coarsening to a liquid phase. However, soft-core repulsions on SPPs are insufficient for these clusters to avoid jamming at these $\rho_0$ values because high densities necessitate small $U$. 
For sufficiently high temperatures, the system is always in a gaseous phase that is unaffected by the control parameters, and we find the critical temperature $T_c \sim 2.3$ is also independent of the nature of restriction [see Fig.~\ref{fig6}(b) and Fig.~\ref{fig6}(d)]. A comparison of the phase diagram in Fig.~\ref{fig10}(d) and the $T-\rho_0$ phase diagram of the unrestricted APM \cite{APM} reveals that restriction has shifted the unrestricted phase diagram to the low-density region (similar for the hard-core diagrams too) where the jamming phase emerges as a fourth phase at low temperatures, keeping the $T_c$ the same.
%The findings also indicate that a strong repulsive interaction $(U)$ is required to generate a glass-like jammed state at low densities.\textcolor{red}{~\cite{paoluzzi2022}.}

In Fig.~\ref{fig11}, we show the phase snapshots and phase diagrams in the $\epsilon-\rho_0$ plane. The qualitative nature of the snapshots shown in Fig.~\ref{fig11}(a) is similar to our findings for the hard-core restriction shown in Fig.~\ref{fig7}(a) where at large velocity, jams emerge when density is increased. At small velocities, the system sequentially shows the gas, coexistence and liquid phases as density is increased.  
%The steady-state snapshots as a function of $\epsilon$ and $\rho_0$ for $\beta=0.7$ ($T=1.42$) and $U=0.07$ are shown in Fig.~\ref{fig11}(a). At small velocity $\epsilon=0.6$, we observe a gas phase at small density and a liquid phase of $\sigma=1$ at large density. For large velocity, jams are seen as we increased density $\rho_0=5$, whereas, for the $\rho_0=2$, a transition from the gas phase to the coexistence region is observed. 

The corresponding phase diagram is shown in Fig.~\ref{fig11}(b). The system behaves nearly diffusively for very small $\epsilon$; therefore, jamming does not exist at this limit. However, with high activity, due to the restriction on propulsion, a jammed phase occurs even at small and intermediate densities where the cutoff $\epsilon$ value to jamming decreases with $\rho_0$. This physically signifies that for small densities, high activity is needed to create a jam whereas for large densities, jamming can happen at less speed. The conventional $\epsilon-\rho_0$ phase diagram of the unrestricted APM \cite{APM} breaks down at the large velocity-density limit in Fig.~\ref{fig11}(b) whereas in the $\epsilon \leqslant 1$ limit, the two diagrams are analogous. At the zero activity limit ($\epsilon=0$), the system behaves like the unrestricted APM at small repulsion where we see a direct liquid-gas transition around $\rho_0 \sim 3.5$ whereas, at strong repulsion, orientationally disordered self-segregated domains are observed instead of the ordered liquid phase (see Appendix \ref{appendix_diffusive} for details).
%$\epsilon-\rho_0$ phase diagram in Fig.~\ref{fig11}(b) for $\epsilon \leqslant 1$ is analogous to the $\epsilon-\rho_0$ phase diagram of the unrestricted APM \cite{APM}. At the zero activity limit ($\epsilon=0$), the system behaves like the unrestricted APM at small $U$ where we see a direct liquid-gas transition with density whereas, at larger $U$, orientationally disordered self-segregated domains are observed instead of the ordered liquid phase (see Appendix \ref{appendix_diffusive}).

Figure~\ref{fig12} is analogous to Fig.~\ref{fig8} but for soft-core repulsion. Similar to Fig.~\ref{fig8}(a), $m_{\rm max}$ versus temperature in Fig.~\ref{fig12}(a) shows two first-order transitions (jammed to coexistence and coexistence to gas) as temperature is increased. $m_{\rm max}$ versus $\epsilon$ in Fig.~\ref{fig12}(b) shows a jamming transition from an ordered liquid phase. At small velocities, due to slow propulsion and large density, the system exhibits a fully ordered liquid phase, but at large velocities, particles quickly accumulate locally, and due to large restriction ($U$), particle movements are heavily restricted and therefore we observe a jammed state. Figure~\ref{fig12}(c) shows variation of $m_{\rm max}$ with $U$ where small $U$ facilitates particle hopping, which together with slow particles and a relatively large density gives rise to an ordered liquid phase. As restriction is enhanced, the system makes an unsurprised transition to the jammed phase. Similar to Fig.~\ref{fig8}, the jamming transition with the soft-core repulsion is also first-order, signifying that the specific origin of the restriction imposed on the particle movement does not alter the nature of the phase transition.

%%%%%%%%%%%%%%%%%%%%%%%%%%%%%%
%%%% Hydrodynamic theory %%%%%
%%%%%%%%%%%%%%%%%%%%%%%%%%%%%%

\section{Hydrodynamic theory}
\label{secHydro}

In this section, we formulate the hydrodynamic continuum theory for the microscopic rAPM. From the microscopic hopping and flipping rates, we derive the equation for the probability density function $\rho_\sigma({\bf x};t) \sim \langle n_i^\sigma(t) \rangle$ for a particle to be at the position ${\bf x}$, corresponding to the site $i$, and in the state $\sigma$ at the time $t$. We keep only the first-order terms in the $|{\bf m}_i| \ll \rho_i$ expansion of the flipping rate~\eqref{flipeq}. To represent the different hopping restrictions, we introduce a function $f(\rho)$ where $\rho$ is the total density at the arrival position. The form of this function is $f(\rho) = 1 - \ess \rho$ for $\ess = 1/{\rm MPS}$, and $f(\rho) = \exp(-s\rho)$ for the soft-core rAPM where $s=2\beta U$. In Appendix~\ref{hydro}, we derive the hydrodynamic equations:
\begin{equation}
\partial_t \rho_\sigma = - \partial_\parallel J_{\sigma \parallel} - \partial_\perp J_{\sigma \perp} + \sum_{\sigma' \ne \sigma} K_{\sigma \sigma'} (\rho_\sigma - \rho_{\sigma'}) \label{eqhydro},
\end{equation}
where the current is
\begin{gather}
J_{\sigma \parallel} = -D_\parallel [f(\rho)\partial_\parallel \rho_\sigma -f'(\rho) \rho_\sigma \partial_\parallel \rho] + vf(\rho) \rho_\sigma,\\
J_{\sigma \perp} = -D_\perp [f(\rho)\partial_\perp \rho_\sigma -f'(\rho) \rho_\sigma \partial_\perp \rho],
\end{gather}
with $D_\parallel = D(1+\epsilon/3)$, $D_\perp=D(1-\epsilon/3)$, and $v=4D\epsilon/3$, and the flipping interaction term $K_{\sigma \sigma'} = - \gamma$ for ${\rm MPS}=1$ and
\begin{equation}
K_{\sigma \sigma'} = \frac{4\beta J}{\rho}(\rho_\sigma + \rho_{\sigma'}) - 1 - \frac{r}{\rho} - \alpha \frac{(\rho_\sigma - \rho_{\sigma'})^2}{\rho^2}, \label{Iflip}
\end{equation}
with $\alpha = 8(\beta J)^2(1-2\beta J/3)$, for ${\rm MPS}>1$ and soft-core rAPM. The different expressions of $f(\rho)$ and $K_{\sigma \sigma'}$ lead to three different hydrodynamic equations for the three different studied models, discussed in the next subsections. Without any restriction for the particle density, i.e. setting $f(\rho) =1$ in Eq.~(\ref{eqhydro}), the unrestricted APM hydrodynamic equations derived in Ref.~\cite{APM} are recovered. Without any loss of generality, we set $D=1$, $r=1$ and $J=1$ defining the scales of time, density, and temperature, respectively.

We solve Eq.~(\ref{eqhydro}) numerically using FreeFEM++~\cite{freefem}, a software package based on the finite element method~\cite{fem}. The equations are integrated over discrete time $t_n = n \Delta t$, at which the density is denoted as $\rho_\sigma^{(n)}({\bf x})$. The initial density $\rho^{(0)}({\bf x})$ is taken as a high density bubble or stripe on a low density background. The final time is denoted as $t_{\rm max}$. The weak formulation of Eq.~(\ref{eqhydro}) is the integral equation:
\begin{gather}
\int_\Omega d{\bf x} \  \sum_\sigma  \left[ w_\sigma \rho_\sigma^{(n+1)} - \Delta t \left(\partial_\parallel w_\sigma \partial_\parallel J_{\sigma \parallel} + \partial_\perp w_\sigma \partial_\perp J_{\sigma \perp} \right) \right] \nonumber\\
- \Delta t \sum_{\sigma' > \sigma } (w_\sigma - w_{\sigma'}) K_{\sigma \sigma'} \left[\rho_\sigma^{(n+1)}-\rho_{\sigma'}^{(n+1)} \right]\nonumber\\
 = \int_\Omega d{\bf x} \ \sum_\sigma w_\sigma \rho_\sigma^{(n)},
\end{gather}
where $\rho_\sigma^{(n)}({\bf x})$ is the known particle density at time $t_n$, $\rho_\sigma^{(n+1)}({\bf x})$ is the unknown particle density at time $t_{n+1}$, and $w_\sigma({\bf x})$ is a test function. $K_{\sigma \sigma'}$ and the restriction terms in $J_{\sigma \parallel}$ and $J_{\sigma \perp}$ are calculated at time $t_n$ to have a linear equation of $\rho_\sigma^{(n+1)}({\bf x})$. This integral equation is solved over a triangular mesh-grid with ${\cal N}$ vertices on the boundaries. The densities are calculated on the nodes of the mesh-grid and interpolated linearly over the space with Lagrange polynomials. The precision of the numerical solution is increased for a narrow grid (${\cal N}\gg 1$) and small time increments ($\Delta t \ll 1$), and the computational time has a complexity proportional to ${\cal N}^2/\Delta t$. It takes about $48$ hours for ${\cal N}=75$ and $t_{\rm max}/\Delta t = 50000$ time steps, on a 4 GHz processor. The FreeFEM++ codes used to compute the numerical solutions are available in Ref.~\cite{zenodo}.

\subsection{MPS = 1 (ALG version of the rAPM)}

\begin{figure}[t]
\centering
\includegraphics[width=\columnwidth]{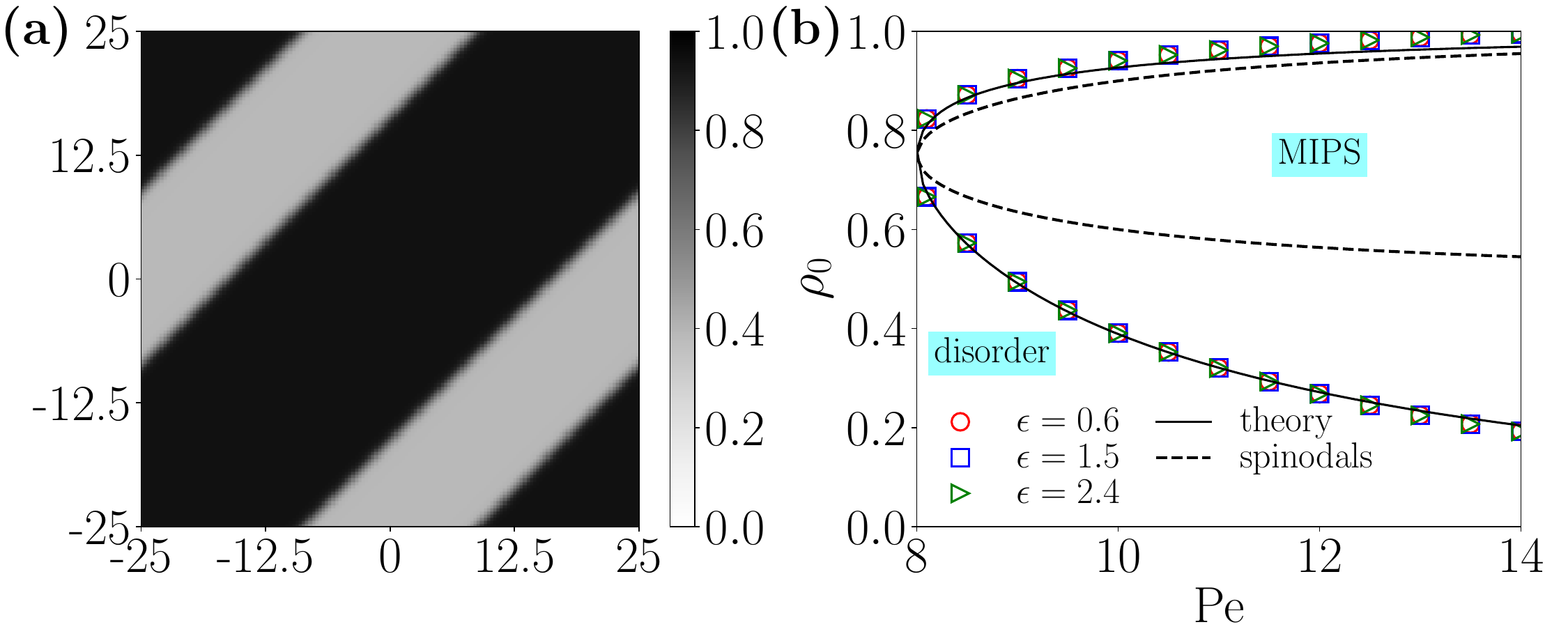}
\caption{(color online) (a)~MIPS density profile for ${\rm MPS}=1$, $\rho_0=0.75$ and ${\rm Pe}=10$, obtained numerically with FreeFEM++ for ${\cal N}=100$, $\Delta t = 0.01,$ and $t_{\rm max}=1000$. The density of the two disordered phases is $\rho_{\rm low}\simeq 0.391$ and $\rho_{\rm high}\simeq 0.941$. (b)~Velocity-density phase diagram for ${\rm MPS}=1$, computed with the numerical solutions of Eq.~\eqref{eqmps1}, in a square domain of linear size $L=50$, for three values of $\epsilon$. The dotted line shows the spinodals given by Eq.~\eqref{spinodal}, and the solid line represents the theoretical value of the binodals calculated in Appendix~\ref{ls_mps1}.}
\label{fig_hydro_MPS1}
\end{figure}

For ${\rm MPS} = 1$, with $f(\rho) = 1 - \rho$ and $K_{\sigma \sigma'} = - \gamma$, Eq.~(\ref{eqhydro}) becomes:
\begin{align}
\partial_t \rho_\sigma &= D_\parallel \partial_\parallel[ (1-\rho) \partial_\parallel \rho_\sigma + \rho_\sigma \partial_\parallel \rho ] \nonumber\\
&+ D_\perp \partial_\perp [ (1-\rho) \partial_\perp \rho_\sigma + \rho_\sigma \partial_\perp \rho ] \nonumber\\
&- v \partial_\parallel \left[ (1 - \rho) \rho_\sigma \right] - \gamma(4\rho_\sigma - \rho) \label{eqmps1}.
\end{align}
The only homogeneous solution is $\rho_\sigma = \rho_0/4$, corresponding to a disordered phase. In Appendix~\ref{ls_mps1}, we perform the linear stability analysis of this solution, leading to the spidodals:
\begin{equation}
\label{spinodal}
\varphi_\pm = \frac{3}{4} \pm \frac{1}{4}\sqrt{1 - \frac{64}{{\rm Pe}^2}},
\end{equation}
where ${\rm Pe} = v/\sqrt{D\gamma}$ is the P\'eclet number. Note that the P\'eclet number must be larger than ${\rm Pe}_c = 8$, to observe MIPS. The MIPS state is a high-density diagonal band on a low density phase, whose densities are denoted by $\rho_{\rm high}$ and $\rho_{\rm low}$, respectively. Figure~\ref{fig_hydro_MPS1}(a) shows the numerically obtained phase-separated density profile for $\rho_0=0.75$ and ${\rm Pe} = 10$. The right boundary of the diagonal high-density band is mainly populated by state $\sigma=2$ (top) and $\sigma=3$ (left) and the left boundary by state $\sigma=1$ (right) and $\sigma=4$ (down). This leads to the relation $\rho_1+\rho_3 = \rho_2 + \rho_4$, irrespective of the direction in which the diagonal band is formed.

In Appendix~\ref{ls_mps1}, we also derive two relations linking implicitly $\rho_{\rm low}$ and $\rho_{\rm high}$, and demonstrate that the binodals are independent of $\epsilon$. The demonstration is similar to the one made in Ref.~\cite{ALG}, for an active lattice gas with a slightly different hydrodynamic equation. From these two relations and for ${\rm Pe} = 10$, we get $\rho_{\rm low}^{\rm th}\simeq 0.389$ and $\rho_{\rm high}^{\rm th}\simeq 0.927$, comparable to the densities numerically obtained in Fig.~\ref{fig_hydro_MPS1}(a). At the large P\'eclet limit, we derive the asymptotic behavior: $\rho_{\rm low} \simeq -(8/{\rm Pe}^2) \ln (16/3{\rm Pe}^2)$ and $\rho_{\rm high} \simeq 1 - 16/3{\rm Pe}^2$. Figure~\ref{fig_hydro_MPS1}(b) shows the velocity-density phase diagram computed with the phase-separated density profiles, validating the independence of the binodal densities $\rho_{\rm low}$ and $\rho_{\rm high}$ on $\epsilon$. The spinodal and binodal lines obtained analytically are also represented.

\subsection{Hard-core restriction (${\rm MPS} > 1$)}
\label{hydrohc}

\begin{figure}[t]
\centering
\includegraphics[width=\columnwidth]{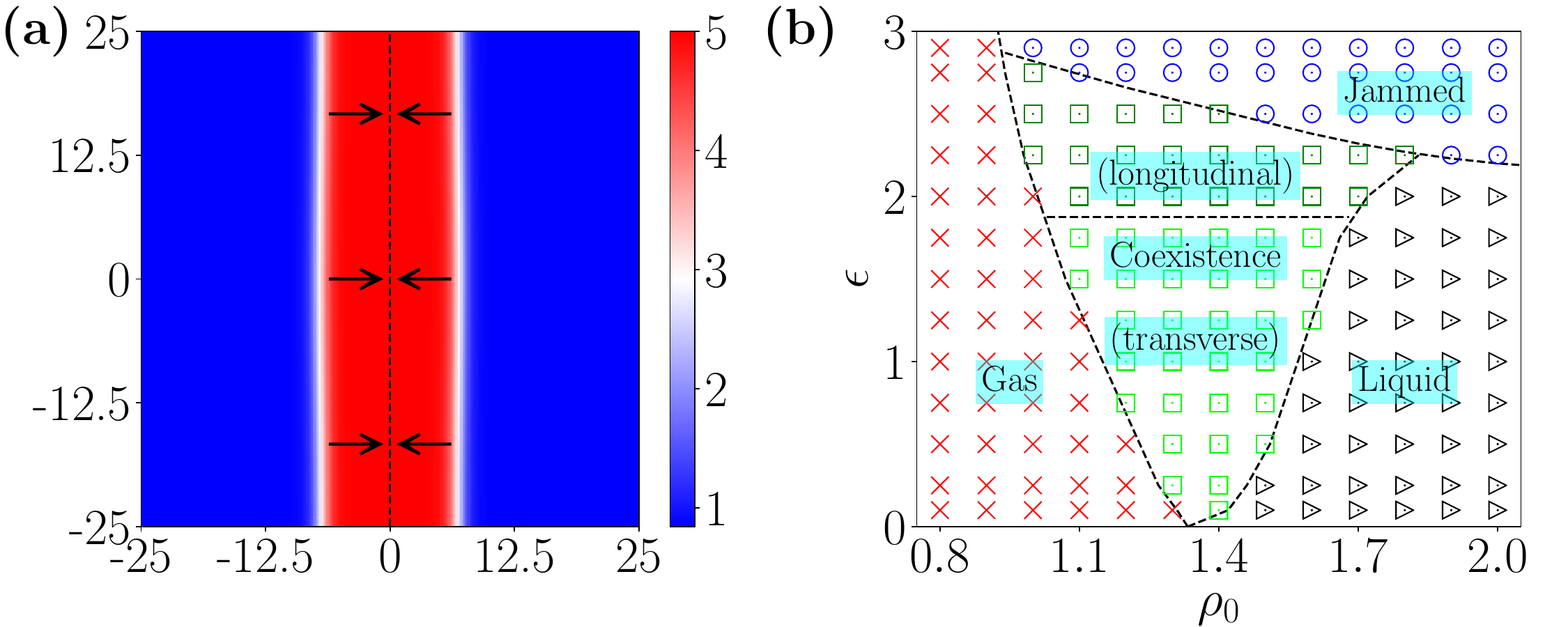}
\caption{(color online) (a)~Jammed density profile for the hard-core rAPM
with ${\rm MPS}=5$, for $\beta=0.75$, $\rho_0=2$, $\epsilon=2.7$, obtained numerically with FreeFEM++ for ${\cal N}=75$, $\Delta t = 0.02$ and $t_{\rm max}=1000$. (b)~Velocity-density phase diagram for the hard-core rAPM, computed with the numerical solutions of Eq.~\eqref{eqmps}, for $\beta=0.75$ and ${\rm MPS}=6$ in a square domain of linear size $L=50$.}
\label{fig_hydro_MPS}
\end{figure}

For ${\rm MPS} > 1$, with $f(\rho) = 1 - \ess\rho$, Eq.~(\ref{eqhydro}) becomes:
\begin{align}
\partial_t \rho_\sigma &= D_\parallel \partial_\parallel \left[ (1-\ess\rho) \partial_\parallel \rho_\sigma + \ess \rho_\sigma \partial_\parallel \rho \right] \nonumber \\
&+ D_\perp \partial_\perp \left[(1-\ess\rho) \partial_\perp \rho_\sigma + \ess \rho_\sigma \partial_\perp \rho  \right]\nonumber\\
&- v \partial_\parallel \left[ (1-\ess\rho) \rho_\sigma \right] + \sum_{\sigma'\ne\sigma} K_{\sigma \sigma'} (\rho_\sigma - \rho_{\sigma'}),\label{eqmps}
\end{align}
with $K_{\sigma \sigma'}$ defined in Eq.~(\ref{Iflip}) and $\ess=1/{\rm MPS}$. The homogeneous solutions, given by the solutions of $K_{\sigma \sigma'} (\rho_\sigma - \rho_{\sigma'}) = 0$, are those of the unrestricted APM derived in Refs.~\cite{APM,APM-2}. The disordered homogeneous solution $\rho_\sigma = \rho_0 / 4$ corresponds to a gas phase. The ordered homogeneous solution is given by the relation $K_{\sigma \sigma'} = 0$, corresponding to a liquid phase moving in a given direction. For a polar liquid along the state $\sigma = 1$, the densities are $\rho_1 = \rho_0 (1+3M) / 4$ and $\rho_{2,3,4} = \rho_0 (1-M) / 4$ with the magnetization $M$:
\begin{equation}
M = \frac{\beta J}{\alpha} \left[ 1 \pm \sqrt{1 + \frac{\alpha \mu_0}{\beta J}} \right],
\end{equation}
with $\mu_0 = 2 \beta J -1 -r/\rho_0$. This ordered homogeneous solution exists only when $\alpha \mu_0 + (\beta J)^2>0$ i.e. for density larger than
\begin{equation}
\rho_* = \frac{8(1-2\beta J/3) r}{1 + 8(2\beta J - 1)(1-2\beta J/3)}.
\end{equation}
Additionally, the temperature must be below $T_c = (1 - \sqrt{22}/8)^{-1} \simeq 2.417$. For $\epsilon=0$, the transition between the gas phase ($M=0$) and the liquid phase ($M>0$) is discontinuous at density $\rho_*$. For $\epsilon>0$, a first-order liquid-gas phase transition occurs, with a phase-separation made by a liquid stripe on a gas background. Two kind of phase-separated profiles can be observed: a transverse or longitudinal band motion for which the liquid phase is mainly populated by one state, and a jammed state formed by two locked liquid bands mainly populated by oppositely moving states ($\sigma=1,3$ or $\sigma=2,4$). Figure~\ref{fig_hydro_MPS}(a) shows a numerically obtained jammed density profile, for $\beta=0.75$, $\rho_0=2$, $\epsilon=2.7,$ and ${\rm MPS}=5$. For $x<0$, the liquid band is mainly populated by the right state $\sigma=1$, while for $x>0$ it is mainly populated by the left state $\sigma=3$, explaining why the band is jammed. 

Figure~\ref{fig_hydro_MPS}(b) shows the velocity-density phase diagram determined  by the numerical solutions of Eq.~\eqref{eqmps} for ${\rm MPS}=6$ and $\beta=0.75$. The topology of this phase diagram agrees well with the what we obtained for the microscopic model shown in Fig.~\ref{fig7}(b), with jammed state at large velocities and densities, and flocking band motion at low velocities and densities (transverse band motion for $\epsilon<1.9$ and longitudinal band motion for $\epsilon>1.9$). Note that the difference in the relevant density $\rho_0$ region is due to the different  MPS values used [MPS=20 in Fig.~\ref{fig7}(b), MPS=6 here].

In Appendix~\ref{ls_hc}, we perform the linear stability analysis of the disordered and ordered homogeneous solutions of Eq.~\eqref{eqmps}. These eigenvalues allow the determination of the velocity for which the reorientation transition occurs, denoted $\epsilon_*$, as a function of ${\rm MPS}$ and $T<T_c$. $\epsilon_*$ is a decreasing function of ${\rm MPS}$, meaning that $\epsilon_*$ increases with the restriction. However, the existence of the jammed state cannot be derived from the linear stability analysis of a homogeneous solution.

\subsection{Soft-core restriction}

\begin{figure}[t]
\centering
\includegraphics[width=\columnwidth]{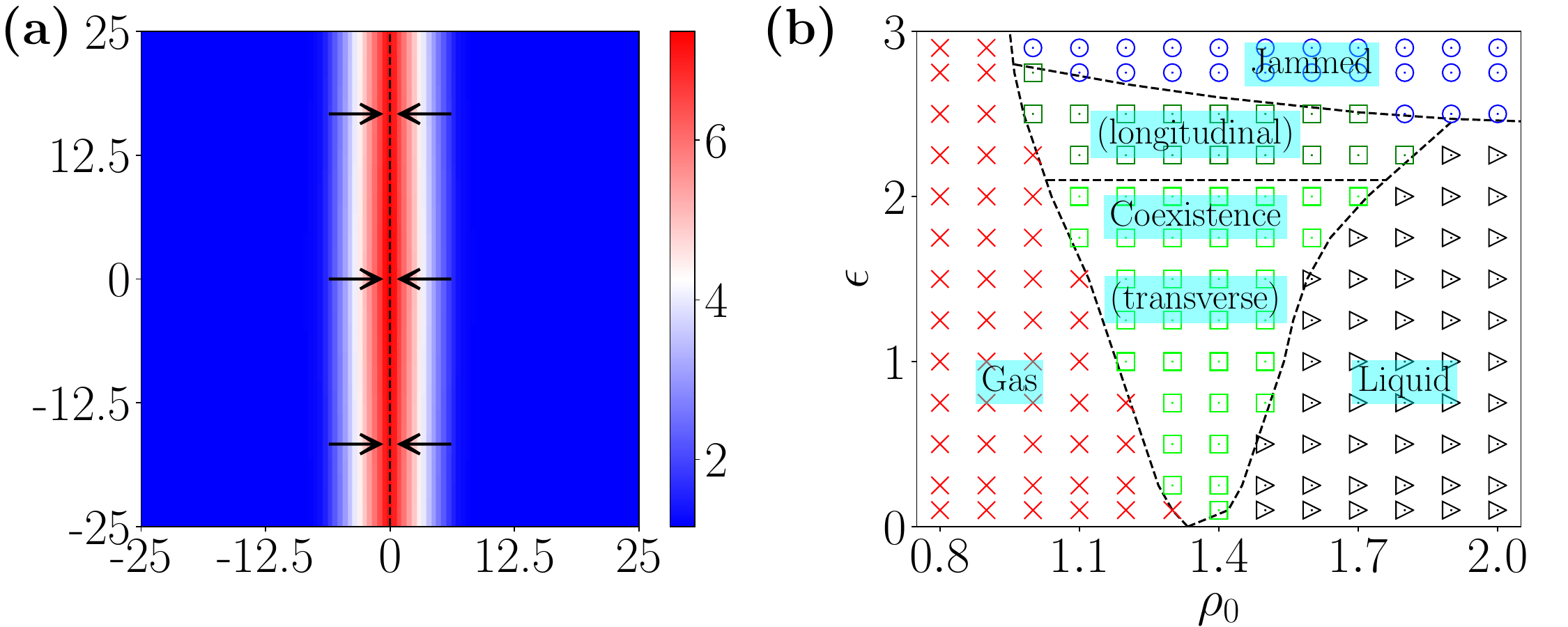}
\caption{(color online) (a)~Jammed density profile for the soft-core rAPM
with $U=1$, for $\beta=0.75$, $\rho_0=2$, $\epsilon=2.7$, obtained numerically with FreeFEM++ for ${\cal N}=75$, $\Delta t = 0.1,$ and $t_{\rm max}=5000$. (b)~Velocity-density phase diagram for the soft-core rAPM, computed with the numerical solutions of Eq.~\eqref{eqsc}, for $\beta=0.75$ and $U=0.25$ in a square domain of linear size $L=50$.}
\label{fig_hydro_softcore}
\end{figure}

For soft-core rAPM, with $f(\rho) = \exp(-s \rho)$, Eq.~(\ref{eqhydro}) becomes:
\begin{align}
\partial_t \rho_\sigma &= D_\parallel \partial_\parallel \left[ \exp(-s\rho) \left( \partial_\parallel \rho_\sigma + s \rho_\sigma \partial_\parallel \rho \right) \right] \nonumber \\
&+ D_\perp \partial_\perp \left[ \exp(-s\rho) \left( \partial_\perp \rho_\sigma + s \rho_\sigma \partial_\perp \rho \right) \right]\nonumber\\
&- v \partial_\parallel \left[ \exp(-s\rho) \rho_\sigma \right] + \sum_{\sigma'\ne\sigma} K_{\sigma \sigma'} (\rho_\sigma - \rho_{\sigma'}),\label{eqsc}
\end{align}
with $K_{\sigma \sigma'}$ defined in Eq.~(\ref{Iflip}) and $s=2\beta U$. The homogeneous solutions, given by the solutions of $K_{\sigma \sigma'} (\rho_\sigma - \rho_{\sigma'}) = 0$, are those of the unrestricted APM derived in Refs.~\cite{APM,APM-2}, as well as those explained for the hard-core rAPM in Sec.~\ref{hydrohc}. Similarly, a first-order liquid-gas phase transition occurs, with a phase-separation made by a liquid stripe on a gas background. The same two kinds of phase-separated profiles can be observed: a transverse or longitudinal band motion for which the liquid phase is mainly populated by one state, and a jammed state formed by two locked liquid bands mainly populated by oppositely moving states ($\sigma=1,3$ or $\sigma=2,4$). Figure~\ref{fig_hydro_softcore}(a) shows a numerically obtained jammed density profile, for $U=1$ $\beta=0.75$, $\rho_0=2$, $\epsilon=2.7$. For $x<0$, the liquid band is mainly populated by the right state $\sigma=1$, while for $x>0$ it is mainly populated by the left state $\sigma=3$, explaining why the band is jammed. 

Figure~\ref{fig_hydro_softcore}(b) shows the velocity-density phase diagram determined by the numerical solutions of Eq.~\eqref{eqsc}, for $\beta=0.75$ and $U=0.25$. The topology of this phase diagram agrees well with what we obtained for the microscopic model shown in Fig.~\ref{fig11}(b), with a jammed state at large velocities and densities, and flocking band motion at low velocities and densities (transverse band motion for $\epsilon<2.1$ and longitudinal band motion for $\epsilon>2.1$). Note that the difference in the relevant density $\rho_0$ region is due to the different $U$ values used ($U=0.07$ in Fig.~\ref{fig11}(b), $U=0.25$ here).

In Appendix~\ref{ls_sc}, we perform the linear stability analysis of the disordered and ordered homogeneous solutions of Eq.~\eqref{eqsc}. These eigenvalues allow the determination of the velocity for which the reorientation transition occurs, denoted $\epsilon_*$, as a function of $U$ and $T<T_c$. $\epsilon_*(U)$ increases for small $U$ and decreases to zero at large $U$. Again, the existence of the jammed state cannot be derived from the linear stability analysis of a homogeneous solution.

%%%%%%%%%%%%%%%%%%%%%%%%%%%%%%%%%%%%%
%%%%%%%%%%%% SUMMARY %%%%%%%%%%%%%%%%
%%%%%%%%%%%%%%%%%%%%%%%%%%%%%%%%%%%%%

\section{Summary and discussion}
\label{s4}
To summarize, we have analyzed a discretized flocking model
with volume exclusion and showed that the interplay between
alignment and on-site repulsion produces a vast spectrum of self-organized patterns ranging from jammed clusters or bands, and we have argued that they are a manifestation of MIPS,
which relies on the reduction of particle velocity with increasing local density~\cite{geyer2019}. Generally, it has been observed that velocity alignment interactions promote MIPS~\cite{sese2018,sese2021}. In the rAPM considered here, alignment is even necessary for MIPS to occur, since the jammed clusters disappear in the gas phase for $T \to \infty$, i.e., vanishing alignment. For increasing alignment, i.e., small $T$, either orientationally ordered domains appear in the jammed clusters, arranged in such a way that the cluster configuration is kinetically arrested (up to fluctuations), or, depending on density and self-propulsion strength, the jammed clusters dissolve into a orientationally ordered liquid phase, both manifestations of flocking.

The phase diagrams of the rAPM with ${\rm MPS}=1$ and those
for ${\rm MPS}>1$ or soft-core repulsion turn out to be different
due to the absence of alignment interactions for ${\rm MPS}=1$.
The latter model is equivalent to an active lattice gas
with persistent walkers instead of diffusing particles.
Consequently, the rAPM with ${\rm MPS}=1$ is always in an
orientationally disordered (gas) phase in which various
MIPS or jammed states occur. For ${\rm MPS}>1$ or soft-core
repulsion in addition to various jammed states the three
typical flocking phases occur: the orientationally
disordered gas, liquid-gas coexistence (flocking phase),
and the orientationally ordered liquid.

Part of the phenomenology we describe in our paper has been seen in an experiment of a colloidal-roller system~\cite{geyer2019}, where the system undergoes a phase transition from a gas phase to a solid jam phase via flocking as the packing fraction of the rollers is increased.
It would be interesting to study the effects of volume exclusion on the transport and jamming of active particles in disordered landscapes. Quenched disorder is abundant in all natural systems and is known to reduce the effect of local interactions. The altered composition may have an impact on the universal behavior of unrestricted systems, in both equilibrium and out-of-equilibrium conditions \cite{Weinrib1983,Viktor1995,Kumar2017}. It is also known that active matter systems with random quenched disorder undergo activity-induced jamming~\cite{Reichhardt2021,Reichhardt2014}. Preliminary results for the restricted and the unrestricted APM with quenched disorder reveal interesting emergent collective behaviors~\cite{RFrAPM}. Another extension of our research would be to look into the influence of soft-core constraints on active systems with continuous symmetry, such as the Vicsek model where initial investigation suggests an arrest of the flocking state with the emergence of MIPS jammed clusters. 

%%%%%%%%%%%%%%%%%%%%%%%%%%%%%%%%%%%%%%%%%%%%
%%%%%%%%%%%% ACKNOWLEDGMENT %%%%%%%%%%%%%%%%
%%%%%%%%%%%%%%%%%%%%%%%%%%%%%%%%%%%%%%%%%%%%
\begin{acknowledgments}
M.K. would like to acknowledge financial support in the form of a research fellowship from CSIR, goverment of India (Award No. 09/080(1106)/2019-EMR-I). R.P. and M.K. thank the Indian Association for the Cultivation of Science (IACS) for the computational facility. R.P. thanks SFB 1027 for supporting his visit to Saarland University. S.C., M.M., and H.R. are financially supported by the German Research Foundation (DFG) within the Collaborative Research Center SFB 1027. 
\end{acknowledgments}
%%%%%%%%%%%%%%%%%%%%%%%%%%%%%%%%%%%%%%
%%%%%%%%%%%% APPENDIX %%%%%%%%%%%%%%%%
%%%%%%%%%%%%%%%%%%%%%%%%%%%%%%%%%%%%%%

\newpage
\onecolumngrid
\appendix
\section{Mean-squared displacements (MSD) of particles in the jammed state}
\label{appendix_msd}
\begin{figure}[t]
\centering
\includegraphics[width=0.8\columnwidth]{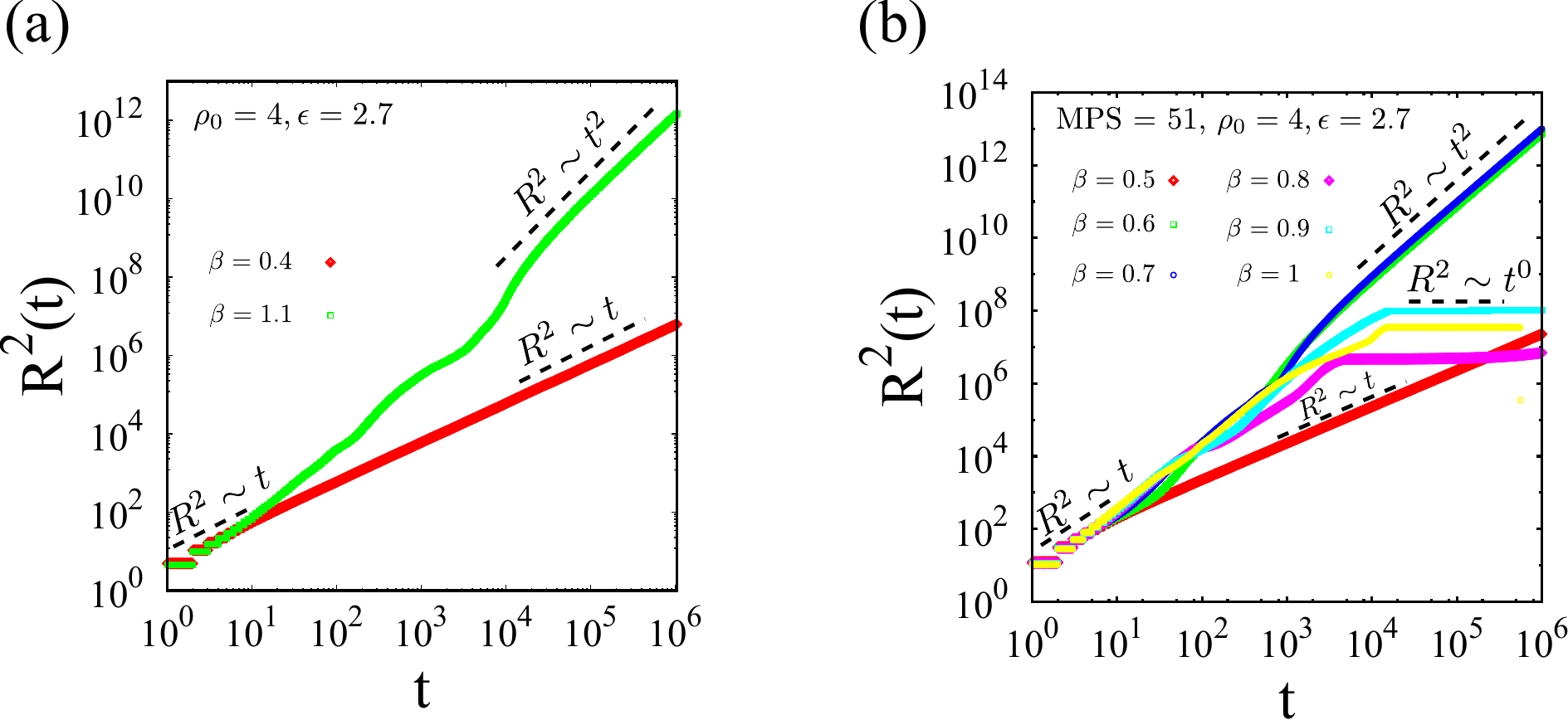}
\caption{(color online) MSD $R^2(t)$ vs $t$ (on a log-log scale) for several $\beta$. (a) Unrestricted APM and (b) rAPM with hard-core repulsion.}
\label{appfig1} 
\end{figure}
Here, we explore the appearance of arrested states through measurements of the MSD of individual particles. The MSD of $N$ number of particles in the system at time $t$ (which quantifies how the particles move from their initial positions under various volume exclusion effects) is defined as
\begin{align}
R^2(t)=\frac{1}{N}\sum_{i=1}^N |\mathbf{r}_i(t)-\mathbf{r}_i(0)|^2 \, ,
\end{align}
where $\mathbf{r}_i(t)$ is the instantaneous position of the $i$-th particle at time $t$. For ballistic motion, $R^2 \sim t^2$ while for diffusive motion $R^2 \sim t$. For an arrested or jammed state, however, MSD is proportional to $t^x$ with $x \sim 0$~\cite{Merrigan2020}.

In Fig.~\ref{appfig1}(a), we show $R^2$ vs $t$ (on a log-log scale) for the unrestricted APM~\cite{APM} as a function of $\beta$ ($T^{-1}$). At small $\beta$, which physically signifies the gaseous phase, the system obeys the diffusive growth $R^2 \sim t$, whereas, for small temperature, where the system exhibits the liquid phase, we observe two distinct regimes in the MSD. The small $t$ limit is characterized by a diffusive regime with $R^2 \sim t$ growth, whereas a ballistic growth regime characterized by the power law $R^2 \sim t^2$ is observed at large $t$. In the liquid phase, advective force (self-propulsion) plays a crucial role as the system exhibits the ballistic growth regime, which is a collision-free regime in which particles travel freely after a majority of the particles switch in the same state.

MSD for the rAPM with hard-core repulsion is shown in Fig.~\ref{appfig1}(b). At small $\beta$, the MSD shows a diffusive growth as the system is in the gas phase. For large $\beta$, the system exhibits the liquid phase, and we observe a crossover from the diffusive growth at small $t$ to a ballistic growth at large $t$. For intermediate $\beta$, the small $t$ diffusive growth regime is followed by a plateau in the MSD at large $t$, which signifies a jammed or arrested state. A similar crossover in MSD can be seen for a random walker confined in a box, where the MSD crosses over from a diffusive growth ($R^2 \sim t$) to a plateau once the walker sticks to the walls~\cite{Merrigan2020}.

The MSD plots signifying the jammed phase describe the fact that initially the particles (the system is initially prepared homogeneous) do not feel the effect of the hopping restrictions and move diffusively but as the system coarsens with time, particles feel the restricted environment and finally, the system reaches the steady state jammed phase.

\section{Structural characteristics and transformation of a jammed phase}
\label{appendix_jam_structure}
\begin{figure}[t]
\centering
\includegraphics[width=0.9\columnwidth]{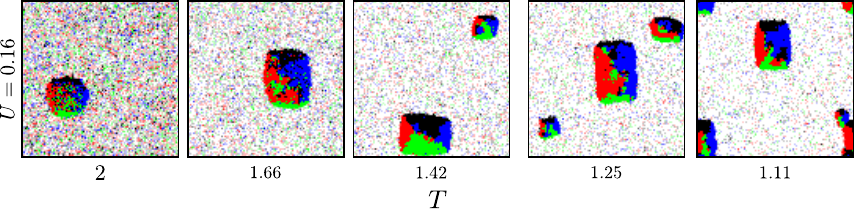}
\caption{(color online) Morphological characteristics and transformation of jammed clusters in the rAPM with soft-core interactions as a function of temperature. Parameters: $U=0.16$, $\epsilon=2.7$, and $\rho_0=4$.}
\label{appfig2}
\end{figure}

Fig.~\ref{appfig2} displays the temperature-dependent structural changes of a jammed cluster. The jammed phase observed in this study is a kinetically arrested phase due to MIPS, where the particles cease to move due to steric repulsion. The active particles in this study interact only repulsively (when the MPS is reached or when dealing with soft-core), and as local density increases, the speed of the motile particles decreases, which results in a phase-separated state with a dilute active gas coexisting with a dense jammed cluster. At low temperatures, this jammed cluster comprises four orientationally ordered sub-domains in a completely gridlocked position, but as temperature is increased, the overall area of the jammed cluster shrinks and the internal structure of the cluster becomes orientationally disordered. This happens because as the on-site ferromagnetic alignment strength between the particles decreases with temperature, the flipping probability increases, which helps the particles to change their predominant hopping direction and break from the gridlock.

\section{Effect of soft-core repulsion on active models in continuum}
\label{racm_vm}
\begin{figure}[t]
\centering
\includegraphics[width=0.9\columnwidth]{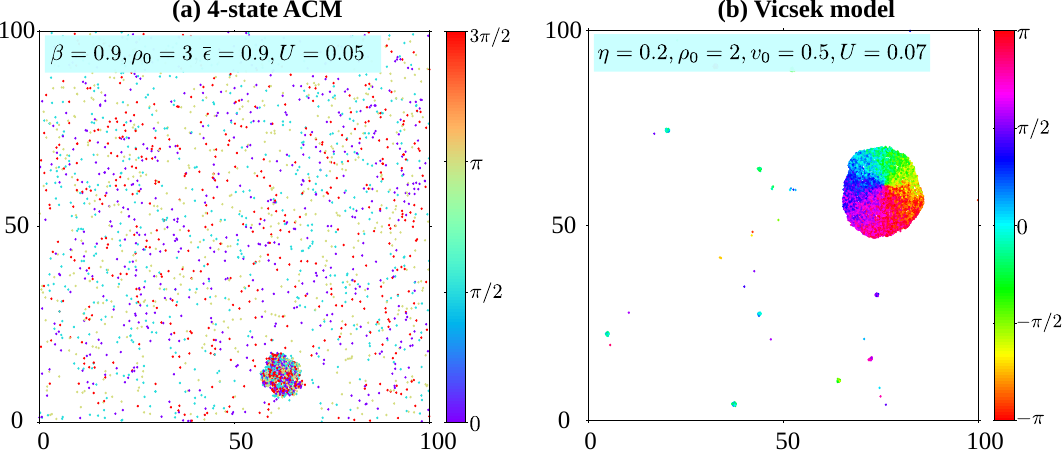}
\caption{(color online) 
Jammed clusters in the active clock model and Vicsek model with soft-core repulsion: (a) four-state ACM ~\cite{ACM} ($\beta=0.9$, $\rho_0=3$, $\bar{\epsilon}=0.9$, and $U=0.05$) and (b) Vicsek model \cite{solon-vicsek} (noise $\eta=0.2$, $\rho_0=2$, $v_0=0.5$, and $U=0.07$). Colorbar represents particle orientation, which for four-state ACM particles can have only four discrete orientations: 0, $\pi/2$, $\pi$, and 3$\pi/2$, whereas for VM, a particle can have any orientation in $[-\pi,\pi]$.}
\label{appfig3}
\end{figure}

Our results are not artifacts of the discrete space we consider, and to establish that we investigate the effect of soft-core repulsion on active models which are defined on an off-lattice geometry such as the ACM~\cite{ACM} and the VM~\cite{solon-vicsek}. We take the four-state ACM with density-dependent motility where particle hopping is allowed with a probability $\exp(-\beta U n_i)$ ($U$ is the strength of the steric repulsion and $n_i$ is the number of particles in the unit neighborhood of particle $i$ which is trying to propel). For such a set up, a jamming transition reminiscent of the rAPM is observed from a flocking phase to a MIPS induced jammed phase as the repulsion strength $U$ is increased. The MIPS state of four-state ACM with a soft-core repulsion is shown in Fig.~\ref{appfig3}(a), where the particle orientations signify that the internal structure of the cluster is disordered. With an off-lattice geometry, the domain boundaries of the jammed cluster are not as sharp as one observes with the lattice geometry, but apart from this apparent difference, the underlying physics of jamming is the same.

Our findings can also be extended to continuum to the active XY model ($q \to \infty$)~\cite{ACM}, or the VM \cite{solon-vicsek} with soft-core repulsion. For instance, our study of the VM with soft-core repulsion, where the particle velocity  is modified as $v_0 \to v_0\exp(-Un_i)$, demonstrates that such a configuration produces circular or elliptical high-density jammed clusters with particle orientation continuously distributed between $[-\pi,\pi]$ [see Fig.~\ref{appfig3}(b)]. This jammed phase is a MIPS cluster typical of repulsively interacting active Brownian particles (ABP)~\cite{puglisi2020} and happens above a threshold value of $U$, below which the system exhibits flocking behavior. A similar kind of density-dependent velocity was introduced in a model of active Brownian particles with alignment interaction \cite{farrell2012}, where at large restriction, the system was shown to exhibit an aster-like jammed stationary phase. In both cases, the origin lies in the slowdown of particles due to crowding jamming.

\section{Zero activity ($\epsilon=0$) limit of the rAPM}
\label{appendix_diffusive}
\begin{figure}[t]
\centering
\includegraphics[width=0.6\columnwidth]{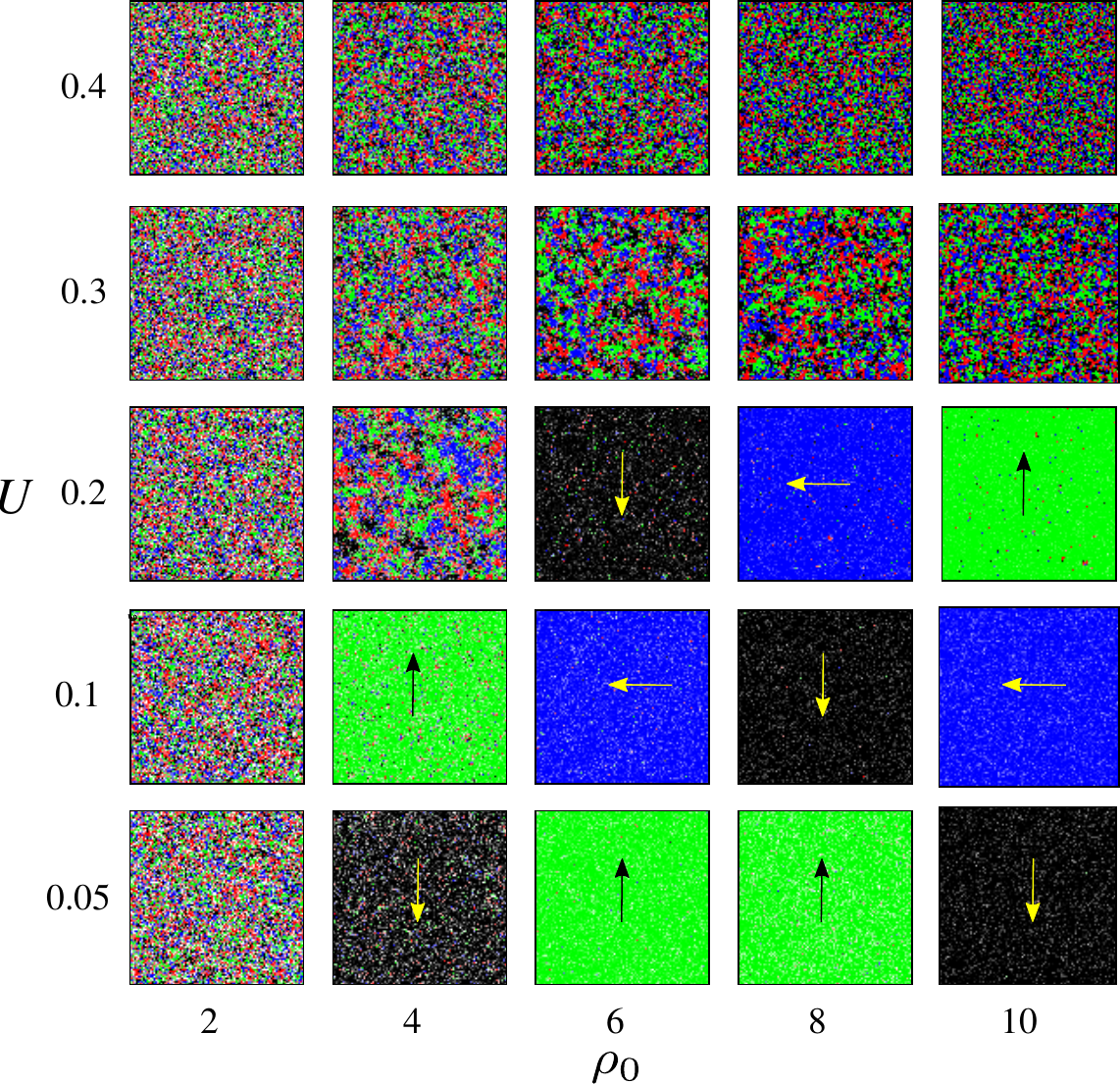}
\caption{(color online) $U-\rho_0$ phase diagram of the rAPM with soft-core
repulsion with snapshots at $t=10^5$ for $\epsilon=0$ and $\beta=0.7$. For small $U$, similar to the unrestricted APM, we observe a phase transition from a disordered gaseous phase to an ordered liquid phase as density is increased. Due to $q=4$, a steady-state liquid phase can be formed by any of the four states, and the color code for the liquid phase is red ($q=1$), green ($q=2$), blue ($q=3$), and black ($q=4$). Arrows indicate the direction of motion. At large $U$, particles can segregate only locally into small clusters forming a orientationally disordered phase at large spatial scales.}
\label{appfig4}
\end{figure}

In Fig.~\ref{appfig4}, we show the late-stage representative snapshots of the rAPM with soft-core repulsion in the $U-\rho_0$ plane for the zero velocity ($\epsilon=0$) limit where particle movement is controlled by diffusion (particles hop without any bias). For small repulsion, the system goes through a direct gas-liquid phase transition without a coexistence regime similar to the unrestricted APM \cite{APM}. With strong repulsion, however, the system exhibits orientational disorder at high densities, where the system forms a domain state without any long-range order because at large density and strong repulsion, particle hopping is impeded (``jammed") and particles align their states in small clusters. The average size of these domains does not grow in time, i.e., the system reaches a stationary state with a specific average domain size which can be quantified following a characteristic length scale analysis.

Note that at large $U$ and the $\rho_0$ limit, the active particles become nearly immobile, and at this limit our model is equivalent to the equilibrium four-state Potts model on a random 2d graph, which should show a long-range order, but our
model clearly does not exhibit that. This is because the equilibrium four-state Potts model is defined with nearest-neighbor interaction, which helps in domain coarsening, whereas in the soft-core rAPM with large $U$ and $\rho_0$, the probability that an edge is present between nodes in the network following the soft-core hopping acceptance probability $\exp(-2\beta U \rho_0)$ is nearly zero. Therefore, at this limit, the system behaves as a disconnected random 2D graph and does not display long-range order.

We have also checked that the order parameter at $\epsilon=0$ for small $U$ values ($U = 0.05$, 0.1, 0.15) shows a discontinuous jump at a critical density ($\rho_* \sim 3.5$ for $\beta = 0.7$) similar to the unrestricted APM.

%%%%%%%%%%%%%%%%%%%%%%%%%%%%%%%%%%%%%%%%%%%%%%
%%%% Derivation of hydrodynamic equations %%%%
%%%%%%%%%%%%%%%%%%%%%%%%%%%%%%%%%%%%%%%%%%%%%%

\section{Derivation of hydrodynamic equations}
\label{hydro}
In this section, we will derive the hydrodynamic equations for the $q$-state rAPM. In Ref.~\cite{APM-2}, we presented the detailed hydrodynamic description of the unrestricted APM and their numerical solutions. In the rAPM, as we have modified only the rule of hopping dynamics of the particles (keeping the flipping rule unchanged), we present here the derivation of the hydrodynamic equations only for the hopping term in detail. To represent the different hopping restrictions, we introduce a function $f(n_i)$ where $i$ denotes the arrival site. The form of this function is $f(\rho) = 1 - \ess\rho$ for ${\rm MPS} = 1/\ess$, and $f(\rho) = \exp(-s \rho)$ for the soft-core rAPM where $s=2\beta U$.

The master equation writes
\begin{gather}
n_i^\sigma(t+dt) = n_i^\sigma(t) \left[ 1 - dt \sum_p W_{\rm hop}(\sigma,p) f(n_{i+p}) - dt \sum_{\sigma \ne \sigma'} W_{\rm flip} (\sigma \to \sigma') \right] \nonumber\\
+ dt \sum_p W_{\rm hop}(\sigma,p) n_{i-p}^\sigma(t) f(n_i) + dt \sum_{\sigma \ne \sigma'} n_i^{\sigma'}(t) W_{\rm flip} (\sigma' \to \sigma).
\end{gather}
Taking the limit $dt \to 0$, we get
\begin{equation}
\partial_t n_i^\sigma = \sum_p W_{\rm hop}(\sigma,p) \left[ n_{i-p}^\sigma f(n_i) - n_i^\sigma f(n_{i+p}) \right] + \sum_{\sigma' \ne \sigma} \left[ n_i^{\sigma'} W_{\rm flip}(\sigma', \sigma)- n_i^{\sigma} W_{\rm flip}(\sigma, \sigma') \right].
\end{equation}
In the following, we decompose the r.h.s of this equation into two terms: the hopping term $I_{\rm hop}$ and the flipping term $I_{\rm flip}$, unchanged under hopping restrictions. We then have $\partial_t n_i^\sigma = I_{\rm hop} + I_{\rm flip}$. Using the definition of $W_{\rm hop}$, given by Eq.~\eqref{whop}, we obtain
\begin{equation}
I_{\rm hop} = D\left(1- \frac{\epsilon}{q-1} \right)\sum_{p} \left[ n_{i-p}^\sigma f(n_i) - n_i^\sigma f(n_{i+p}) \right] + \frac{qD\epsilon}{q-1} \left[ n_{i-\sigma}^\sigma f(n_i) - n_i^\sigma f(n_{i+\sigma}) \right].\label{Ihop1}
\end{equation}

We consider the Taylor expansion of the function $G_{i+p}$ as
\begin{equation}
G_{i+p} = G_i + a \partial_p G_i + \frac{a^2}{2} \partial_p^2 G_i + {\cal O}(a^3),
\end{equation}
which leads to the expression
\begin{equation}
G_{i-p} H_i - G_i H_{i+p} = -a \partial_p[G_i H_i] + \frac{a^2}{2} [H_i \partial_p^2 G_i - G_i \partial_p^2 H_i] + {\cal O}(a^3).
\end{equation}
Knowing that the sum of the derivatives of any function $F$ are
\begin{gather}
\sum_{p=1}^q \partial_p F = \sum_{p=1}^q ({\bf e_p} \cdot {\bf e_x}) \partial_x F + \sum_{p=1}^q ({\bf e_p} \cdot {\bf e_y}) \partial_y F = 0, \\
\sum_{p=1}^q \partial_p^2 F = \sum_{p=1}^q ({\bf e_p} \cdot {\bf e_x})^2 \partial_x^2 F + 2\sum_{p=1}^q ({\bf e_p} \cdot {\bf e_x})({\bf e_p} \cdot {\bf e_y}) \partial_x \partial_y F + \sum_{p=1}^q ({\bf e_p} \cdot {\bf e_y})^2 \partial_y^2 F = \frac{q}{2} \nabla^2 F,
\end{gather}
we finally get
\begin{equation}
\sum_{p=1}^q [G_{i-p} H_i - G_i H_{i+p}] = \frac{qa^2}{4} [H_i \nabla^2 G_i - G_i \nabla^2 H_i] + {\cal O}(a^3).
\end{equation}

Equation.~\eqref{Ihop1} becomes
\begin{equation}
I_{\rm hop} = D\left(1- \frac{\epsilon}{q-1} \right) a^2 \left[ f(n_i) \nabla^2 n_i^\sigma - n_i^\sigma \nabla^2 f(n_i) \right] + \frac{qD\epsilon}{q-1} \left\{ - a \partial_\sigma\left[ f(n_i)n_i^\sigma \right] + \frac{a^2}{2} \left[ f(n_i) \partial_\sigma^2 n_i^\sigma - n_i^\sigma \partial_\sigma f(n_i) \right] \right\}.
\end{equation}
We can decompose $a^2 \nabla^2 = \partial_\parallel^2 + \partial_\perp^2$ and $a \partial_\sigma = \partial_\parallel$, and denote $\rho_\sigma = \langle n_i^\sigma \rangle$ as well as $\rho = \langle \rho_i \rangle$, which leads to the expression
\begin{equation}
\langle I_{\rm hop} \rangle = D_\parallel \left[ f(\rho) \partial_\parallel^2 \rho_\sigma - \rho_\sigma \partial_\parallel^2 f(\rho) \right] + D_\perp \left[ f(\rho) \partial_\perp^2 \rho_\sigma - \rho_\sigma \partial_\perp^2 f(\rho) \right] - v \partial_\parallel \left[ f(\rho) \rho_\sigma \right],
\end{equation}
with 
\begin{equation}
D_\parallel = D\left(1+ \frac{\epsilon}{q-1} \right), \quad D_\perp = D\left(1- \frac{\epsilon}{q-1} \right) \quad  {\rm and} \quad  v= \frac{qD\epsilon}{q-1}.
\end{equation}
Note that we can decompose
\begin{equation}
f(\rho) \partial_i^2 \rho_\sigma - \rho_\sigma \partial_i^2 f(\rho) = \partial_i \left[ f(\rho) \partial_i \rho_\sigma - \rho_\sigma \partial_i f(\rho) \right],
\end{equation}
and $\partial_i f(\rho) = f'(\rho)\partial_i \rho$, which yields the drift term of the rAPM equation
\begin{equation}
\langle I_{\rm hop} \rangle = D_\parallel \partial_\parallel \left[ f(\rho) \partial_\parallel \rho_\sigma - f'(\rho)\rho_\sigma \partial_\parallel \rho \right] + D_\perp \partial_\perp \left[ f(\rho) \partial_\perp \rho_\sigma - f'(\rho) \rho_\sigma \partial_\perp \rho \right] - v \partial_\parallel \left[ f(\rho) \rho_\sigma \right].
\end{equation}

The current is then 
\begin{gather}
{J_\sigma}_\parallel = -D_\parallel \left[ f(\rho) \partial_\parallel \rho_\sigma - f'(\rho)\rho_\sigma \partial_\parallel \rho \right] + v f(\rho) \rho_\sigma, \label{Jpara} \\
{J_\sigma}_\perp = -D_\perp \left[ f(\rho) \partial_\perp \rho_\sigma - f'(\rho) \rho_\sigma \partial_\perp \rho \right], \label{Jperp}
\end{gather}
or in the vectorial form, $J_{\sigma i} = f(\rho) J_{\sigma i}^0 - \lambda_i \rho_\sigma \partial_i \rho$,
where ${\bf J}_\sigma^0$ is the current without restriction, and $\lambda_i$ a positive constant since $f(\rho)$ is a strictly decreasing function. The first term corresponds to the current without restriction multiplied by $f(\rho)$, while the second term corresponds to an additional current from high to low densities.

For ${\rm MPS} = 1$, the flipping term $I_{\rm flip}$ is calculated according to $W_{\rm flip}(\sigma \to \sigma') = \gamma$, which gives
\begin{equation}
\langle I_{\rm flip} \rangle = \gamma \sum_{\sigma' \ne \sigma} (\rho_{\sigma'} - \rho_{\sigma}).
\end{equation}
For ${\rm MPS} > 1$ and soft-core rAPM, the flipping term $I_{\rm flip}$ is calculated according to Eq.~\eqref{flipeq}, and equal to the flipping term of the unrestricted APM derived in Ref.~\cite{APM-2}:
\begin{equation}
\langle I_{\rm flip} \rangle = \sum_{\sigma' \ne \sigma} \left[ \frac{4\beta J}{\rho}(\rho_\sigma + \rho_{\sigma'}) - 1 - \frac{r}{\rho} - \alpha \frac{(\rho_\sigma - \rho_{\sigma'})^2}{\rho^2} \right] (\rho_{\sigma} - \rho_{\sigma'}),
\end{equation}
with $\alpha = 8(\beta J)^2(1-2\beta J/3)$.

The hydrodynamic equation for the rAPM is then	
\begin{equation}
\partial_t \rho_\sigma = - \partial_\parallel J_{\sigma \parallel} - \partial_\perp J_{\sigma \perp} + \sum_{\sigma' \ne \sigma} K_{\sigma \sigma'} (\rho_\sigma - \rho_{\sigma'}),
\end{equation}
where $J_{\sigma \parallel}$ and $J_{\sigma \perp}$ are given by Eqs.~\eqref{Jpara} and~\eqref{Jperp}, respectively, and the flipping interaction term is $K_{\sigma \sigma'} = - \gamma$ for ${\rm MPS}=1$ and
\begin{equation}
K_{\sigma \sigma'} = \frac{4\beta J}{\rho}(\rho_\sigma + \rho_{\sigma'}) - 1 - \frac{r}{\rho} - \alpha \frac{(\rho_\sigma - \rho_{\sigma'})^2}{\rho^2},
\end{equation}
for ${\rm MPS}>1$ and soft-core rAPM.

%%%%%%%%%%%%%%%%%%%%%%%%%%%%%%%%%%%%%%%
%%%% hydrodynamic theory for MPS=1 %%%%
%%%%%%%%%%%%%%%%%%%%%%%%%%%%%%%%%%%%%%%

\section{Linear stability analysis and binodal calculus for ${\rm MPS}=1$}
\label{ls_mps1}

For ${\rm MPS} = 1$, the hydrodynamic equation is
\begin{equation}
\partial_t \rho_\sigma = D_\parallel \partial_\parallel \left[ (1 - \rho) \partial_\parallel \rho_\sigma + \rho_\sigma \partial_\parallel \rho \right] + D_\perp \partial_\perp \left[ (1 - \rho) \partial_\perp \rho_\sigma + \rho_\sigma \partial_\perp \rho \right] - v \partial_\parallel \left[ (1 - \rho) \rho_\sigma \right] - \gamma(4\rho_\sigma - \rho).
\end{equation}

Using dimensionless coordinates $\tau = \gamma t$ and ${\bf X}=\sqrt{\gamma/D} {\bf x}$, the hydrodynamic equation rewrites
\begin{equation}
\partial_\tau \rho_\sigma = D_\parallel \partial_\parallel \left[ (1 - \rho) \partial_\parallel \rho_\sigma + \rho_\sigma \partial_\parallel \rho \right] + D_\perp \partial_\perp \left[ (1 - \rho) \partial_\perp \rho_\sigma + \rho_\sigma \partial_\perp \rho \right] - {\rm Pe} \partial_\parallel \left[ (1 - \rho) \rho_\sigma \right] - (4\rho_\sigma - \rho), \label{eqmps1app}
\end{equation}
with $D_\parallel = 1+\epsilon/3$, $D_\perp = 1-\epsilon/3$, and the P\'eclet number ${\rm Pe} = v / \sqrt{D\gamma}$. The only homogeneous solution is $\rho_\sigma = \rho_0/4$ for all states $\sigma$.

{\bf Linear stability analysis.} We consider a linear stability analysis for $\rho_\sigma = \rho_0/4 + \delta \rho_\sigma$, where $\delta \rho_\sigma \ll \rho_0$ is a small perturbation. Keeping only the first-order terms in $\delta \rho_\sigma$, the hydrodynamic equation becomes
\begin{gather}
\partial_\tau \delta \rho_\sigma = \left[\left( 1 - \frac{3\rho_0}{4} \right)(D_\parallel \partial_\parallel^2 + D_\perp \partial_\perp^2 ) - {\rm Pe} \left( 1 - \frac{5\rho_0}{4} \right) \partial_\parallel - 3 \right] \delta \rho_\sigma \nonumber \\
+ \left[\frac{\rho_0}{4} (D_\parallel \partial_\parallel^2 + D_\perp \partial_\perp^2 ) + {\rm Pe} \frac{\rho_0}{4} \partial_\parallel + 1 \right] \sum_{\sigma' \ne \sigma} \delta \rho_{\sigma'}.
\end{gather}

Performing a Fourier transform in space, we obtain
\begin{equation}
\partial_\tau \delta \rho_\sigma = A(k_\parallel,k_\perp) \delta \rho_\sigma + B(k_\parallel,k_\perp) \sum_{\sigma' \ne \sigma} \delta \rho_{\sigma'},
\end{equation}
with
\begin{gather}
A(k_\parallel,k_\perp) = \left( 1 - \frac{3\rho_0}{4} \right)(-D_\parallel k_\parallel^2 - D_\perp k_\perp^2 ) + \imath {\rm Pe} \left( 1 - \frac{5\rho_0}{4} \right) k_\parallel - 3 , \\
B(k_\parallel,k_\perp) = \frac{\rho_0}{4} (-D_\parallel k_\parallel^2 - D_\perp k_\perp^2 ) - \imath {\rm Pe} \frac{\rho_0}{4} k_\parallel + 1 .
\end{gather}

The stability of the homogeneous solution is then given by the eigenvalues of the matrix
\begin{equation}
M = \begin{pmatrix}
A(k_x,k_y) & B(k_x,k_y) & B(k_x,k_y) & B(k_x,k_y) \\
B(k_y,-k_x) & A(k_y,-k_x) & B(k_y,-k_x) & B(k_y,-k_x) \\
B(-k_x,-k_y) & B(-k_x,-k_y) & A(-k_x,-k_y) & B(-k_x,-k_y) \\
B(-k_y,k_x) & B(-k_y,k_x) & B(-k_y,k_x) & A(-k_y,k_x)
\end{pmatrix}.
\end{equation}

With the help of \textit{Mathematica}~\cite{mathematica}, we get that three eigenvalues are always negative and the fourth eigenvalue writes
\begin{equation}
\lambda = \frac{1}{8} \left[ -4 (D_\parallel + D_\perp) + {\rm Pe}^2(1-\rho_0)(2\rho_0-1) \right](k_x^2 + k_y^2) + {\cal O}(k_x^2,k_y^2).
\end{equation}
Using $D_\parallel + D_\perp=2$, the homogeneous solution is then stable if and only if $(1-\rho_0)(2\rho_0-1) < 8/{\rm Pe}^2$, leading to the spinodals $\varphi_\pm$:
\begin{equation}
\varphi_\pm = \frac{3}{4} \pm \frac{1}{4}\sqrt{1 - \frac{64}{{\rm Pe}^2}}, \label{spinodalapp}
\end{equation}
and a critical P\'eclet number ${\rm Pe}_c = 8$, to observe the MIPS.

\medskip

{\bf Derivation of the binodals.} Now we derive the expression of the binodals, denoted as $\rho_{\rm low}$ and $\rho_{\rm high}$, following the demonstration made in Ref.~\cite{ALG}, for an active lattice gas with a slightly different hydrodynamic equation. At steady state, Eq.~\eqref{eqmps1app} writes
\begin{equation}
0 = D_\parallel \partial_\parallel \left[ (1 - \rho) \partial_\parallel \rho_\sigma + \rho_\sigma \partial_\parallel \rho \right] + D_\perp \partial_\perp \left[ (1 - \rho) \partial_\perp \rho_\sigma + \rho_\sigma \partial_\perp \rho \right] - {\rm Pe} \partial_\parallel \left[ (1 - \rho) \rho_\sigma \right] - (4\rho_\sigma - \rho). \label{eqmps1ss}
\end{equation}
We define $\rho_x = \rho_1 + \rho_3$, $\rho_y = \rho_2 + \rho_4$, and the magnetization vector ${\bf m} = (m_x,m_y)$ with $m_x = \rho_1 - \rho_3$ and $m_y = \rho_2 - \rho_4$. From Eq.~\eqref{eqmps1ss}, the equation for $\rho$ writes
\begin{align}
0 &= D_\parallel \left\{ \partial_x \left[ (1 - \rho) \partial_x \rho_x + \rho_x \partial_x \rho \right] + \partial_y \left[ (1 - \rho) \partial_y \rho_y + \rho_y \partial_y \rho \right] \right\} \nonumber \\
&+ D_\perp \left\{ \partial_y \left[ (1 - \rho) \partial_y \rho_x + \rho_x \partial_y \rho \right] + \partial_x \left[ (1 - \rho) \partial_x \rho_y + \rho_y \partial_x \rho \right] \right\} \nonumber \\
&- {\rm Pe} \left\{ \partial_x \left[ (1 - \rho) m_x \right] + \partial_y \left[ (1 - \rho) m_y \right] \right\} .
\end{align}
From microscopic simulations and numerical solutions of Eq.~\eqref{eqmps1app}, we suppose the relation $\rho_x = \rho_y= \rho/2$. After simplifications and using the relation $D_\parallel + D_\perp = 2$, the equation for $\rho$ becomes
\begin{equation}
0 = \nabla^2 \rho - {\rm Pe} \nabla \cdot \left[ (1-\rho){\bf m} \right] = - \nabla \cdot {\bf J}.
\end{equation}
Since we observe a MIPS state, there is no steady current, ${\bf J} = 0$, and then the steady-state magnetization is
\begin{equation}
{\bf m}  = \frac{\nabla \rho}{{\rm Pe}(1-\rho)}.
\end{equation}

From Eq.~\eqref{eqmps1ss}, the equation for $m_x$ writes
\begin{equation}
0 = D_\parallel \partial_x \left[ (1 - \rho) \partial_x m_x + m_x \partial_x \rho \right] + D_\perp \partial_y \left[ (1 - \rho) \partial_y m_x + m_x \partial_y \rho \right] - \frac{{\rm Pe}}{2} \partial_x \left[ (1 - \rho) \rho \right] - 4 m_x.
\end{equation}
Using $m_x = \partial_x \rho/{\rm Pe}(1-\rho)$, we get
\begin{equation}
0 =  \partial_x \left[ \frac{D_\parallel}{{\rm Pe}} \partial_{xx} \rho + \frac{2D_\parallel(\partial_x\rho)^2}{{\rm Pe(1-\rho)}} - \frac{{\rm Pe}}{2} (1 - \rho) \rho + \frac{4}{{\rm Pe}} \ln(1-\rho) \right] + \partial_y \left[ \frac{D_\perp}{{\rm Pe}} \partial_{xy} \rho + \frac{2D_\perp(\partial_x\rho) (\partial_y\rho)}{{\rm Pe(1-\rho)}} \right]. \label{eqmx}
\end{equation}

From Eq.~\eqref{eqmps1ss}, the equation for $m_y$ writes
\begin{equation}
0 = D_\parallel \partial_y \left[ (1 - \rho) \partial_y m_y + m_y \partial_y \rho \right] + D_\perp \partial_x \left[ (1 - \rho) \partial_x m_y + m_y \partial_x \rho \right] - \frac{{\rm Pe}}{2} \partial_y \left[ (1 - \rho) \rho \right] - 4 m_y.
\end{equation}
Using $m_y = \partial_y \rho/{\rm Pe}(1-\rho)$, we get
\begin{equation}
0 =  \partial_y \left[ \frac{D_\parallel}{{\rm Pe}} \partial_{yy} \rho + \frac{2D_\parallel(\partial_y\rho)^2}{{\rm Pe(1-\rho)}} - \frac{{\rm Pe}}{2} (1 - \rho) \rho + \frac{4}{{\rm Pe}} \ln(1-\rho) \right] + \partial_x \left[ \frac{D_\perp}{{\rm Pe}} \partial_{xy} \rho + \frac{2D_\perp(\partial_x\rho) (\partial_y\rho)}{{\rm Pe(1-\rho)}} \right]. \label{eqmy}
\end{equation}

Performing $\partial_x$\eqref{eqmx}$-\partial_y$\eqref{eqmy}, we get
\begin{align}
0 &=  (\partial_{xx} - \partial_{yy}) \left\{ \frac{D_\parallel}{2{\rm Pe}} (\partial_{xx}+\partial_{yy}) \rho + \frac{D_\parallel}{{\rm Pe(1-\rho)}} [(\partial_x\rho)^2 + (\partial_y\rho)^2] - \frac{{\rm Pe}}{2} (1 - \rho) \rho + \frac{4}{{\rm Pe}} \ln(1-\rho) \right\} \nonumber \\
&+ (\partial_{xx} + \partial_{yy}) \left\{ \frac{D_\parallel}{2{\rm Pe}} (\partial_{xx}-\partial_{yy}) \rho + \frac{D_\parallel}{{\rm Pe(1-\rho)}} [(\partial_x\rho)^2 - (\partial_y\rho)^2] \right\}.
\end{align}
We take the new coordinates $u=x+y$ and $v=x-y$, for which $\partial_x = (\partial_u + \partial_v)/2$ and $\partial_y = (\partial_u - \partial_v)/2$. We get
\begin{align}
0 &=  \partial_u\partial_v \left\{ \frac{D_\parallel}{8{\rm Pe}} (\partial_{uu}+\partial_{vv}) \rho + \frac{D_\parallel}{4{\rm Pe(1-\rho)}} [(\partial_u\rho)^2 + (\partial_v\rho)^2] - \frac{{\rm Pe}}{2} (1 - \rho) \rho + \frac{4}{{\rm Pe}} \ln(1-\rho) \right\} \nonumber \\
&+ \frac{1}{4}(\partial_{uu} + \partial_{vv}) \left\{ \frac{D_\parallel}{2{\rm Pe}} \partial_{uv} \rho + \frac{D_\parallel}{{\rm Pe(1-\rho)}} [\partial_u\rho\partial_v\rho] \right\}.
\end{align}
The solution is then symmetric under the transformation $u \leftrightarrow v$, as observed from numerical solutions. If we choose a solution such that $\partial_v \rho = 0$, the quantity
\begin{equation}
g(u)= - \frac{D_\parallel}{8{\rm Pe}} \partial_{uu} \rho - \frac{D_\parallel}{4{\rm Pe(1-\rho)}} (\partial_u\rho)^2 + \frac{{\rm Pe}}{2} (1 - \rho) \rho - \frac{4}{{\rm Pe}} \ln(1-\rho),
\end{equation}
is constant. We define the quantities $\kappa(\rho)$, $\Lambda(\rho)$ and $g_0(\rho)$ such that
\begin{equation}
g= - \kappa(\rho) \partial_{uu} \rho + \Lambda(\rho) (\partial_u\rho)^2 + g_0(\rho).\label{defgu}
\end{equation}
Since $g$ is constant, we have a first relation between the two binodals: $g= g_0(\rho_{\rm high}) = g_0(\rho_{\rm low})$. A second relation has to be found to determine the values of $\rho_{\rm high}$ and $\rho_{\rm low}$. We consider the quantity
\begin{equation}
I = \int_{u_{\rm low}}^{u_{\rm high}} du g(u) \partial_u R(u)
\end{equation}
integrated between the regions of low densities and high densities, where $R(\rho)$ is a monotonic function to determine. Since $R(\rho)$ is monotonic, we can calculate the quantity $I$ by doing a change of variable, and we get $I= g [R(\rho_{\rm high}) - R(\rho_{\rm low})]$. Using the definition of $g(u)$ given by Eq.~\eqref{defgu}, we obtain
\begin{equation}
I = \Phi(R_{\rm high}) - \Phi(R_{\rm low}) + \frac{1}{2}\int_{u_{\rm low}}^{u_{\rm high}} du \left[ 2R'(\rho)\Lambda(\rho) + \kappa(\rho) R''(\rho) + R'(\rho)\kappa'(\rho)\right] (\partial_{uu} \rho)^3,
\end{equation}
with $\Phi'(R) = g_0(R)$. Choosing the function $R(\rho)$ such that $(\kappa R')' = -2R' \Lambda$, the remaining integral vanishes, and $I = \Phi(R_{\rm high}) - \Phi(R_{\rm low})$. We may then define a second relation $h_0(\rho_{\rm high}) = h_0(\rho_{\rm low})$ with $h_0(R) = g_0(R) R - \Phi(R)$. Using the definitions $\kappa(\rho) = D_\parallel/8 {\rm Pe}$ and $\Lambda(\rho) = - D_\parallel/4 {\rm Pe} (1-\rho)$, we can take the monotonic function
\begin{equation}
R(\rho) = \frac{1}{(1-\rho)^3}.
\end{equation}

\begin{figure}[t]
\centering
\includegraphics[width=\columnwidth]{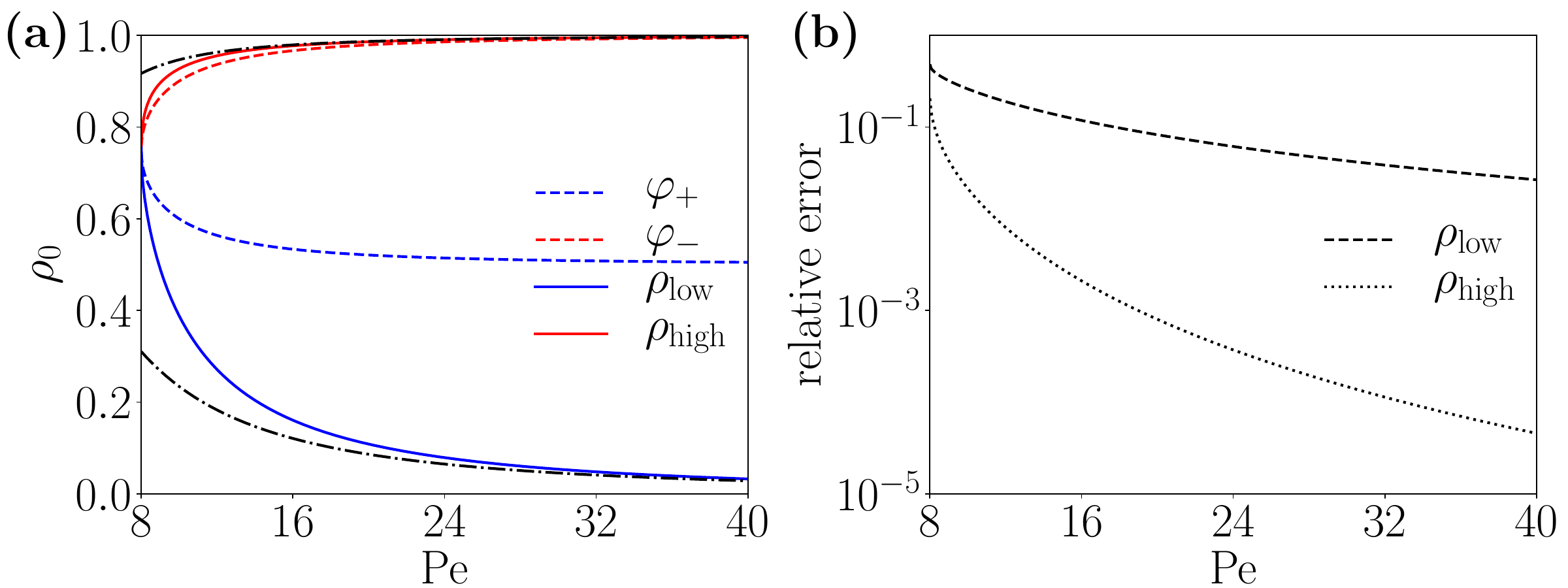}
\caption{(color online) Spinodals and binodals of the rAPM with MPS=1. (a)~The dotted lines show the spinodal lines given by Eq.~\eqref{spinodalapp}, and the solid lines show the binodals calculated from the relations $g_0(\rho_{\rm high}) = g_0(\rho_{\rm low})$ and $h_0(\rho_{\rm high}) = h_0(\rho_{\rm low})$, with Eqs.~\eqref{defg0} and~\eqref{defh0}. The dash-dotted lines display the asymptotic value of the binodals at large P\'eclet number, given by Eqs.~\eqref{rhighlim} and~\eqref{rlowlim}. (b)~Relative error between the binodal densities and their asymptotic values at large P\'eclet number.} 
\label{fig_MPS1_coex}
\end{figure}

After simplifications, we get the relations $g_0(\rho_{\rm high}) = g_0(\rho_{\rm low})$ and $h_0(\rho_{\rm high}) = h_0(\rho_{\rm low})$, with
\begin{gather}
g_0(\rho) = \frac{{\rm Pe}}{2} (1 - \rho) \rho - \frac{4}{{\rm Pe}} \ln(1-\rho),\label{defg0}\\
h_0(\rho) = \frac{{\rm Pe}}{4} \frac{3-4\rho}{(1-\rho)^2} + \frac{4}{3{\rm Pe}} \frac{1}{(1-\rho)^3} \label{defh0}.
\end{gather}

\begin{table}[H]
\begin{center}
\begin{tabular}{ |c|c|c|c|c|c|c|c|c| } 
\hline
Pe & $8$ & $9$ & $10$ & $11$ & $12$ & $13$ & $14$ \\
\hline
$\rho_{\rm low}$ & $0.75$ & $0.491$ & $0.389$ & $0.321$ & $0.271$ & $0.234$ & $0.204$ \\
\hline
$\rho_{\rm high}$ & $0.75$ & $0.896$ & $0.927$ & $0.944$ & $0.955$ & $0.963$ & $0.969$ \\
\hline
\end{tabular}
\caption{Theoretical values of the binodals for MPS=1, for several P\'eclet numbers.\label{tabmps1}}
\end{center}
\end{table}

Table~\ref{tabmps1} shows the values of $\rho_{\rm low}$ and $\rho_{\rm high}$ for small ${\rm Pe}$ values. At the large P\'eclet number, we derive the asymptotic expression of $\rho_{\rm low}$ and $\rho_{\rm high}$ as follows. We know the leading order: $\rho_{\rm low}=0$ and $\rho_{\rm high}=1$. For $\rho_{\rm low}=0$, we get the value $h_0(0) \simeq 3{\rm Pe}/4$. We search the sub-leading order as $\rho_{\rm high}\simeq 1-\xi$, with $\xi \ll 1$. At leading order, we get
\begin{equation}
h_0(1-\xi) \simeq -\frac{{\rm Pe}}{4\xi^2} + \frac{4}{3{\rm Pe} \xi^3} \simeq \frac{3{\rm Pe}}{4},
\end{equation}
which says that the diverging terms in $\xi$ must cancel: $\xi = 16/3{\rm Pe}^2$. Then we have:
\begin{equation}
\rho_{\rm high}\simeq 1-\frac{16}{3{\rm Pe}^2}. \label{rhighlim}
\end{equation}
For $\rho_{\rm low} \ll 1$, we find that $g_0(\rho_{\rm low}) \simeq {\rm Pe} \rho_{\rm low}/2$. Since $g_0(\rho_{\rm high}) \simeq -(4/{\rm Pe}) \ln (16/3{\rm Pe}^2)$, we get
\begin{equation}
\rho_{\rm low}\simeq-\frac{8}{{\rm Pe}^2} \ln \frac{16}{3{\rm Pe}^2}. \label{rlowlim}
\end{equation}

Figure~\ref{fig_MPS1_coex}(a) shows the spinodals given by Eq.~\eqref{spinodalapp}, and the binodals calculated with Eqs.~\eqref{defg0} and~\eqref{defh0}. The asymptotic value of the binodals at large P\'eclet number, given by Eqs.~\eqref{rhighlim} and~\eqref{rlowlim}, is also represented. Figure~\ref{fig_MPS1_coex}(b) shows the relative error between the binodal densities and their asymptotic values at large P\'eclet number, decreasing to zero when ${\rm Pe} \to \infty$.

%%%%%%%%%%%%%%%%%%%%%%%%%%%%%%%%%%%%%%%
%%%% hydrodynamic theory for MPS>1 %%%%
%%%%%%%%%%%%%%%%%%%%%%%%%%%%%%%%%%%%%%%

\section{Linear stability analysis for ${\rm MPS}>1$}
\label{ls_hc}

For soft-core rAPM, the hydrodynamic equation is
\begin{gather}
\partial_t \rho_\sigma = D_\parallel \partial_\parallel \left[ (1-\ess \rho) \partial_\parallel \rho_\sigma + \ess \rho_\sigma \partial_\parallel \rho \right] + D_\perp \partial_\perp \left[ (1-\ess\rho)  \partial_\perp \rho_\sigma + \ess \rho_\sigma \partial_\perp \rho  \right] - v \partial_\parallel \left[ (1-\ess\rho) \rho_\sigma \right] \nonumber \\
+ \sum_{\sigma'\ne\sigma} \left[ \frac{4\beta J}{\rho}(\rho_\sigma + \rho_{\sigma'}) - 1 - \frac{r}{\rho} - \alpha \frac{(\rho_\sigma - \rho_{\sigma'})^2}{\rho^2} \right](\rho_\sigma - \rho_{\sigma'}),
\end{gather}
with $\alpha = 8(\beta J)^2(1-2\beta J/3)$ and $\ess = 1/{\rm MPS}$. The homogeneous solutions are given by:
\begin{equation}
I_{\rm flip}(\sigma,\sigma') = \left[ \frac{4\beta J}{\rho}(\rho_\sigma + \rho_{\sigma'}) - 1 - \frac{r}{\rho} - \alpha \frac{(\rho_\sigma - \rho_{\sigma'})^2}{\rho^2} \right](\rho_\sigma - \rho_{\sigma'}) = 0,
\end{equation}
and are then those of the unrestricted rAPM~\cite{APM-2}. The disordered homogeneous solution is $\rho_\sigma = \rho_0 / 4$, and the ordered homogeneous solution (supposed along state $\sigma = 1$) is $\rho_1 = \rho_0 (1+3M) / 4$ and $\rho_{2,3,4} = \rho_0 (1-M) / 4$ with the magnetization $M$ following the equation:
\begin{equation}
\label{eqMag}
2\beta J (1+M) - 1 - \frac{r}{\rho_0} - \alpha M^2 = 0,
\end{equation}
or $M=M_0 \pm M_1 \delta$ with $M_0 = \beta J / \alpha$, $M_1 = \sqrt{r/\alpha \rho_*}$ and $\delta = \sqrt{(\rho_0-\rho_*)/\rho_0}$, where $\rho_*$ defined by
\begin{equation}
\rho_* = \frac{8(1-2\beta J/3) r}{1 + 8(2\beta J - 1)(1-2\beta J/3)},
\end{equation}
is the critical density below which the ordered homogeneous solution does not exist, for a temperature below $T_c = (1 - \sqrt{22}/8)^{-1} \simeq 2.417$.

\medskip

{\bf Linear stability analysis for the disordered homogeneous solution.} We take $\rho_\sigma = \rho_0/4 + \delta \rho_\sigma$ and $\rho = \rho_0 + \delta \rho$, with $\delta \rho = \sum_\sigma \delta \rho_\sigma$. The hopping term writes
\begin{gather}
I_{\rm hop} \simeq \left[ \left( 1 - \frac{3\ess\rho_0}{4}\right) (D_\parallel \partial_\parallel^2 + D_\perp \partial_\perp^2) - v \left( 1 - \frac{5\ess\rho_0}{4}\right) \partial_\parallel \right] \delta \rho_\sigma \nonumber \\
+ \frac{\ess\rho_0}{4} \left[ D_\parallel \partial_\parallel^2 + D_\perp \partial_\perp^2 + v \partial_\parallel \right] \sum_{\sigma' \ne \sigma} \delta \rho_{\sigma'}, \label{IhopMPS}
\end{gather}
and the flipping term writes $I_{\rm flip}(\sigma,\sigma') \simeq \mu_0 (\delta \rho_\sigma - \delta \rho_{\sigma'})$, with $\mu_0 = 2\beta J - 1 - r/\rho_0$. Then, in the Fourier space, the hydrodynamic equation becomes
\begin{equation}
\partial_t \delta\rho_\sigma = \left[A(k_\parallel,k_\perp) + 3 \mu_0 \right] \delta \rho_\sigma + \left[B(k_\parallel,k_\perp)- \mu_0\right] \sum_{\sigma' \ne \sigma} \delta \rho_{\sigma'},
\end{equation}
with
\begin{gather}
A(k_\parallel,k_\perp) = \left( 1 - \frac{3\ess\rho_0}{4}\right) (-D_\parallel k_\parallel^2 - D_\perp k_\perp^2) + \imath k_\parallel v \left( 1 - \frac{5\ess\rho_0}{4}\right)  \\
B(k_\parallel,k_\perp) = \frac{\ess\rho_0}{4} \left[ -D_\parallel k_\parallel^2 - D_\perp k_\perp^2 - \imath k_\parallel v \right] 
\end{gather}

The stability of the homogeneous disordered solution is then given by the eigenvalues of the matrix
\begin{equation}
M_{\rm gas} = \begin{pmatrix}
A(k_x,k_y) + 3 \mu_0 & B(k_x,k_y) - \mu_0 & B(k_x,k_y) - \mu_0 & B(k_x,k_y) - \mu_0\\
B(k_y,-k_x) - \mu_0 & A(k_y,-k_x) + 3 \mu_0 & B(k_y,-k_x) - \mu_0 & B(k_y,-k_x) - \mu_0\\
B(-k_x,-k_y) - \mu_0 & B(-k_x,-k_y) - \mu_0 & A(-k_x,-k_y) + 3 \mu_0 & B(-k_x,-k_y) - \mu_0\\
B(-k_y,k_x) - \mu_0 & B(-k_y,k_x) - \mu_0 & B(-k_y,k_x) - \mu_0 & A(-k_y,k_x) + 3 \mu_0
\end{pmatrix}.
\end{equation}

Supposing $k_x=k$ and $k_y=0$ (since no preferred direction), at leading order in $k \ll 1$, the eigenvalues, calculated with \textit{Mathematica}~\cite{mathematica}, are: $\lambda_{\rm gas}^{1,2,3} \simeq 4 \mu_0$ and
\begin{equation}
\lambda_{\rm gas}^4 \simeq \left[-\frac{D_\parallel +D_\perp}{2} + (1- \ess\rho_0) (1-2\ess\rho_0) \frac{v^2}{8\mu_0} \right] k^2.
\end{equation}
The disordered homogeneous solution is then stable if $\mu_0<0$ and
\begin{equation}
\label{lambdaGasMPS}
\lambda_{\rm gas} = -D + (1- \ess \rho_0)(1-2\ess \rho_0) \frac{v^2}{8\mu_0}<0  .
\end{equation}

\begin{figure}[t]
\centering
\includegraphics[width=\columnwidth]{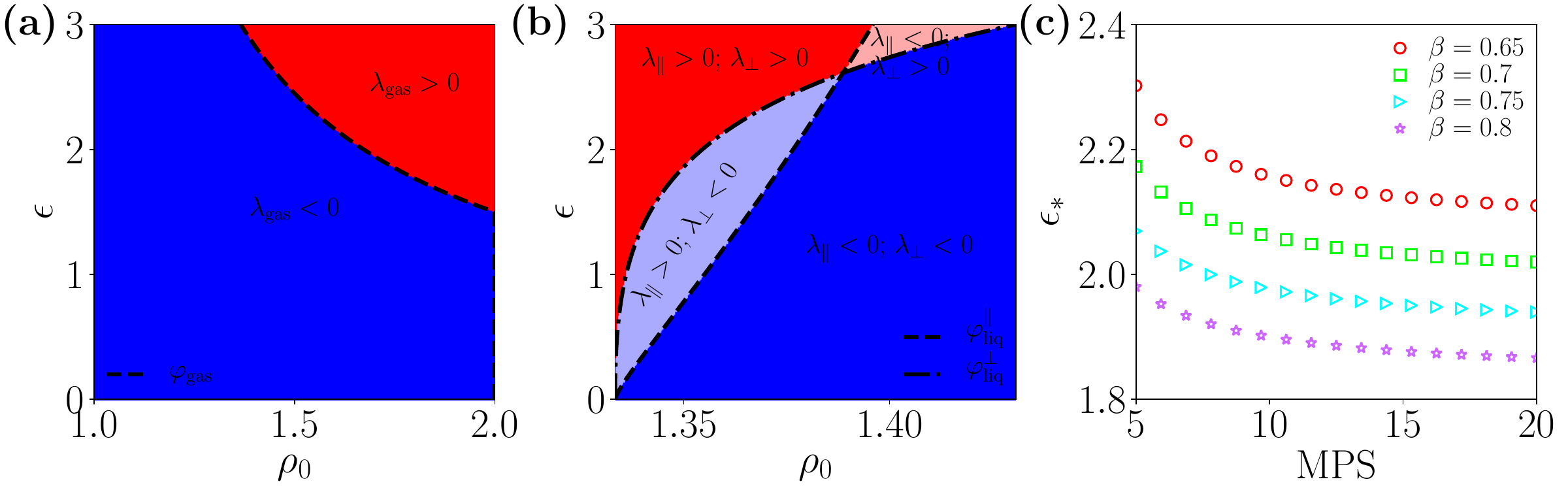}
\caption{(color online) (a,b)~Velocity-density stability diagram of the rAPM with hard-core repulsion for the disordered and ordered homogeneous solution for $\beta = 0.75$ for ${\rm MPS}=2$. (a) The stability region of the disordered solution is plotted in blue, according to the eigenvalue $\lambda_{\rm gas}$ given by Eq.~\eqref{lambdaGasMPS}. (b) The stability region of the ordered solution is plotted in blue, according to the eigenvalues $\lambda_\parallel$ and $\lambda_\perp$ given by Eqs.~\eqref{lambdaParaMPS} and~\eqref{lambdaPerpMPS}, respectively. The ordered solution is unstable only under longitudinal perturbations ($\lambda_\parallel > 0$) in the light blue region, and under transverse perturbations ($\lambda_\perp > 0$) in the light red region. (c)~$\epsilon_*$ value for which the reorientation transition occurs in the rAPM with MPS$>$, as a function of MPS for several temperatures $T=\beta^{-1}$.} 
\label{fig_MPS_stab}
\end{figure}

In Fig.~\ref{fig_MPS_stab}(a), we have represented the velocity-density stability diagram for $\beta=0.75$ and ${\rm MPS}=2$ according to the sign of this eigenvalue. Note that the last inequality is always fulfilled when $\ess=0$, meaning that $\mu_0<0$ impacts only the spinodals: $\varphi_{\rm gas}(\ess=0)=r/(2\beta J -1)$, independent of $\epsilon$. However, we cannot extract an analytical expression for the spinodal $\varphi_{\rm gas}$ for all ${\rm MPS}$ values.

\medskip

{\bf Linear stability analysis for the ordered homogeneous solution.} We consider the ordered solution along the right state, and we take $\rho_1 = \rho_0 (1+3M)/4 + \delta \rho_1$, $\rho_{2,3,4} = \rho_0 (1+3M)/4 + \delta \rho_{2,3,4}$ and $\rho = \rho_0 + \delta \rho$, with $\delta \rho = \sum_\sigma \delta \rho_\sigma$. The hopping term of the right state $\sigma=1$ writes
\begin{gather}
I_{\rm hop}^{(1)} \simeq \left\{ \left[ 1 - \frac{3\ess\rho_0}{4}(1-M)\right] (D_\parallel \partial_\parallel^2 + D_\perp \partial_\perp^2) - v \left[ 1 - \frac{\ess\rho_0}{4}(5+3M)\right] \partial_\parallel \right\} \delta \rho_\sigma \nonumber \\
+ \frac{\ess \rho_0}{4}(1+3M) \left[ D_\parallel \partial_\parallel^2 + D_\perp \partial_\perp^2 + v \partial_\parallel \right] \sum_{\sigma' \ne \sigma} \delta \rho_{\sigma'},
\end{gather}
and the hopping term of the other states $\sigma\ne 1$ writes
\begin{gather}
I_{\rm hop}^{(2)} \simeq \left\{ \left[ 1 - \frac{\ess\rho_0}{4}(3+M)\right] (D_\parallel \partial_\parallel^2 + D_\perp \partial_\perp^2) - v \left[ 1 - \frac{\ess\rho_0}{4}(5-M)\right] \partial_\parallel \right\} \delta \rho_\sigma \nonumber \\
+ \frac{\ess\rho_0}{4}(1-M) \left[ D_\parallel \partial_\parallel^2 + D_\perp \partial_\perp^2 + v \partial_\parallel \right] \sum_{\sigma' \ne \sigma} \delta \rho_{\sigma'}.
\end{gather}
We may note that for $M=0$, we recover the expression of $I_{\rm hop}$, given by Eq.~\eqref{IhopMPS}, calculated for the disordered solution.

The flipping terms implying the right state write
\begin{equation}
I_{\rm flip}(1,\sigma') \simeq M\left\{ (4\beta J - 2\alpha M) \delta \rho_1 + (4\beta J + 2\alpha M) \delta \rho_{\sigma'} - \left[ 2 \beta J (1+M) - \frac{r}{\rho_0} - 2 \alpha M^2\right] \delta \rho \right\}.
\end{equation}
Using Eq.~\eqref{eqMag}, we get
\begin{equation}
I_{\rm flip}(1,\sigma') \simeq M\left[ (4\beta J - 2\alpha M) \delta \rho_1 + (4\beta J + 2\alpha M) \delta \rho_{\sigma'} + (\alpha M^2 -1) \delta \rho \right] \equiv \gamma_1 \delta \rho_1 + \gamma_2 \delta \rho_{\sigma'} + \gamma_3 \delta \rho.
\end{equation}
The flipping terms which do not imply the right state write
\begin{equation}
I_{\rm flip}(\sigma,\sigma') \simeq M ( \alpha M - 4 \beta J) (\delta \rho_\sigma - \delta \rho_{\sigma'}) \equiv \gamma_4 (\delta \rho_\sigma - \delta \rho_{\sigma'}).
\end{equation}
Then we have the terms:
\begin{equation}
\sum_{\sigma' \ne \sigma} I_{\rm flip}(\sigma,\sigma') \simeq
\begin{dcases}
3(\gamma_1 + \gamma_3) \delta \rho_\sigma + (\gamma_2 + 3 \gamma_3) \sum_{\sigma' \ne \sigma} \rho_{\sigma'}, \qquad &{\rm if}~\sigma=1,\\
- (\gamma_1 + \gamma_3) \delta \rho_1 + (-\gamma_2-\gamma_3+2 \gamma_4) \delta \rho_{\sigma} - (\gamma_3+\gamma_4)\sum_{\sigma' \ne \{1,\sigma\}} \rho_{\sigma'}, \qquad & {\rm if}~\sigma\ne 1.
\end{dcases}
\end{equation}

Then, in the Fourier space, the hydrodynamic equation becomes
\begin{equation}
\partial_t \delta\rho_\sigma =
\begin{dcases}
\left[ A_1(k_\parallel,k_\perp) + 3\mu \right] \delta \rho_\sigma + \left[ B_1(k_\parallel,k_\perp) + \nu \right] \sum_{\sigma' \ne \sigma} \delta \rho_{\sigma'}, \qquad &{\rm if}~\sigma=1,\\
\left[ A_2(k_\parallel,k_\perp) + \kappa \right] \delta \rho_{\sigma} + \left[ B_2(k_\parallel,k_\perp) - \mu \right]\delta \rho_1 + \left[ B_2(k_\parallel,k_\perp) - \frac{\kappa + \nu}{2} \right] \sum_{\sigma' \ne \{1,\sigma\}} \delta \rho_{\sigma'}, \qquad & {\rm if}~\sigma\ne 1,
\end{dcases}
\end{equation}
with $\mu = \gamma_1 + \gamma_3 = M(4\beta J - 2 \alpha M + \alpha M^2 - 1)$, $\nu = \gamma_2 + 3 \gamma_3 = M(4\beta J + 2 \alpha M + 3\alpha M^2 - 3)$, $\kappa = - \gamma_2 - \gamma_3 + 2 \gamma_4 = M(-12\beta J - \alpha M^2 +1)$, and
\begin{gather}
A_1(k_\parallel,k_\perp) = \left[ 1 - \frac{3\ess\rho_0}{4}(1-M)\right] (-D_\parallel k_\parallel^2 - D_\perp k_\perp^2) + \imath k_\parallel v \left[ 1 - \frac{\ess\rho_0}{4}(5+3M)\right], \\
B_1(k_\parallel,k_\perp) = \frac{\ess\rho_0}{4}(1+3M) \left[ -D_\parallel k_\parallel^2 - D_\perp k_\perp^2 - \imath k_\parallel v \right], \\
A_2(k_\parallel,k_\perp) =  \left[ 1 - \frac{\ess\rho_0}{4}(3+M)\right] (-D_\parallel k_\parallel^2 - D_\perp k_\perp^2) + \imath k_\parallel v \left[ 1 - \frac{\ess\rho_0}{4}(5-M)\right], \\
B_2(k_\parallel,k_\perp) = \frac{\ess\rho_0}{4}(1-M) \left[ -D_\parallel k_\parallel^2 - D_\perp k_\perp^2 - \imath k_\parallel v \right].
\end{gather}

The stability of the homogeneous ordered solution is then given by the eigenvalues of the matrix
\begin{equation}
M_{\rm liq} = \begin{pmatrix}
A_1(k_x,k_y) + 3\mu & B_1(k_x,k_y)+\nu & B_1(k_x,k_y)+\nu & B_1(k_x,k_y)+\nu \\
B_2(k_y,-k_x)-\mu & A_2(k_y,-k_x)+\kappa & B_2(k_y,-k_x)- (\kappa + \nu)/2 & B_2(k_y,-k_x)- (\kappa + \nu)/2 \\
B_2(-k_x,-k_y)-\mu & B_2(-k_x,-k_y)- (\kappa + \nu)/2 & A_2(-k_x,-k_y)+\kappa & B_2(-k_x,-k_y)- (\kappa + \nu)/2 \\
B_2(-k_y,k_x)-\mu & B_2(-k_y,k_x)- (\kappa + \nu)/2 & B_2(-k_y,k_x)- (\kappa + \nu)/2 & A_2(-k_y,k_x)+\kappa
\end{pmatrix}.
\end{equation}

First, we consider a perturbation in the $x$ direction ($k_y=0$). At leading order in $k_x \ll 1$, the real part of the eigenvalues, calculated with \textit{Mathematica}~\cite{mathematica}, is $\lambda_{\rm liq,x}^{1,2} = (3\kappa+\nu)/2$, $\lambda_{\rm liq,x}^3 \simeq (3\mu+\nu)$, and
\begin{gather}
\lambda_{{\rm liq}, x}^4 \simeq  - \left[ \frac{(D_\parallel + 2D_\perp) \mu - D_\parallel \nu}{3\mu - \nu} + \frac{D_\parallel-D_\perp}{2(3\mu-\nu)}[\mu(1+3M)+\nu(1-M)] \ess \rho_0 + \frac{(c_1+c_2 \ess \rho_0)(1-\ess \rho_0)}{(3\mu-\nu)^3(3\kappa+\nu)} v^2\right] k_x^2,
\end{gather}
with $c_1=4\mu[-3\mu^2+\nu(2\mu+4\kappa+\nu)]$, and
\begin{equation}
c_2=\kappa\nu^2(1-M) + \mu[3\mu^2(7-3M)-2\kappa\nu(11+3M)-\nu^2(3+5M)+9\kappa\mu(1+3M)+ 2\mu\nu(-7+9M)].\nonumber
\end{equation}

Now we look at a perturbation in the $y$ direction ($k_x=0$). At leading order in $k_y \ll 1$, the real part of the eigenvalues, calculated with \textit{Mathematica}~\cite{mathematica}, is $\lambda_{\rm liq,y}^{1,2} = (3\kappa+\nu)/2$, $\lambda_{\rm liq,y}^3 \simeq (3\mu+\nu)$, and
\begin{gather}
\lambda_{{\rm liq}, y}^4 \simeq \left[ \frac{-(2 D_\parallel + D_\perp) \mu + D_\perp \nu}{3\mu - \nu} + \frac{D_\parallel-D_\perp}{2(3\mu-\nu)}[\mu(1+3M)+\nu(1-M)] \ess \rho_0 +  \frac{(d_1+d_2 \ess \rho_0)(1-\ess \rho_0)}{(3\mu-\nu)(3\kappa+\nu)}v^2 \right] k_y^2,
\end{gather}
with $d_1=4\mu$, and $d_2=\mu(-7+3M)+\nu(1-M)$.

The ordered homogeneous solution is then stable if $3\kappa+\nu<0$ and $3\mu-\nu<0$ for the two different perturbations. This result was already observed for the unrestricted APM~\cite{APM}, and allows the selection of the position magnetization solution: $M=M_0 + M_1 \delta$. However, the stability of the two different perturbations differs from $\lambda_{{\rm liq}, x}^4$ and $\lambda_{{\rm liq}, y}^4$. The perturbation along $x$ is stable only if
\begin{gather}
\lambda_\parallel = \frac{\lambda_{{\rm liq}, x}^4}{k_x^2} = -D + \left[\frac{\mu+\nu}{3\mu-\nu}(1-\ess \rho_0) - M\ess \rho_0 \right] \frac{D\epsilon}{3} - \frac{(c_1+c_2 \ess \rho_0)(1-\ess \rho_0)}{(3\mu-\nu)^3(3\kappa+\nu)}  \left(\frac{4D\epsilon}{3} \right)^2 \label{lambdaParaMPS}
\end{gather}
is negative and the perturbation along $y$ is stable only if
\begin{equation}
\label{lambdaPerpMPS}
\lambda_\perp = \frac{\lambda_{{\rm liq}, y}^4}{k_y^2} = -D - \left[\frac{\mu+\nu}{3\mu-\nu}(1-\ess \rho_0) - M\ess \rho_0 \right] \frac{D\epsilon}{3} +  \frac{(d_1+d_2 \ess \rho_0)(1-\ess \rho_0)}{(3\mu-\nu)(3\kappa+\nu)} \left(\frac{4D\epsilon}{3} \right)^2
\end{equation}
is negative. We may note that these eigenvalues are those obtained in Ref.~\cite{APM} for $\ess=0$. In Fig.~\ref{fig_MPS_stab}(b), we have represented the velocity-density stability diagram for $\beta=0.75$ and ${\rm MPS}=2$ according to the sign of these two eigenvalues. Here, we have not derived an analytical expression of $\epsilon_*$, for which the reorientation transition occurs, but we have computed a numerical estimation in Fig.~\ref{fig_MPS_stab}(c). $\epsilon_*$ is an decreasing function of MPS, meaning that $\epsilon_*$ increases with a strong repulsion.

%%%%%%%%%%%%%%%%%%%%%%%%%%%%%%%%%%%%%%%%%%%%%%%%%%%
%%%% hydrodynamic theory for the softcore rAPM %%%%
%%%%%%%%%%%%%%%%%%%%%%%%%%%%%%%%%%%%%%%%%%%%%%%%%%%

\section{Linear stability analysis for the soft-core rAPM}
\label{ls_sc}

For soft-core rAPM, the hydrodynamic equation is
\begin{gather}
\partial_t \rho_\sigma = D_\parallel \partial_\parallel \left[ \exp(-s\rho) \left( \partial_\parallel \rho_\sigma + s \rho_\sigma \partial_\parallel \rho \right) \right] + D_\perp \partial_\perp \left[ \exp(-s\rho) \left( \partial_\perp \rho_\sigma + s \rho_\sigma \partial_\perp \rho \right) \right] - v \partial_\parallel \left[ \exp(-s\rho) \rho_\sigma \right] \nonumber \\
+ \sum_{\sigma'\ne\sigma} \left[ \frac{4\beta J}{\rho}(\rho_\sigma + \rho_{\sigma'}) - 1 - \frac{r}{\rho} - \alpha \frac{(\rho_\sigma - \rho_{\sigma'})^2}{\rho^2} \right](\rho_\sigma - \rho_{\sigma'}),
\end{gather}
with $\alpha = 8(\beta J)^2(1-2\beta J/3)$ and $s=2\beta U$. The homogeneous solutions are given by:
\begin{equation}
I_{\rm flip}(\sigma,\sigma') = \left[ \frac{4\beta J}{\rho}(\rho_\sigma + \rho_{\sigma'}) - 1 - \frac{r}{\rho} - \alpha \frac{(\rho_\sigma - \rho_{\sigma'})^2}{\rho^2} \right](\rho_\sigma - \rho_{\sigma'}) = 0,
\end{equation}
and are then those of the unrestricted rAPM~\cite{APM-2}. The disordered homogeneous solution is $\rho_\sigma = \rho_0 / 4$, and the ordered homogeneous solution (supposed along state $\sigma = 1$) is $\rho_1 = \rho_0 (1+3M) / 4$ and $\rho_{2,3,4} = \rho_0 (1-M) / 4$ with the magnetization $M$ following the equation:
\begin{equation}
\label{eqMag2}
2\beta J (1+M) - 1 - \frac{r}{\rho_0} - \alpha M^2 = 0,
\end{equation}
or $M=M_0 \pm M_1 \delta$ with $M_0 = \beta J / \alpha$, $M_1 = \sqrt{r/\alpha \rho_*}$ and $\delta = \sqrt{(\rho_0-\rho_*)/\rho_0}$, where $\rho_*$ defined by
\begin{equation}
\rho_* = \frac{8(1-2\beta J/3) r}{1 + 8(2\beta J - 1)(1-2\beta J/3)},
\end{equation}
is the critical density below which the ordered homogeneous solution does not exist, for a temperature below $T_c = (1 - \sqrt{22}/8)^{-1} \simeq 2.417$.

\medskip

{\bf Linear stability analysis for the disordered homogeneous solution.} We take $\rho_\sigma = \rho_0/4 + \delta \rho_\sigma$ and $\rho = \rho_0 + \delta \rho$, with $\delta \rho = \sum_\sigma \delta \rho_\sigma$. The hopping term writes
\begin{gather}
I_{\rm hop} \simeq \exp(-s\rho_0) \left[ \left( 1 + \frac{s\rho_0}{4}\right) (D_\parallel \partial_\parallel^2 + D_\perp \partial_\perp^2) - v \left( 1 - \frac{s\rho_0}{4}\right) \partial_\parallel \right] \delta \rho_\sigma \nonumber \\
+ \exp(-s\rho_0) \frac{s\rho_0}{4} \left[ D_\parallel \partial_\parallel^2 + D_\perp \partial_\perp^2 + v \partial_\parallel \right] \sum_{\sigma' \ne \sigma} \delta \rho_{\sigma'}, \label{Ihop}
\end{gather}
and the flipping term writes $I_{\rm flip}(\sigma,\sigma') \simeq \mu_0 (\delta \rho_\sigma - \delta \rho_{\sigma'})$, with $\mu_0 = 2\beta J - 1 - r/\rho_0$. Then, in the Fourier space, the hydrodynamic equation becomes
\begin{equation}
\partial_t \delta\rho_\sigma = \left[A(k_\parallel,k_\perp) + 3 \mu_0 \right] \delta \rho_\sigma + \left[B(k_\parallel,k_\perp)- \mu_0\right] \sum_{\sigma' \ne \sigma} \delta \rho_{\sigma'},
\end{equation}
with
\begin{gather}
A(k_\parallel,k_\perp) = \exp(-s\rho_0) \left[ \left( 1 + \frac{s\rho_0}{4}\right) (-D_\parallel k_\parallel^2 - D_\perp k_\perp^2) + \imath k_\parallel v \left( 1 - \frac{s\rho_0}{4}\right) \right] \\
B(k_\parallel,k_\perp) = \exp(-s\rho_0) \frac{s\rho_0}{4} \left[ -D_\parallel k_\parallel^2 - D_\perp k_\perp^2 - \imath k_\parallel v \right] 
\end{gather}

The stability of the homogeneous disordered solution is then given by the eigenvalues of the matrix
\begin{equation}
M_{\rm gas} = \begin{pmatrix}
A(k_x,k_y) + 3 \mu_0 & B(k_x,k_y) - \mu_0 & B(k_x,k_y) - \mu_0 & B(k_x,k_y) - \mu_0\\
B(k_y,-k_x) - \mu_0 & A(k_y,-k_x) + 3 \mu_0 & B(k_y,-k_x) - \mu_0 & B(k_y,-k_x) - \mu_0\\
B(-k_x,-k_y) - \mu_0 & B(-k_x,-k_y) - \mu_0 & A(-k_x,-k_y) + 3 \mu_0 & B(-k_x,-k_y) - \mu_0\\
B(-k_y,k_x) - \mu_0 & B(-k_y,k_x) - \mu_0 & B(-k_y,k_x) - \mu_0 & A(-k_y,k_x) + 3 \mu_0
\end{pmatrix}.
\end{equation}

\begin{figure}[t]
\centering
\includegraphics[width=\columnwidth]{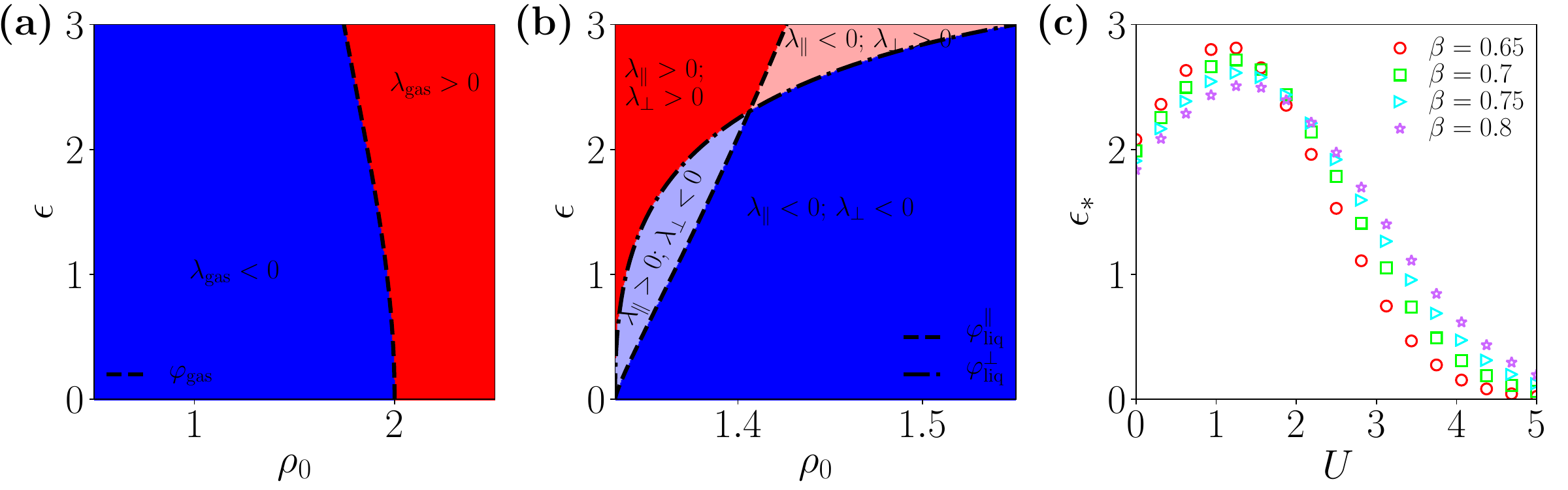}
\caption{(color online) {\bf (a)}~and {\bf (b)}~Velocity-density stability diagram of the rAPM with soft-core repulsion for the disordered and ordered homogeneous solution for $\beta = 0.75$ and $U=0.5$. {\bf (a)} The stability region of the disordered solution is plotted in blue, according to the eigenvalue $\lambda_{\rm gas}$ given by Eq.~\eqref{lambdaGas}. {\bf (b)} The stability region of the ordered solution is plotted in blue, according to the eigenvalues $\lambda_\parallel$ and $\lambda_\perp$ given by Eqs.~\eqref{lambdaPara} and~\eqref{lambdaPerp}, respectively. The ordered solution is only unstable under longitudinal perturbations ($\lambda_\parallel > 0$) in the light blue region, and under transverse perturbations ($\lambda_\perp > 0$) in the light red region. {\bf (c)}~$\epsilon_*$ value for which the reorientation transition occurs, as a function of $U$ for several temperatures $T=\beta^{-1}$.} 
\label{fig_softcore_stab}
\end{figure}

Supposing $k_x=k$ and $k_y=0$ (since no preferred direction), at leading order in $k \ll 1$, the eigenvalues, calculated with \textit{Mathematica}~\cite{mathematica}, are: $\lambda_{\rm gas}^{1,2,3} \simeq 4 \mu_0$ and
\begin{gather}
\lambda_{\rm gas}^4 \simeq \exp(-s\rho_0) \left[-(1+s\rho_0)\frac{D_\parallel +D_\perp}{2} + (1- s\rho_0) \exp(-s\rho_0) \frac{v^2}{8\mu_0} \right] k^2.
\end{gather}
The disordered homogeneous solution is then stable if $\mu_0<0$ and
\begin{equation}
\label{lambdaGas}
\lambda_{\rm gas} = -(1+s\rho_0)D + (1- s\rho_0) \exp(-s\rho_0) \frac{v^2}{8\mu_0}<0.
\end{equation}
In Fig.~\ref{fig_softcore_stab}(a), we have represented the velocity-density stability diagram for $\beta=0.75$ and $U=0.5$ according to the sign of this eigenvalue. Note that the last inequality is always fulfilled when $U=0$, meaning that $\mu_0<0$ only impacts the spinodals: $\varphi_{\rm gas}(U=0)=r/(2\beta J -1)$, independent of $\epsilon$. However, we cannot extract an analytical expression for the spinodal $\varphi_{\rm gas}$ for all $U$ values.

\medskip

{\bf Linear stability analysis for the ordered homogeneous solution.} We consider the ordered solution along the right state, and we take $\rho_1 = \rho_0 (1+3M)/4 + \delta \rho_1$, $\rho_{2,3,4} = \rho_0 (1+3M)/4 + \delta \rho_{2,3,4}$ and $\rho = \rho_0 + \delta \rho$, with $\delta \rho = \sum_\sigma \delta \rho_\sigma$. The hopping term of the right state $\sigma=1$ writes
\begin{gather}
I_{\rm hop}^{(1)} \simeq \exp(-s\rho_0) \left\{ \left[ 1 + \frac{s\rho_0}{4}(1+3M)\right] (D_\parallel \partial_\parallel^2 + D_\perp \partial_\perp^2) - v \left[ 1 - \frac{s\rho_0}{4}(1+3M)\right] \partial_\parallel \right\} \delta \rho_\sigma \nonumber \\
+ \exp(-s\rho_0) \frac{s\rho_0}{4}(1+3M) \left[ D_\parallel \partial_\parallel^2 + D_\perp \partial_\perp^2 + v \partial_\parallel \right] \sum_{\sigma' \ne \sigma} \delta \rho_{\sigma'},
\end{gather}
and the hopping term of the other states $\sigma\ne 1$ writes
\begin{gather}
I_{\rm hop}^{(2)} \simeq \exp(-s\rho_0) \left\{ \left[ 1 + \frac{s\rho_0}{4}(1-M)\right] (D_\parallel \partial_\parallel^2 + D_\perp \partial_\perp^2) - v \left[ 1 - \frac{s\rho_0}{4}(1-M)\right] \partial_\parallel \right\} \delta \rho_\sigma \nonumber \\
+ \exp(-s\rho_0) \frac{s\rho_0}{4}(1-M) \left[ D_\parallel \partial_\parallel^2 + D_\perp \partial_\perp^2 + v \partial_\parallel \right] \sum_{\sigma' \ne \sigma} \delta \rho_{\sigma'}.
\end{gather}
We may note for $M=0$, we recover the expression of $I_{\rm hop}$, given by Eq.~\eqref{Ihop}, calculated for the disordered solution.

The flipping terms implying the right state write
\begin{equation}
I_{\rm flip}(1,\sigma') \simeq M\left\{ (4\beta J - 2\alpha M) \delta \rho_1 + (4\beta J + 2\alpha M) \delta \rho_{\sigma'} - \left[ 2 \beta J (1+M) - \frac{r}{\rho_0} - 2 \alpha M^2\right] \delta \rho \right\}.
\end{equation}
Using Eq.~\eqref{eqMag2}, we get
\begin{equation}
I_{\rm flip}(1,\sigma') \simeq M\left[ (4\beta J - 2\alpha M) \delta \rho_1 + (4\beta J + 2\alpha M) \delta \rho_{\sigma'} + (\alpha M^2 -1) \delta \rho \right] \equiv \gamma_1 \delta \rho_1 + \gamma_2 \delta \rho_{\sigma'} + \gamma_3 \delta \rho.
\end{equation}
The flipping terms which does not imply the right state write
\begin{equation}
I_{\rm flip}(\sigma,\sigma') \simeq M ( \alpha M - 4 \beta J) (\delta \rho_\sigma - \delta \rho_{\sigma'}) \equiv \gamma_4 (\delta \rho_\sigma - \delta \rho_{\sigma'}).
\end{equation}
Then we have the terms:
\begin{equation}
I_\sigma = \sum_{\sigma' \ne \sigma} I_{\rm flip}(\sigma,\sigma') \simeq
\begin{dcases}
3(\gamma_1 + \gamma_3) \delta \rho_\sigma + (\gamma_2 + 3 \gamma_3) \sum_{\sigma' \ne \sigma} \rho_{\sigma'}, \qquad &{\rm if}~\sigma=1,\\
- (\gamma_1 + \gamma_3) \delta \rho_1 + (-\gamma_2-\gamma_3+2 \gamma_4) \delta \rho_{\sigma} - (\gamma_3+\gamma_4)\sum_{\sigma' \ne \{1,\sigma\}} \rho_{\sigma'}, \qquad & {\rm if}~\sigma\ne 1.
\end{dcases}
\end{equation}

Then, in the Fourier space, the hydrodynamic equation becomes
\begin{equation}
\partial_t \delta\rho_\sigma =
\begin{dcases}
\left[ A_1(k_\parallel,k_\perp) + 3\mu \right] \delta \rho_\sigma + \left[ B_1(k_\parallel,k_\perp) + \nu \right] \sum_{\sigma' \ne \sigma} \delta \rho_{\sigma'}, \qquad &{\rm if}~\sigma=1,\\
\left[ A_2(k_\parallel,k_\perp) + \kappa \right] \delta \rho_{\sigma} + \left[ B_2(k_\parallel,k_\perp) - \mu \right]\delta \rho_1 + \left[ B_2(k_\parallel,k_\perp) - \frac{\kappa + \nu}{2} \right] \sum_{\sigma' \ne \{1,\sigma\}} \delta \rho_{\sigma'}, \qquad & {\rm if}~\sigma\ne 1,
\end{dcases}
\end{equation}
with $\mu = \gamma_1 + \gamma_3 = M(4\beta J - 2 \alpha M + \alpha M^2 - 1)$, $\nu = \gamma_2 + 3 \gamma_3 = M(4\beta J + 2 \alpha M + 3\alpha M^2 - 3)$, $\kappa = - \gamma_2 - \gamma_3 + 2 \gamma_4 = M(-12\beta J - \alpha M^2 +1)$, and
\begin{gather}
A_1(k_\parallel,k_\perp) = \exp(-s\rho_0) \left\{ \left[ 1 + \frac{s\rho_0}{4}(1+3M)\right] (-D_\parallel k_\parallel^2 - D_\perp k_\perp^2) + \imath k_\parallel v \left[ 1 - \frac{s\rho_0}{4}(1+3M)\right] \right\}, \\
B_1(k_\parallel,k_\perp) = \exp(-s\rho_0) \frac{s\rho_0}{4}(1+3M) \left[ -D_\parallel k_\parallel^2 - D_\perp k_\perp^2 - \imath k_\parallel v \right], \\
A_2(k_\parallel,k_\perp) = \exp(-s\rho_0) \left\{ \left[ 1 + \frac{s\rho_0}{4}(1-M)\right] (-D_\parallel k_\parallel^2 - D_\perp k_\perp^2) + \imath k_\parallel v \left[ 1 - \frac{s\rho_0}{4}(1-M)\right] \right\}, \\
B_2(k_\parallel,k_\perp) = \exp(-s\rho_0) \frac{s\rho_0}{4}(1-M) \left[ -D_\parallel k_\parallel^2 - D_\perp k_\perp^2 - \imath k_\parallel v \right].
\end{gather}

The stability of the homogeneous ordered solution is then given by the eigenvalues of the matrix
\begin{equation}
M_{\rm liq} = \begin{pmatrix}
A_1(k_x,k_y) + 3\mu & B_1(k_x,k_y)+\nu & B_1(k_x,k_y)+\nu & B_1(k_x,k_y)+\nu \\
B_2(k_y,-k_x)-\mu & A_2(k_y,-k_x)+\kappa & B_2(k_y,-k_x)- (\kappa + \nu)/2 & B_2(k_y,-k_x)- (\kappa + \nu)/2 \\
B_2(-k_x,-k_y)-\mu & B_2(-k_x,-k_y)- (\kappa + \nu)/2 & A_2(-k_x,-k_y)+\kappa & B_2(-k_x,-k_y)- (\kappa + \nu)/2 \\
B_2(-k_y,k_x)-\mu & B_2(-k_y,k_x)- (\kappa + \nu)/2 & B_2(-k_y,k_x)- (\kappa + \nu)/2 & A_2(-k_y,k_x)+\kappa
\end{pmatrix}.
\end{equation}

In the following, we will denote $\overline{D_\parallel} = D_\parallel \exp(-s\rho_0)$, $\overline{D_\perp} = D_\perp \exp(-s\rho_0)$ and $\overline{v} = v \exp(-s\rho_0)$. First, we consider a perturbation in the $x$ direction ($k_y=0$). At leading order in $k_x \ll 1$, the real part of the eigenvalues, calculated with \textit{Mathematica}~\cite{mathematica}, are: $\lambda_{\rm liq,x}^{1,2} \simeq (3\kappa+\nu)/2$, $\lambda_{\rm liq,x}^3 \simeq (3\mu+\nu)$, and
\begin{gather}
\lambda_{{\rm liq}, x}^4 \simeq - \left[ \frac{(\overline{D_\parallel} + 2\overline{D_\perp}) \mu - \overline{D_\parallel} \nu}{3\mu - \nu} + \frac{\overline{D_\parallel}(1+M)+\overline{D_\perp}(1-M)}{2} s\rho_0 +\frac{c_1+c_2 s\rho_0}{(3\mu-\nu)^3(3\kappa+\nu)} \overline{v}^2\right] k_x^2,
\end{gather}
with $c_1 = 4\mu[-3\mu^2+\nu(2\mu+4\kappa+\nu)]$, and $c_2 = (3\mu-\nu) \{3\mu[\kappa(1+3M)+\mu(1-M)] - \nu[\kappa(1-M)+\mu(1-5M)] \}$.

Now we look at a perturbation in the $y$ direction ($k_x=0$). At leading order in $k_y \ll 1$, the real part of the eigenvalues, calculated with \textit{Mathematica}~\cite{mathematica}, are: $\lambda_{\rm liq,y}^{1,2} \simeq (3\kappa+\nu)/2$, $\lambda_{\rm liq,y}^3 \simeq (3\mu+\nu)$, and
\begin{equation}
\lambda_{{\rm liq}, y}^4 \simeq \left[ \frac{-(2\overline{D_\parallel} + \overline{D_\perp}) \mu + \overline{D_\perp} \nu}{3\mu - \nu} - \frac{\overline{D_\parallel}(1-M)+\overline{D_\perp}(1+M)}{2} s\rho_0 + \frac{d_1+d_2 s\rho_0}{(3\mu-\nu)(3\kappa+\nu)}  \overline{v}^2 \right] k_y^2,
\end{equation}
with $d_1 = 4\mu$, and $d_2=-(1-M)(3\mu-\nu)$.

The ordered homogeneous solution is then stable if $3\kappa+\nu<0$ and $3\mu-\nu<0$ for the two different perturbations. This result was already observed for the unrestricted APM~\cite{APM}, and allows the selection of the position magnetization solution: $M=M_0 + M_1 \delta$. However, the stability of the two different perturbations differs from $\lambda_{{\rm liq}, x}^4$ and $\lambda_{{\rm liq}, y}^4$. The perturbation along $x$ is stable only if
\begin{gather}
\lambda_\parallel = \frac{\lambda_{{\rm liq}, x}^4}{k_x^2 \exp(-s\rho_0)} = -D(1+s\rho_0) + \left[\frac{\mu+\nu}{3\mu-\nu} - M s\rho_0 \right] \frac{D\epsilon}{3} -  \frac{(c_1 + c_2 s\rho_0)\exp(-s\rho_0)}{(3\mu-\nu)^3(3\kappa+\nu)} \left(\frac{4D\epsilon}{3} \right)^2\label{lambdaPara}
\end{gather}
is negative and the perturbation along $y$ is stable only if
\begin{equation}
\label{lambdaPerp}
\lambda_\perp = \frac{\lambda_{{\rm liq}, y}^4}{k_y^2 \exp(-s\rho_0)} = -D(1+s\rho_0) - \left[\frac{\mu+\nu}{3\mu-\nu} - M s\rho_0 \right] \frac{D\epsilon}{3} +  \frac{(d_1+d_2 s\rho_0)\exp(-s\rho_0)}{(3\mu-\nu)(3\kappa+\nu)} \left(\frac{4D\epsilon}{3} \right)^2 
\end{equation}
is negative. We may note that these eigenvalues are those obtained in Ref.~\cite{APM} for $s=0$. In Fig.~\ref{fig_softcore_stab}(b), we have represented the velocity-density stability diagram for $\beta=0.75$ and $U=0.5$ according to the sign of these two eigenvalues. Here, we have not derived an analytical expression of $\epsilon_*$, for which the reorientation transition occurs, but we have computed a numerical estimation in Fig.~\ref{fig_softcore_stab}(c). $\epsilon_*(U)$ increases for small $U$ and decreases to zero at large $U$.

%%%%%%%%%%%%%%%%%%%%%%%%%%%%%%%%%%%%%%%%%%
%%%%%%%%%%%% BIBLIOGRAPHY %%%%%%%%%%%%%%%%
%%%%%%%%%%%%%%%%%%%%%%%%%%%%%%%%%%%%%%%%%%

%%%%%%%%%%%%%%%%%%%%%%%%%%%%%%%%%%%%%%%%%%%%%%%%%%%%%%%%%%%%%%%%%%%%%%%%%%%%%%%%

\begin{thebibliography}{62}

\bibitem{Marchetti2013} M. C. Marchetti, J. F. Joanny, S. Ramaswamy, T. B. Liverpool, J. Prost, M. Rao, and R. A. Simha, {\it Hydrodynamics of soft active matter}, Rev. Mod. Phys. {\bf 85}, 1143 (2013).

\bibitem{Shaebani2020}
M. R. Shaebani, A. Wysocki, R. G. Winkler, G. Gompper, and H. Rieger, \textit{Computational models for active matter,} Nat. Rev. Phys. {\bf 2}, 181 (2020). 

\bibitem{Bottinelli2016} A. Bottinelli, D. T. J. Sumpter, and J. L. Silverberg, {\it Emergent Structural Mechanisms for High-Density Collective Motion Inspired by Human Crowds}, Phys. Rev. Lett. {\bf 117}, 228301 (2016).

\bibitem{Helbing1995} D. Helbing and P. Molnar, {\it Social force model for pedestrian dynamics}, Phys. Rev. E {\bf 51}, 4282 (1995).

\bibitem{Garcimartin2015} A. Garcimartin, J. M. Pastor, L. M. Ferrer, J. J. Ramos, C. Martin-Gomez, and I. Zuriguel, {\it Flow and clogging of a sheep herd passing through a bottleneck}, Phys. Rev. E {\bf 91}, 022808
(2015).

\bibitem{Ballerini2008} M. Ballerini, N. Cabibbo, R. Candelier, A. Cavagna, E. Cisbani, I. Giardina, V. Lecomte, A. Orlandi, G. Parisi, A. Procaccini, \textit{et al.}, {\it Interaction ruling animal collective behavior depends on topological rather than metric distance: Evidence from a field study}, Proc. Natl. Acad. Sci. U. S. A {\bf 105}, 1232 (2008).

\bibitem{Beccoa2006} C. Beccoa, N. Vandewallea, J. Delcourtb, and P. Poncinb, {\it Experimental evidences of a structural and dynamical transition in fish school}, Physica A {\bf 367}, 487 (2006).

\bibitem{Calovi2014} D. S. Calovi, U. Lopez, S. Ngo, C. Sire, H. Chat\'e, and G. Theraulaz, {\it Swarming, schooling, milling: Phase diagram of a data-driven fish school model}, New J. Phys. {\bf 16}, 015026 (2014).

\bibitem{Steager2008} E. B. Steager, C.-B. Kim and M. J. Kim, {\it Dynamics of pattern formation in bacterial swarms}, Phys. Fluids {\bf 20}, 073601 (2008).

\bibitem{Peruani2012} F. Peruani, J. Starruss, V. Jakovljevic, L. Sogaard-Andersen, A. Deutsch, and M. B\"ar, {\it Collective Motion and Nonequilibrium Cluster Formation in Colonies of Gliding Bacteria}, Phys. Rev. Lett. {\bf 108}, 098102 (2012).

\bibitem{Giavazzi2018} F. Giavazzi, M. Paoluzzi, M. Macchi, D. Bi, G. Scita, M. Lisa Manning, R. Cerbino and M. Cristina Marchetti, {\it Flocking transitions in confluent tissues}, Soft Matter {\bf 14}, 3471 (2018).

\bibitem{Schaller2010} V. Schaller, C. Weber, C. Semmrich, E. Frey, and A. R. Bausch, {\it Polar patterns of driven filaments}, Nature {\bf 467}, 73 (2010).

\bibitem{Sumino2012} Y. Sumino, K. H. Nagai, Y. Shitaka, D. Tanaka, K. Yoshikawa, H. Chate, and K. Oiwa, {\it Large-scale vortex lattice emerging from collectively moving microtubules}, Nature {\bf 483}, 448 (2012).

\bibitem{Sanchez2012} T. Sanchez, D. T. N. Chen, S. J. DeCamp, M Heymann, and Z. Dogic, {\it Spontaneous motion in hierarchically assembled active matter}, Nature 491, {\bf 431} (2012).

\bibitem{viscek} T. Vicsek, A. Czir\'ok, E. Ben-Jacob, I. Cohen, and O. Shochet, \textit{Novel Type of Phase Transition in a System of Self-Driven Particles,} Phys. Rev. Lett. {\bf 75}, 1226 (1995).


\bibitem{toner} J. Toner and Y. Tu, \textit{Long-Range Order in a Two-Dimensional Dynamical Model: How Birds Fly Together,} Phys. Rev. Lett. {\bf 75}, 4326 (1995).


\bibitem{toner-2} J. Toner and Y. Tu, \textit{Flocks, herds, and schools: A quantitative theory of flocking,} Phys. Rev. E {\bf 58}, 4828 (1998).


\bibitem{toner-3} J. Toner, \textit{Reanalysis of the hydrodynamic theory of fluid, polar-ordered flocks,} Phys. Rev. E {\bf 86}, 031918 (2012).


\bibitem{solon-vicsek} A. P. Solon, H. Chat\'e, and J. Tailleur, \textit{From Phase to Microphase Separation in Flocking Models: The Essential Role of Nonequilibrium Fluctuations,} Phys. Rev. Lett. {\bf 114}, 068101 (2015).

\bibitem{Chate2004} G. Gr\'egoire, and H. Chat\'e, {\it Onset of Collective and
Cohesive Motion}, Phys. Rev. Lett. {\bf 92}, 025702 (2004); H. Chat\'e, F. Ginelli, G. Gr\'egoire, and F. Raynaud, {\it Collective motion of self-propelled particles interacting without cohesion}, Phys. Rev. E {\bf 77}, 046113 (2008).

\bibitem{solon-vm2}
A. P. Solon, Jean-Baptiste Caussin, D. Bartolo, H. Chat\'e and J. Tailleur, \textit{Pattern formation in flocking models: A hydrodynamic description,} Phys. Rev. E {\bf 92}, 062111 (2015).

\bibitem{ihle} R. K\"ursten and T. Ihle, {\it Dry Active Matter Exhibits a Self-organized Cross Sea Phase}, Phys. Rev. Lett. {\bf 125}, 188003 (2020).

\bibitem{AIM} A. P. Solon and J. Tailleur, \textit{Revisiting the
Flocking Transition Using Active Spins,} Phys. Rev. Lett. {\bf 111}, 078101 (2013).

\bibitem{AIM-2} A. P. Solon and J. Tailleur, \textit{Flocking with discrete symmetry: The two-dimensional active Ising model,} Phys. Rev. E {\bf 92}, 042119 (2015).

\bibitem{ishibashi2022} K. Ishibashi and H. Sakaguchi, \textit{Solitary wave states maintained by stochastic direction changes in a population of self-propelled particles,} J. Phys. Soc. Jpn. {\bf 91}, 034003 (2022).

\bibitem{APM} S. Chatterjee, M. Mangeat, R. Paul and H. Rieger, \textit{Flocking and reorientation transition in the 4-state active Potts model,} EPL {\bf 130}, 66001 (2020).


\bibitem{APM-2} M. Mangeat, S. Chatterjee, R. Paul, and H. Rieger, \textit{Flocking with a q-fold discrete symmetry: Band-to-lane transition in the active Potts model,} Phys. Rev. E {\bf 102}, 042601 (2020).

\bibitem{ACM} S. Chatterjee, M. Mangeat, and H. Rieger, \textit{Polar flocks with discretized directions: The active clock model approaching the Vicsek model,} EPL {\bf 138}, 41001 (2022).


\bibitem{ACM-2} A. Solon, H. Chaté, J. Toner, and J. Tailleur, \textit{Susceptibility of Polar Flocks to Spatial Anisotropy,} Phys. Rev. Lett. {\bf 128}, 208004 (2022).

\bibitem{Romanczuk2012} P. Romanczuk, M. B\"ar, W. Ebeling, B. Lindner,
and L. Schimansky-Geier, {\it Active Brownian particles: From individual to collective stochastic dynamics}, Eur. Phys. J. Special Topics {\bf 202}, 1 (2012).

\bibitem{mips}
M. E. Cates and J. Tailleur, \textit{Motility-Induced Phase Separation,} Annu. Rev. Condens. Matter Phys. {\bf 6}, 219 (2015). 

\bibitem{chate2003} G. Gr\'egoire, H. Chat\'e, and Y. Tu, {\it Moving and staying together without a leader}, Physica D {\bf 181}, 157 (2003).

\bibitem{ignacio} A. Mart\'in-G\'omez, D. Levis, A. D\'iaz-Guilera, and I. Pagonabarraga, {\it Collective motion of active Brownian particles with polar alignment}, Soft Matter {\bf 14}, 2610 (2018).

\bibitem{sese2018} E. Ses\'e-Sansa, I. Pagonabarraga and D. Levis, {\it Velocity alignment promotes motility-induced phase separation}, EPL {\bf 124}, 30004 (2018).

\bibitem{peruani2011} F. Peruani, T. Klauss, A. Deutsch, and A. Voss-Boehme, \textit{Traffic Jams, Gliders, and Bands in the Quest for Collective Motion of Self-Propelled Particles,} Phys. Rev. Lett. {\bf 106}, 128101 (2011).

\bibitem{Berthier2019} L. Berthier, E. Flenner, and G. Szamel, \textit{Glassy dynamics in dense systems of active particles,} J. Chem. Phys. {\bf 150}, 200901 (2019).

\bibitem{ALG} M. Kourbane-Houssene, C. Erignoux, T. Bodineau, and J. Tailleur, \textit{Exact Hydrodynamic Description of Active Lattice Gases,} Phys. Rev. Lett. {\bf 120}, 268003 (2018).

\bibitem{kolmogorov} A. N. Kolmogorov, {\it On the theory of Markov chains}, Math. Ann. {\bfseries 112}, 155 (1936).

\bibitem{SM} See Supplemental Material for movies.

\bibitem{nagel2010} A. J. Liu and S. R. Nagel, {\it The jamming transition and the marginally jammed solid,} Annu. Rev. Condens. Matter Phys. {\bf 1}, 347 (2010).

\bibitem{odmips} P. Digregorio, D. Levis, A. Suma, L. F. Cugliandolo, G. Gonnella, and I. Pagonabarraga, {\it Full Phase Diagram of Active Brownian Disks: From Melting to Motility-Induced Phase Separation}, Phys. Rev. Lett. {\bf 121}, 098003 (2018); M. N. van der Linden, L. C. Alexander, D. G. A. L. Aarts, and O. Dauchot, {\it Interrupted Motility Induced Phase Separation in Aligning Active Colloids}, Phys. Rev. Lett. {\bf 123}, 098001 (2019); C. B. Caporusso, P. Digregorio, D. Levis, L. F. Cugliandolo, and G. Gonnella, {\it Motility-Induced Microphase and Macrophase Separation in a Two-Dimensional Active Brownian Particle System}, Phys. Rev. Lett. {\bf 125}, 178004 (2020).

\bibitem{geyer2019} D. Geyer, D. Martin, J. Tailleur, and D. Bartolo, \textit{Freezing a Flock: Motility-Induced Phase Separation in Polar Active Liquids,} Phys. Rev. X {\bf 9}, 031043 (2019).

\bibitem{sese2021} E. Ses\'e-Sansa, D. Levis, and I. Pagonabarraga, {\it Phase separation of self-propelled disks with ferromagnetic and nematic alignment}, Phys. Rev. E {\bf 104}, 054611 (2021).

\bibitem{levis2017} D. Levis, J. Codina, and I. Pagonabarraga , \textit{Active Brownian equation of state: Metastability and phase coexistence,} Soft Matter {\bf 13}, 8113 (2017).


\bibitem{kerner1997} B. S. Kerner and H. Rehborn, \textit{Experimental Properties of Phase Transitions in Traffic Flow,} Phys. Rev. Lett. {\bf 79}, 4030, (1997).

\bibitem{Marchetti2011} S. Henkes, Y. Fily, and M. Cristina Marchetti, \textit{Active jamming: Self-propelled soft particles at high density,} Phys. Rev. E {\bf 84}, 040301(R) (2011).

\bibitem{freefem} F. Hecht, \textit{New development in FreeFem++}, J. Numer. Math. 20, 251 (2012).


\bibitem{fem} O. C. Zienkiewicz, R. L Taylor, P. Nithiarasu and J. Z. Zhu, The Finite Element Method, (McGraw-Hill, London, 1977).

\bibitem{zenodo} M. Karmakar, S. Chatterjee, M. Mangeat, H. Rieger, and R. Paul, {\itshape Jamming and flocking in the restricted active Potts model}, \href{https://doi.org/10.5281/zenodo.7942534}{https://doi.org/10.5281/zenodo.7942534} (2023).

\bibitem{Weinrib1983} A. Weinrib and B. I. Halperin, \textit{Critical phenomena in systems with long-range-correlated quenched disorder,} Phys. Rev. B {\bf 27}, 413 (1983).


\bibitem{Viktor1995} V. S. Dotsenko, \textit{Critical phenomena and quenched disorder,} Phys. Usp. {\bf 38}, 457 (1995).


\bibitem{Kumar2017} M. Kumar, S. Chatterjee, R. Paul, and S. Puri, \textit{Ordering kinetics in the random-bond XY model,} Phys. Rev. E {\bf 96}, 042127 (2017).

\bibitem{Reichhardt2021} P. Forg\'acs, A. Lib\'al, C. Reichhardt, and C.J.O Reichhardt, \textit{Active matter shepherding and clustering in inhomogeneous environments,} Phys. Rev. E {\bf 104}, 044613 (2021).

\bibitem{Reichhardt2014} C. Reichhardt and C. J. O. Reichhardt, \textit{Active matter transport and jamming on disordered landscapes}, Phys. Rev. E {\bf 90}, 012701 (2014).


\bibitem{RFrAPM} M. Karmakar, S. Chatterjee, M. Mangeat, H. Rieger, and R. Paul, \textit{Disordered media and its impact on self-propelled particles: a study
using active Potts model} (unpublished).

\bibitem{Merrigan2020} C. Merrigan, K. Ramola, R. Chatterjee, N. Segall, Y. Shokef, and B. Chakraborty, \textit{Arrested states in persistent active matter: Gelation without attraction,} Phys. Rev. R {\bf 2}, 013260 (2020).

\bibitem{puglisi2020} L. Caprini, U. Marini Bettolo Marconi, and A. Puglisi, \textit{Spontaneous Velocity Alignment in Motility-Induced Phase Separation,} Phys. Rev. Lett. {\bf 124}, 078001 (2020).

\bibitem{farrell2012} F. D. C. Farrell, M. C. Marchetti, D. Marenduzzo, and J. Tailleur, \textit{Pattern Formation in Self-Propelled Particles with Density-Dependent Motility,} Phys. Rev. Lett. {\bf 108}, 248101 (2012).

\bibitem{mathematica} Wolfram Research, Inc., \textit{Mathematica}, Version 11.2, Champaign, IL (2017).

\end{thebibliography}
\end{document}